\documentclass[a4paper,11pt]{article}
\pdfoutput=1 
\usepackage{jcappub} % for details on the use of the package, please
\usepackage[T1]{fontenc} % if needed
\usepackage{verbatim}
\usepackage{amssymb}
\usepackage{xcolor}
\usepackage{booktabs}
\usepackage{soul}

\title{\boldmath Quantum signatures and decoherence during inflation from deep subhorizon perturbations}

\author[a,b,c]{Francescopaolo Lopez,}
\author[d,e,f]{Nicola Bartolo}

\affiliation[a]{SISSA, International School for Advanced Studies,
Via Bonomea 265, 34136 Trieste, Italy}
\affiliation[b]{INFN, Sezione di Trieste,
Via Valerio 2, 34127 Trieste, Italy}
\affiliation[c]{IFPU, Istitute for Fundamental Physics of the Universe,
Via Beirut 2, 34014 Trieste, Italy}
\affiliation[d]{Dipartimento di Fisica e Astronomia ``Galileo Galilei'',
Universit\`a degli Studi di Padova, via Marzolo 8, I-35131, Padova, Italy}
\affiliation[e]{INFN, Sezione di Padova, via Marzolo 8, I-35131, Padova, Italy}
\affiliation[f]{INAF-Osservatorio Astronomico di Padova, Italy}

\emailAdd{flopez@sissa.it}
\emailAdd{nicola.bartolo@pd.infn.it}

\abstract{In order to shed light on the quantum-to-classical transition of the
primordial perturbations in single field inflation, we investigate the decoherence and associated quantum corrections to the correlation functions  of large-scale (superhorizon) scalar curvature perturbations. The latter are considered as an open quantum system which undergoes quantum decoherence 
induced by a \textit{time-dependent} environment of deep subhorizon tensorial
modes through the trilinear interactions predicted by General Relativity.  A complete calculation is a complex one and still to be fully performed. We advance in that direction, and we elaborate on various aspects that have been overlooked in the literature.
We first prove that, in full generality,  a time dependent subhorizon environment can be relevant for decoherence during inflation, by considering  derivativeless interactions, which, in our case, give the most important results.
For the first time, the time dependence of the environment is properly taken into account by modifying the quantum master equation.

Important non-Markovian effects pop up, instead, when dealing with derivative interactions. We adopt a possible way to treat them which has been recently proposed and seems well suited for our case. Our results show that when considering the interplay between derivativeless and derivative interactions, decoherence is slowed down. This underlines the importance of accounting for all the interactions in open quantum-system calculations in an inflationary setting. 

We finally compute the quantum corrections to cosmological correlation functions, by solving the transport equations induced by the quantum master equation.  We also compare the results to the solutions obtained by an alternative method previously used in the literature. We observe a resummation of the quantum corrections to the power-spectrum, which is a general property of quantum master equations. We extend these results to the bispectrum, showing a decay of this correlation function in time which is analogous to the one found, previously, for the power-spectrum. 
}

\begin{document}
\maketitle
\flushbottom
\newcommand{\red}{\textcolor{red}}

\section{Introduction}
\label{sec:intro}
Inflation~\cite{mukhanov_quantum_1981,guth_inflationary_1981,linde_chaotic_1983,Starobinsky:1980te,Albrecht:1982wi,Linde:1981mu} is currently the most widely accepted paradigm to describe the first instants of the history of the universe. A variety of cosmological data, first of all Cosmic Microwave Background data~\cite{akrami_planck_2020,Planck:2018jri} on temperature anisotropies and polarization are in agreement with the predictions of a primordial accelerated expansion driven by a single slowly rolling scalar field. For example, they are in accordance with a Gaussian stochastic distribution of primordial perturbations. However, given the upper bounds on primordial non-Gaussianity a long way is still open down to the minimal level of primordial non-Gaussianity predicted by the simplest models~\cite{Acquaviva:2002ud,bartolo_non-gaussianity_2004,maldacena_non-gaussian_2003}, whose observation is one of the future primary goals in early universe cosmology. In particular, such an observable gives crucial information on non linearities of gravity, the self-interactions of the scalar field driving inflation, and on the other (spectator) fields and the interactions with them, which should be present at least for the reheating process to be efficient.  

Despite the experimental success, the precise mechanism underlying inflation is still not known, even though various models have been already ruled out. Also, on the theoretical side, there are still some questions which are left open. This paper addresses in particular the ``quantum-to-classical transition'' problem. 

One of the most attractive features of the inflationary paradigm is the mechanism through which the anisotropies and inhomonogeneities in the present universe are generated. They should be born, actually, as tiny quantum fluctuations of the scalar field driving inflation. These fluctuations are then stretched by the accelerated expansion, in such a way to first cross the Hubble horizon, and then to correspond to scales of cosmological relevance now when they re-enter the horizon after inflation. The cosmological history of the universe,  as inferred from various observations,  is compatible with the hypotheses that these quantum fluctuations could be the seeds of the anisotropies and inhomogeneities corresponding to the present cosmological structures. However, e.g. CMB temperature anisotropies or clusters of galaxies are classical objects: all their physical parameters appear single valued, and there is no sign of a quantum coherent superposition between different values (e.g. a macroscopic Schrodinger's cat). So, how can these quantum inflationary fluctuations have become classical objects?

The importance of this question has been recognized since the first papers on inflation~\cite{guth_quantum_1985}. The general approach, in the past, has been dictated by a series of papers by Polarski, Starobinski et al. through the approach of ``decoherence without decoherence'', or intrinsic decoherence (see~\cite{polarski_semiclassicality_1996, lesgourgues_quantum_1997, albrecht_inflation_1994,kiefer_pointer_2007} and Refs therein). In the first papers, the authors used to consider the system of cosmological perturbations as a closed quantum system, unitarily and freely evolving, so neglecting interactions with other fields, and the evolution of perturbations in a squeezed quantum state. \footnote{See also~\cite{Danieli:2023dvu} for more recent works on squeezing of perturbations of spectator fields.}

However, quantum coherent superpositions between different outcomes of the system are not shown to be erased by any unitary evolution (therefore, not from a closed system framework).\footnote{As discussed, in a new series of more recent papers, also by the authors of the ``decoherence without decoherence'' papers, see e.g.
\cite{kiefer_why_2009} and references therein.} This is supported by the well-known property of unitary evolution, which naturally preserves symmetries, and cannot so evolve the initial state of the universe, assumed to be Bunch-Davies vacuum  (so naturally isotropic and homogeneous) into  a classical, inhomogeneous and anisotropic state that we observe now (even if these inhomogeneities are small).

The erasure of quantum coherent superposition can instead be made possible and explicitly computed by a well established phenomenon, widely studied and observed in laboratories in other contexts, as decoherence.

Consider a quantum system, which could be divided into two parts. If it is possible to ignore the evolution of (in the equations, this is equivalent to tracing over) one of the two parts, called environment, the net effect observed on the observable part, called system, is a non unitary evolution.

The non-unitary evolution of the system is caused by the quantum correlations which are built by interactions between the system and the environmental degrees of freedom, and provide a ``leak'' of information towards the environment.  The system is consequently called an ``open quantum system'', since it now exchanges information with the environment.

For these reasons, many efforts of the community working on the issue of the quantum-to-classical transition of inflationary perturbations have been directed towards decoherence trying to evaluate its effectiveness during inflation.\footnote{Decoherence has been evaluated also for alternative early universe cosmologies, like ekpyrosis \cite{colas_decoherence_2024}, and non-conventional inflationary scenario, as an ultra slow-roll background \cite{Brahma:2024ycc}.}

 The general answer is that even in the case of not considering any additional interaction (or self interaction) apart from gravity, and no other fields besides the inflaton, decoherence can be effective for a system of modes as large as CMB modes, for many of the allowed scales of inflation~\cite{burgess_minimal_2023,nelson_quantum_2016,gong_quantum_2019,martineau_decoherence_2007}. 
A typical environment to be considered in single field inflation can be an environment of short wavelength modes of the inflaton perturbations, keeping into account that we only observe long wavelength modes, which therefore in this case would play the role of the system.

The unavoidable presence of spectator fields (just as the Standar Model ones) can only  (in general) speed up the process of decoherence.\footnote{See however recent studies on the possibility to have a ``recoherence process'' due to some specific couplings~\cite{Colas:2022kfu,Colas:2024ysu}.} Thus, decoherence seems likely to play a crucial role in the quantum-to-classical transition of inflationary perturbations. 

The theory of open quantum systems is now well established in a condensed matter/quantum information context~\cite{Breuer:2007juk}. The ``equation of motion'' that can explain the decoherence of the system is called the ``quantum master equation''. In general, many approximations must be done in order to derive this equation; they are customary in a condensed matter setting,  but their translation to cosmology seems not straightforward. Two major differences, for example, are the fact that in cosmology, contrary to the laboratory situations, environments are not thermal, and the background is not static, but dynamic, being an expanding, quasi de Sitter, universe.

 The most discussed approximation, both because of the dramatic simplification it brings in the calculation and its key theoretical importance, is the Markovian approximation.
 
 The Markovian approximation can be applied if  the environment entails very ``fast'' degrees of freedom, in such a way that the typical time of environmental correlation function is much smaller than the typical time of evolution of the system due to the interactions with the environment. This is pictorially referred to by affirming that the environment has no, or really short, memory. 

In particular, many recent papers~\cite{colas_benchmarking_2022, Burgess:2024eng,burgess_minimal_2023} have struggled in order to translate the Markovian approximation in cosmology, and the community has not given a definitive and general answer. Arguably, one of the main goals of the research in this topic in the next few years is checking the limits of validity of the Markovian approximation, and see, if necessary, how it is possible to go beyond it.  For some works in this direction, see, e.g. {\cite{Shandera:2017qkg,Colas:2022kfu,Zarei:2021dpb}}.

On the other hand, decoherence is not the only quantity which can be investigated through the open quantum system approach. As a consequence of an increasing interest, recent works have started investigating the importance of the consequences of entanglement in an inflationary setting, with respect to many different aspects ~\cite{Pueyo:2024twm,Green:2024cmx,Alicki:2023tfz,Alicki:2023rfv}.  It is natural to expect that the interactions, and the entanglement with the environmental degrees of freedom  which leads to decoherence may also change the value of some observables  related to the system, inducing some quantum signatures of the process of decoherence.  In particular here we are referring to the decoherence-induced contributions to cosmological correlators, such as power-spectra or higher-order correlation functions~\cite{deKruijf:2024ufs,martin_observational_2018, martin_non_2018, daddi_hammou_cosmic_2023}. 
If observed and singled out, these contributions would also be an indirect proof of the quantum origin of perturbations, complementary to other possible probes~\cite{Arkani-Hamed:2015bza,Micheli:2022tld,Maldacena:2015bha,Bhattacharyya:2024duw,Campo:2005qn,martin_real-space_2021,Tejerina-Perez:2024opu,dePutter:2019xxv,Martin:2021znx,Martin:2022kph,Espinosa-Portales:2022yok,gomez_quantum_2023}. 

In particular, the corrections to observables are (at least) second order in the coupling constant between the system and the environment. For this reason, and others which will be clear in the following, we will call these corrections ``loop corrections''. Papers that computed quantum corrections to the power-spectrum of inflationary perturbations  (see, e.g.~\cite{brahma_universal_2022, brahma_quantum_2022, burgess_minimal_2023})
by employing quantum master equations usually found very small corrections, far beyond the present level of observations. However, they looked specifically to the minimal case of cubic non-Gaussian interactions present in the General Relativity (GR) framework (as those that can be found in, e.g.,~\cite{maldacena_non-gaussian_2003}).\footnote{Other works looked for contributions to the power-spectrum due to stronger interactions, e.g. due to some general interactions with some spectator fields, finding signatures that are more promising from an observational point of view~\cite{martin_observational_2018, boyanovsky_effective_2015,colas_benchmarking_2022}.}. The coupling constants are really suppressed in this case, because they are of the order of:
\begin{equation}
\left(\frac{H}{M_{\rm Pl}}\right)^2 \lesssim 10^{-10},
\end{equation}
from {\it Planck} data. Here, and in the following, we consider the reduced Planck Mass $M_{\rm Pl}\simeq2.4 \times10^{18} \rm{GeV}$. In the future, however, it would be a reasonable development to consider specific well-known and well-motivated inflation models characterized by stronger non-Gaussian interactions. 

There is another interesting aspect of the use of quantum master equations. Often loop corrections computed with the so called in-in formalism (see, e.g.,~\cite{Weinberg:2005vy}) used to present some secular growth of perturbations~\cite{Seery:2010kh}. Many resummations techniques have been employed in the past to resum these logarithmics secular divergence or various regularizations have been proposed to cure them (see, e.g.,~\cite{Green:2020txs,Boyanovsky:1998aa, Burgess:2009bs}).

In this respect, the quantum master equations are known  (from condensed matter) for resumming perturbative results which  otherwise would have shown a breaking in perturbation theory. In particular, the resummation is considered by finite renormalization of the energy levels of the system, expressed by a correction of the Hamiltonian due to the dressing by the environment, called Lamb Shift.\footnote{But see ~\cite{Burgess:2024eng} for a discussion on the resummation of purity through the quantum master equation.} This resummation has already been shown to be effective in cosmology in some particular cases~\cite{boyanovsky_effective_2015,colas_benchmarking_2022}. It is remarkable that, by carefully taking into account the interactions (and the non unitary effects, as entanglement) with an unobservable environment when computing observable, we automatically obtain corrections which are resummed (so IR-safe).

The paper is organized as follows. In Section~\ref{sec:Our framew} we present the framework of our work, i.e. system ~\ref{sec:freeevofpert}, environment~\ref{subsection:env}, interactions~\ref{subsec:interaction}. In Section \ref{sec:derivativelesscanonical} we compute the level of decoherence achieved in our scenario and the corrections to the power-spectrum of the scalar perturbations by the derivativeless interactions. We then compute, for this case, the effect of a time dependent cutoff for the environment in section \ref{subsec:timedependentenvironment}. In Section \ref{sec:derivativeinteractions} we include also derivative interactions and consider the interplay between the various interactions.  We give then full results for all of them and we discuss the need for the application of the ``Strong Markovian Approximation'' (SMAR). In section ~\ref{sec:boy} we first discuss the application of the Boyakovsky method~\cite{boyanovsky_effective_2015}, and the comparison of the results with our case. Then, we extend the same method to compute the quantum corrections to the bispectrum. Finally, Section~\ref{Conclusions} contains our conclusions.
In the Appendices, we consider a brief summary of the concept of decoherence and its application in cosmology; then there are some technical calculations for the boundary terms associated to each couple of interactions.

However, with the aim of facilitating the reader below here we make a summary of our main results, comparison with literature, and main (new) aspects addressed. The corresponding various parts are also summarized in Table~\ref{tab:contents}.

\subsection{Our work: novelties and comparison with previous literature}
\label{sec:bulletpoints}
\begin{itemize}  

\item \textbf{Framework and goals}

Our paper aims at computing decoherence and quantum corrections to the power-spectrum and bispectrum of inflationary curvature perturbations, further investigating the minimal, limited to GR, case of non-Gaussian cubic interactions, as explicitly written down in~\cite{maldacena_non-gaussian_2003}. We considered a mixed framework, i.e. an  environment provided by tensor fluctuations acting on a scalar system, with the interactions among one scalar and two gravitons.  This case was only analyzed, in previous works, by~\cite{burgess_minimal_2023}. However, a remarkable difference between the environment considered in their work and in ours is that we consider a {\it time-dependent}  environment, featuring only {\it deep subhorizon modes}. We also, for the first time in this context, evaluate the interplay within the different interactions, showing that we can obtain very different results if we don't consider it.

Although we recognize to be far from a definitive result, we mostly aim to shed light on some aspects of such calculations, and to pose questions on others. We believe this is important, in order to contribute to the collective effort of the community to arrive to a widely accepted, free of ambiguities, way of computing decoherence and quantum corrections, to be compared in some future with possible experiments and data.

In the following we explain the main novelties introduced in our work.

 \item \textbf{Environment and cutoff}
 
  In~\cite{burgess_minimal_2023}, the authors considered a fixed cutoff for the environment at a certain scale $k_{UV}$, which then naturally will cross the horizon during inflation at a certain point (see fig.(\ref{fig:fixedenvironment})). Consequently, the environment features also superhorizon modes. In particular, the authors analyzed the region where the modes of the environment are superhorizon, and argued that this is the most effective one in the environment for decoherence. In this paper we try to address a complementary calculation, explicitly evaluating the effect of an environment of deep subhorizon modes on the decoherence of superhorizon modes.
 
 For simplicity, we decided to consider an environment composed of ``deep subhorizon modes'', i.e.  fluctuations modes which at every instant of conformal time $\eta=-1/aH$ have a momentum which is $k > 10aH$, see fig.(\ref{fig:whatmissing}). The choice of the factor 10 is absolutely arbitrary. As a work hypotheses, we also decided to consider the environmental correlation functions that the fields would have had as in Minkowski, so neglecting the effect of the curvature of de Sitter spacetime (only on the environment), with the idea that they would have been subleading, since the environment is chosen in such a way that its degrees of freedom are always well inside the horizon. 

 One of the consequence of an environment featuring mostly superhorizon modes, as in~\cite{burgess_minimal_2023}, is that only one interaction out of the three in~\cite{maldacena_non-gaussian_2003}, the one with spatial derivatives applied to the environmental field, and without any time derivatives, is dominant, since the latter are suppressed after horizon crossing. Our choice of considering subhorizon modes does not imply a priori any hierarchy between the different interactions, and leads us to consider the interplay between all of them in eq.(\ref{eq:interactions}), also the ones featuring term with time derivatives.

First, we will rewrite the interactions in~\cite{maldacena_non-gaussian_2003} using canonical fields and conformal time in eq.(\ref{eq:interactions}). We will divide the interactions in two type: derivativeless and derivative. The interaction considered by~\cite{burgess_minimal_2023} is a derivative one. 

We will show that, in our framework, the derivativeless interaction in eq.(\ref{eq:interactions}) gives, considered by itself alone, the most important corrections to cosmological correlators and decoherence. This interaction has never been studied explicltly in the context of a  subhorizon environment acting on superhorizon modes in a  single field inflation scenario. However, it has been taken into consideration by many models  in the literature featuring a conformally coupled scalar field as an environment, considered as a proxy for a subhorizon environment~\cite{Boyanovsky:2015xoa,Burgess:2024eng, Brahma:2024yor}. We dedicate section 3 just to study its effects.
\newline

\item \textbf{Discussing non-Markovianity in Inflation}

In order to extract the decoherence rate, we need to compute the real part of the coefficient of the quantum master equation. It is also known that when this coefficient is positive at every time the dynamics is physical for sure. In full generality, in the literature, the presence of a negative coefficient is associated to non-Markovianity~\cite{Breuer_2016,Hall_2014}. However, in some situation, if the coefficient becomes negative the density matrix may evolve in an unphysical way. In particular, this coefficient has been found negative also in a cosmological framework, as noticed first in~\cite{burgess_minimal_2023} for a derivative interaction.\footnote{See eq. (3.33) in~\cite{burgess_minimal_2023}.} One of the reason why this happens could be that the usual Markovian approximation applied in the laboratory is not suitable for the cosmological applications, i.e. it does not neglect the memory effects in a consistent way. This motivated the authors of~\cite{burgess_minimal_2023} (following also the work in~\cite{Kaplanek:2022xrr}) to look for what we call from now on ``strong Markovian approximation'' (SMAR)\footnote{See paragraph 3.2.2. of~\cite{burgess_minimal_2023}.}, as a possible prescription to extract the Markovian contribution in a consistent way, thus obtaining a positive coefficient, and consequently compute the corresponding decoherence rate out of the quantum master equation. 

Considering only the derivativeless interactions, this coefficient turns out to be positive in our calculations (see eq.(\ref{eq:derivativelessresult})), without any further approximation. 

Thus, in the first place, introducing the Strong Markovian approximation is not necessary. We believe that this result encourages more investigation on decoherence from subhorizon modes.
\newline

\item \textbf{Interplay with different interactions}

The conclusion reached in the previous paragraph will change when considering derivative interactions.
We checked that couples of derivative interactions, gave negligible contributions. This is probably due to the approximation we make in considering Minkowski modes in the environment and won't likely happen in a full calculation. 

However, considering also the interplay between different interactions is another novelty of this work. In our calculations, since decoherence is second order in the couplings, we had also terms with two different interactions (inside the two different commutation parenthesis in the quantum master equation, see e.g. eq.(\ref{eq:nonmarkqme})). The mixed derivative-derivativeless terms that appear give a noticeable result, of the order of the derivativeless one, both in the corrections to the cosmological observables and in the decoherence, underlying  the importance of considering also these mixed terms even when we would expect that one interaction gives the most relevant results. 
The results for the real part of the coefficient of the quantum master equation are of opposite sign with respect to the derivativeless interactions. In the end, by considering all of the contributions, we obtain a negative coefficient, thus coming back to the problem of \cite{burgess_minimal_2023}.

At this point, we will apply the ``strong Markovian approximation'': it will be able to extract a positive coefficient when applied to the full calculation. Thus, we benchmark this prescription, showing that it seems to work well also in our case. In this way, we find a real positive coefficient in the end, which guarantees a physically well-behaved quantum master equation in this framework, and also to extract a decoherence rate. Nevertheless, the problem of the adequate treatment of non-Markovianity, in particular for derivative interactions, is still waiting for a full solution in an inflationary setting. 
\newline

\item \textbf{Time dependent environment: computing the effects}

We point out that employing a time-dependent environment in a consistent way requires to modify accordingly the derivation of the quantum master equation, as also suggested by~\cite{burgess_gravity_2023}. Since as an environment we are considering (tensor) fluctuations modes which at {\it every instant} of conformal time $\eta=-1/aH$ have a momentum which is $k > 10aH$, then such an environment unavoidably is time-dependent. To be more specific, the environment is time-dependent in the sense that some modes initially belonging to the environment will leave it after being stretched by inflation and become bigger than the horizon. In section ~\ref{subsec:timedependentenvironment}, we first show how by carefully taking into account the time dependence of the cutoff in the quantum master equation brings to adding additional terms to the quantum master equation. Notice that these additional terms preserve the canonical form of the quantum master equation, thus just modifying the coefficients with respect to the usual derivation. This is not trivial at all: changing the environment continuosly in time does not guarantee to have a quantum map of the same form as with a fixed environment. The final result is so encoded in a slight modification of the coefficients. 
 In this case, in order to obtain this result, we had to use the ``Strong Markovian Approximation''.  At least for the derivativeless interaction, 
 in the end, we show that for our values of the parameters and for our scopes, e.g. extracting the decoherence and the contributions to the power-spectrum, the modifications can be neglected. But to our knwowledge this is the first time that this corrections are evaluated, and in other situations, however, the modifications can have a bigger influence on the final results. 
\newline

\item \textbf{Corrections to cosmological observables}

The other goal of our paper is to study the corrections to the power-spectrum (and higher-order correlators) of the large-scale curvature perturbations. We first consider the Lamb shift contributions.
The corrections we obtain are modifications of the spectral index, without any running (contrarily to the case with an environment with a conformally coupled field in~\cite{boyanovsky_effective_2015}). We employed the method by~\cite{colas_benchmarking_2022}, showing, by summing over all the interaction contributions, an overall blue correction to the scalar spectral index. Also in this case, such a global blue correction is the sum of a blue correction due to derivativeless interactions and a red one of the derivative interactions.

As already pointed out in other works, e.g.~\cite{burgess_minimal_2023, martin_observational_2018,martin_non_2018,daddi_hammou_cosmic_2023,deKruijf:2024ufs},  it is possible to find also different corrections
 which are related to the non-unitary terms in the dynamics; these, instead of simply modifying the tilt, add another contribution to the power-spectrum. We explicitly verify this, by showing that the corrections coming from the derivativeless interaction are of the same form and order as the ones computed in~\cite{burgess_minimal_2023}, notwithstanding the (very) different environment employed and the different interactions. 
 \newline

\item  \textbf{Comparison to the Boyanovsky method and Bispectrum corrections}
 
 We perform again the calculations for the Lamb Shift corrections, in our framework, employing a slight modification of the method applied by Boyanovsky in~\cite{boyanovsky_effective_2015}. The latter is an approximated method, in which the contribution by the decaying mode is neglected. We thus extend the comparison already performed by~\cite{colas_benchmarking_2022, Brahma:2024yor} where, respectively, bilinear and trilinear interactions for an environment of a scalar field has been considered, to a single field inflaton framework, so with a different environment, and also employing derivative interactions. Both of the works have shown little differences in the results for the Lamb Shift computed solving directly the quantum master equation. We confirm the same result for derivativeless interactions, in the case of a time dependent environment. Following this success, we apply for the first time this setting to the bispectrum, finding a slow decay in time of this correlation function analogous to the one obtained for the power-spectrum. This means that in general we should expect similar effects of the Lamb Shift Hamiltonian on bispectra, and motivates further investigations in the direction of the higher-order correlation functions. Bispectrum and trispectrum have already be analyzed in the open quantum system settings by~\cite{ daddi_hammou_cosmic_2023, martin_non_2018}, but looking at the non-unitary corrections, and by~\cite{Salcedo:2024smn}, but with different scopes. 

 We also consider, for the first time, the extension of the Boyanovsky method to the case of derivative interactions. We then compared the results we obtained with those obtained by directly solving the quantum master equation. In this case, the correspondence is not great: we obtained fairly different results, thus revealing the difficulties of the Boyanovsky method with derivativate interactions. 

 Notice that the memory in the quantum master equation depends heavily on the decaying mode. In this sense, Boyanovksy applies a sort of ``Markovian approximation'' by neglecting the decaying mode. Thus, this result confirms that the correct Markovian approximation is not well assessed in the context where derivative interactions are present.
\newline

\end{itemize}

 \medskip

\begin{table}[]
\begin{tabular}{|lllllll|}
\hline
\multicolumn{6}{|l|}{Interactions}                                           & \ref{subsec:interaction} \\ \hline
\multicolumn{6}{|l|}{Discussion on Markovianity}                             & \ref{subsec:the Markovian approximation} \\ \hline
\multicolumn{6}{|l|}{Motivations for a time-dependent environment}           & \ref{subsubsec:timedependentenvironment} \\ \hline
\multicolumn{7}{|l|}{\textbf{Derivativeless interactions}}                               \\ \hline
\multicolumn{6}{|l|}{QME coefficients}                                       & \ref{sec:derivativelesscanonical} \\ \hline
\multicolumn{6}{|l|}{Corrections to the power-spectrum}                      & \ref{sec:derivativelesscanonical} \\ \hline
\multicolumn{6}{|l|}{Decoherence}                                            &\ref{subsec:decoherencederivativeless}  \\ \hline
\multicolumn{6}{|l|}{Time dependent environment corrections (applying SMAR)} &  \ref{subsec:timedependentenvironment}\\ \hline
\multicolumn{7}{|l|}{\textbf{Mixed terms, derivativeless interactions integrated}}    \\ \hline
\multicolumn{6}{|l|}{QME coefficients}                                       & \ref{sec:mixedderivterms123}    \\ \hline
\multicolumn{6}{|l|}{Corrections to the power-spectrum}                          & \ref{sec:mixedderivterms123}    \\ \hline
\multicolumn{7}{|l|}{\textbf{Mixed terms, derivative interactions integrated}}           \\ \hline
\multicolumn{6}{|l|}{QME coefficients}                                       & \ref{sec:mixedterms231} \\ \hline
\multicolumn{7}{|l|}{\textbf{Summing all the contributions: final results}}              \\ \hline
\multicolumn{6}{|l|}{QME coefficients and decoherence}                       & \ref{subsec:finalresultsdeco} \\ \hline
\multicolumn{6}{|l|}{Total Lamb Shift corrections}                           & \ref{subsec:lambshiftqme} \\ \hline
\multicolumn{7}{|l|}{\textbf{Comparison to Boyanovsky method}}                           \\ \hline
\multicolumn{6}{|l|}{Corrections to the power-spectrum}                          & \ref{sec:boyps} \\ \hline
\multicolumn{6}{|l|}{Corrections to the bispectrum}                              & \ref{sec:boybisp}  \\ \hline
\end{tabular}
\caption{Table of the most innovative aspects and discussions of the paper.}
\label{tab:contents}
\end{table}

\section{Our Framework}
\label{sec:Our framew}
\subsection{Free evolution of perturbations}
\label{sec:freeevofpert}
In this paper we will only consider a single field inflation scenario. The dynamics is described by the GR action with a minimally coupled scalar field $\varphi$:
\begin{equation}
S=\int \mathrm{d}^4 x \sqrt{-g}\left[\frac{M_{\mathrm{p}}^2}{2} R-\frac{1}{2} g^{\mu \nu} \partial_\mu \varphi \partial_\nu \varphi-V(\varphi)\right],
\label{eq:singlefieldaction}
\end{equation}
where the potential $V(\varphi)$ is the usual slow roll potential, which is sufficiently flat. Indeed, its derivatives are controlled by the here supposed small 
slow-roll parameters $\varepsilon=-\dot{H}/{H^2}\simeq (1/16 \pi G) (V^{\prime}/V)^2 \ll 1$ and $\eta_V=V^{\prime \prime}/{3 H^2} \simeq 1/(8 \pi G) (V^{\prime \prime}/V) \ll 1$, where $H$ is the Hubble parameter introduced below in eq.(\ref{scalefactor}).
In the usual semiclassical approach of inflation, it is customary to consider a classical background, isotropic and homogeneous, on top of which quantum perturbations are generated, thus separating the inflaton field as $\varphi(t, x)=\phi(t)+\delta \varphi(t, x)$.  This leads first to solve the Einstein Equations of motion for the background with classical field theory methods leading to a metric of FLRW type 
\begin{equation}
\mathrm{d} s^2=-\mathrm{d} t^2+a^2(t) \mathrm{d} x^2=a^2(\eta)\left(-\mathrm{d} \eta^2+\mathrm{d} x^2\right),
\end{equation}
where we have introduced the scale factor $a$, which in a pure de Sitter ($H=const$) is
\begin{equation}
a=e^{Ht}=-\frac{1}{H\eta},
\label{scalefactor}
\end{equation}
$\eta$ being the conformal time $a\mathrm{d}\eta=\mathrm{d} t$, whose domain is $\eta \in ]-\infty,0]$ during inflation.
In the following, we will consider a pure de Sitter background. This is consistent with a slow-roll expansion at lowest-order, since our goal is to compute corrections to cosmological observables from non-Gaussianities which are already proportional to some powers of $\varepsilon$. Every correction to the background metric will thus be subdominant.

In order to consider perturbations to this metric, we follow the notations of~\cite{maldacena_non-gaussian_2003}, by employing the ADM formalism (see, for a much more detailed discussion, appendix A in~\cite{burgess_minimal_2023}): 
\begin{equation}
\mathrm{d} s^2=-N^2 \mathrm{~d} t^2+\gamma_{i j}\left(\mathrm{~d} x^i+N^i \mathrm{~d} t\right)\left(\mathrm{d} x^j+N^j \mathrm{~d} t\right),
\end{equation}
where $N^{i}$ and $N$ are a Lagrange multiplier for the dynamical variable $\gamma_{ij}(t,{x})$, which propagates in space and time. The perturbations to the metric that we are going to consider are the scalar and the tensor ones (i.e. gravitational waves).  

Choosing to work in a comoving gauge, in which $\delta \varphi=0$, we can then write 
\begin{equation}
\gamma_{i j}=a^2 e^{2 \zeta} \hat{h}_{i j} \quad \quad {\rm with} \quad \quad \hat{h}_{i j}=\delta_{i j}+h_{i j}+\frac{1}{2} \delta^{k l} h_{i k} h_{l j}+\dots
\end{equation}
where $\zeta$ represents the comoving curvature perturbation, and $h_{ij}$ the gravitational waves with 
\begin{equation}
    \operatorname{det} \hat{h}_{i j}=1\quad \quad \delta^{i j} \partial_i h_{j k}=\delta^{i j} h_{i j}=0,
\end{equation}
$h_{ij}$ being transverse and traceless. It is important to keep in mind that $\zeta$, under certain specific conditions  (i.e. adiabatic conditions for the perturbations), is conserved on superhorizon scales, from the end of inflation to the radiation dominated epoch when the modes cross for the second time the Hubble horizon. Therefore, its spectrum from the end of inflation is directly connected to the scalar perturbations of the metric (e.g. Bardeen potential) when they reenter the horizon (see e.g.~\cite{Kodama:1984ziu}), and it is not sensitive to the arguably complicated physics happening inside the horizon in the reheating epoch. The $\zeta$ power-spectrum can be directly connected to the large scale perturbations of CMB, and give an initial condition for the computations of the cosmological perturbations. For this reason, it is important to compute the correlation functions of the  quantity $\zeta$ during inflation. 

Its quadratic action is  (see, e.g.,~\cite{maldacena_non-gaussian_2003,burgess_minimal_2023})
\begin{equation} 
     S_{free}=\int \mathrm{d} t \mathrm{~d}^3 x \varepsilon M_{\rm Pl} ^2 \left[a^3 \dot{\zeta}^2-a (\partial \zeta)^2\right].
\end{equation}

When we quantize perturbations, however, we prefer to work with a canonically normalized variable, the Sasaki-Mukhanov variable 
\begin{equation}
\label{v}
v=a \sqrt{2 \varepsilon} M_{\rm Pl}  \zeta\, .
\end{equation}
In this way the free action looks like the one of a scalar field\footnote{Another action can be written by applying a canonical transformation. It differs just for boundary terms from this one, i.e. integrating by parts the $a''$, and it is considered in many papers, such as~\cite{polarski_semiclassicality_1996} and~\cite{colas_benchmarking_2022}. }
\begin{equation}
 S_{free}=\frac{1}{2} \int \mathrm{d}^4 x\left[\left(v^{\prime}\right)^2-(\nabla v)^2+\frac{a^{\prime \prime}}{a} v^2\right].
\end{equation}
By Fourier transforming the field $v(\eta,\boldsymbol{x})$:
\begin{equation}
v(\eta, {\bf x})=\frac{1}{(2 \pi)^{3 / 2}} \int_{\mathbb{R}^3} \mathrm{~d}^3 \boldsymbol{k} v_{\boldsymbol{k}}(\eta) \mathrm{e}^{i \boldsymbol{k} \cdot \boldsymbol{x}},
\label{Fourierseries}
\end{equation}
and considering the conjugate momentum $p=v'$ we can write the free Hamiltonian:
\begin{equation}
    \mathcal{H}(\eta):=\frac{1}{2} \int \mathrm{d}^3 \boldsymbol{k}\left[p_{\boldsymbol{k}}(\eta) p_{\boldsymbol{k}}^*(\eta)+\omega^2(\boldsymbol{k}, \eta) v_{\boldsymbol{k}}(\eta) v_{\boldsymbol{k}}^*(\eta)\right].
    \label{eq:free field Hamiltonian}
\end{equation}
The equations of motion are consequently the ones of a parametric oscillator with frequency $\omega^2=k^2-a^{\prime\prime}/a$:
\begin{equation}
    v_{\boldsymbol{k}}^{\prime \prime}+\omega^2(\eta, \boldsymbol{k}) v_{\boldsymbol{k}}=0.
    \label{eq:freeeom}
\end{equation}
This equation is modified if, instead, we explicitly consider a quasi de Sitter background, with a non negligible $\varepsilon$ or $\eta_V$, i.e. scalar inflaton perturbation have a mass \footnote{This can happen for a non perfectly flat potential, even though the mass must be small.}. In that case it is possible to show that the above formulas change as $\omega^2=k^2-\frac{\nu^2-\frac{1}{4}}{\eta^2}$, where $\nu^2=\frac{9}{4}+{9} \varepsilon-\frac{m^2}{H^2}$.  
We now proceed with canonical quantization of the field $v$ and its momentum:
\begin{equation}
    \left[v_{\boldsymbol{k}}(\eta), p_{\boldsymbol{q}}(\eta)\right]=i \delta(\boldsymbol{k}+\boldsymbol{q}),
    \label{eq:commrel}
\end{equation}
where $v$ can be developed in terms of creation/annihilation operators:
\begin{equation}
v_{\boldsymbol{k}}(\eta)=u_{k}(\eta) c_{\boldsymbol{k}}+u_{-k}^*(\eta) c_{-\boldsymbol{k}}^{\dagger}.
\label{eq:decomposition}
\end{equation}
Notice that since $v$ is real not all the $v_{\boldsymbol{k}}(\eta)$ in \ref{Fourierseries} are independent, but instead we have $v_{-\boldsymbol{k}}=v^{\dagger}_{\boldsymbol{k}}$.

In order to have the right commutation relations also for the creation/annihilation operators, we have also to assume that 
\begin{equation}
u_{k}(\eta) u_{k}^{* \prime}(\eta)-u_{k}^*(\eta) u_{k}^{\prime}(\eta)=i.
\label{eq:wronskian}
\end{equation}
Corrisponding operatorial equations of motion as eq.(\ref{eq:freeeom}) are derived from the Heisenberg equation for the operator $v_{\boldsymbol{k}}$ in  the Heisenberg picture.
Notice that also $u_{k}(\eta)$ solves the eq.(\ref{eq:freeeom}).

In general, the explicit form of the mode functions $u_{k}(\eta)$ is
\begin{equation}
u_{k}(\eta)=\frac{1}{2} e^{i \frac{\pi}{2}\left(\nu+\frac{1}{2}\right)} \sqrt{-\pi \eta} H_{\nu}^{(1)}(-k \eta)\xrightarrow[\nu \to 3/2]{}\frac{e^{-i k \eta}}{\sqrt{2 k}}\left(1-\frac{i}{k \eta}\right),
\label{eq:fullmodefunction}
\end{equation}
where we have reported for future convenience their expression in the limit for a massless perturbation in a pure de Sitter background ($\nu=3/2$). Notice that $u_{k}(\eta)$ depends only on the modulus $k=|\boldsymbol{k}|$, as a consequence of the isotropy of the universe.

These conclusions we have reported for the scalar case can be directly translated into the case of tensor fluctuations of the metric, i.e. primordial gravitational waves. Analogously we can write a free action for the tensor perturbations:
\begin{equation}
S_{\text {free }}^{(t)}=\int d^3 x d \eta \frac{a^2 M_{\rm Pl}^2}{8}\left(h_{i j}^{\prime 2}-\left(\nabla h_{i j}\right)^2\right).
\label{free action for tensorial}
\end{equation}
 Being massless spin two particles, the gravitons include two propagating degrees of freedom, associated to two polarization states, $+$ and $\times$. It can be proved that they propagate as two independent massless minimally coupled scalars~\cite{Ford:1977dj}: 
\begin{equation}
h_{i j}(\boldsymbol{x}, \eta)=\sum_{\lambda=+, \times} \int \frac{d^3 k}{(2 \pi)^{3/2}} e^{i \boldsymbol{k} \cdot \boldsymbol{x}} h_\lambda(\boldsymbol{k}, t) \epsilon_{i j}^\lambda(\boldsymbol{k}),
\end{equation}
where in particular it is possible to define $h_{i j}(\boldsymbol{k}, \eta)=h_{+}(\boldsymbol{k}, \eta) \varepsilon_{i j}^{+}(\boldsymbol{k})+h_{\times}(\boldsymbol{k}, \eta) \varepsilon_{i j}^{\times}(\boldsymbol{k})=\sum_\lambda h_\lambda \epsilon_{i j}^\lambda$, $\varepsilon^{\lambda}_{i j}$ being the polarization tensors with the normalization taken to be $\epsilon_{i j}^P \epsilon^{i j P^{\prime}}=\delta^{P P^{\prime}}$.

\begin{comment}
which can be written in a well known matricial form
\footnote{$\varepsilon_{i j}^{+}=\frac{1}{\sqrt{2}}\left(\begin{array}{ccc}
1 & 0 & 0 \\
0 & -1 & 0 \\
0 & 0 & 0
\end{array}\right)$ and $\varepsilon_{i j}^{\times}=\frac{1}{\sqrt{2}}\left(\begin{array}{ccc}
0 & 1 & 0 \\
1 & 0 & 0 \\
0 & 0 & 0
\end{array}\right)$}
\end{comment}

 We recall that they are traceless $\epsilon_{ii}=0$, symmetric $\epsilon_{i j}=\epsilon_{j i}$ and transverse $k^i \epsilon_{i j}^P(\boldsymbol{k})=0$. Therefore, the $h_{\lambda}$ fields behave just like scalar fields, and we can introduce the canonical variable 
\begin{equation}
\label{thetaij}
\theta_{ij}=\frac{a M_{\rm Pl} h_{ij}}{2}
\end{equation} 
which satisfies similar equations of motion
\begin{equation}
\theta_\lambda^{\prime \prime}(k, \eta)+\left(k^2-\frac{a^{\prime \prime}}{a}\right) \theta_\lambda(k, \eta)=0,
\end{equation}
and can be canonically quantized analogously with eq.(\ref{eq:commrel}) for the scalar perturbations. 

In the Heisenberg picture, evolution of the fields is also considered in another basis in the Hilbert space, instead of the creation/annihilation operators. The considered operators are time independent~\cite{boyanovsky_effective_2015}
\begin{equation}
\begin{split}
&Q_{\boldsymbol{q}}=\frac{1}{\sqrt{2}}\left(b_{\boldsymbol{q}} e^{i \frac{\pi}{2}\left(\nu+\frac{3}{2}\right)}+b_{-\boldsymbol{q}}^{\dagger} e^{-i \frac{\pi}{2}\left(\nu+\frac{3}{2}\right)}\right) \xrightarrow[\nu \to 3/2]{}\frac{i}{\sqrt{2}} \left(-b_{\boldsymbol{q}} +b_{-\boldsymbol{q}}^{\dagger}\right)\\
&P_{\boldsymbol{q}}=\frac{i}{\sqrt{2}}\left(b_{-\boldsymbol{q}}^{\dagger} e^{-i \frac{\pi}{2}\left(\nu+\frac{3}{2}\right)}-b_{\boldsymbol{q}} e^{i \frac{\pi}{2}\left(\nu+\frac{3}{2}\right)}\right) \xrightarrow[\nu \to 3/2]{}\frac{1}{\sqrt{2}} \left(b_{\boldsymbol{q}} +b_{-\boldsymbol{q}}^{\dagger}\right)\\
\end{split}
\end{equation}

where again we have written the general expression valid also for $\nu \neq 3/2$ and then specified the limit for $\nu = 3/2$.
Notice that also $Q$ and $P$ are two conjugate canonical variables:
\begin{equation}
    \left[P_{\boldsymbol{q}}^{\dagger}, Q_{\boldsymbol{k}}\right]=-i \delta(\boldsymbol{q}- \boldsymbol{k}),
    \label{eq:QPcommutators}
\end{equation}
and all the other commutators are null.
In this way it is possible to divide the perturbation fields and their momentum as:
\begin{equation}
\begin{split}
    &v_{\boldsymbol{q}}(\eta)=Q_{\boldsymbol{q}} g_{+}(q ; \eta)+P_{\boldsymbol{q}} g_{-}(q ; \eta),\\
    &p_{\boldsymbol{q}}(\eta)=v_{\boldsymbol{q}}^{\prime}(\eta)=Q_{\boldsymbol{q}} g_{+}^{\prime}(q ; \eta)+P_{\boldsymbol{q}} g_{-}^{\prime}(q ; \eta).\\
    \label{eq:growdecfieldmom}
\end{split}
\end{equation}
where $g_{+}(q,\eta)$ and $g_{-}(q,\eta)$ are called respectively the growing and the decaying mode, and correspond to the Imaginary and Real part of $u_{k}$.\footnote{Notice that we are using the notation by~\cite{boyanovsky_effective_2015}, but effectively this is equivalent at all to e.g.~\cite{polarski_semiclassicality_1996}, where $u_{k}\rightarrow f_{k}$ and $g_{+}$ and $g_{-}$ correspond to $f_{k1}$ and $f_{k2}$.}
For a generic $\nu$ we can then say that:
\begin{equation}
g_{+}(q ; \eta)=\sqrt{\frac{-\pi \eta}{2}} Y_{\nu}(-q \eta) \quad \quad g_{-}(q ; \eta)=\sqrt{\frac{-\pi \eta}{2}} J_{\nu}(-q \eta).
\label{eq:growing decaying definitions}
\end{equation}
This decomposition reveals its usefulness in particular when we deal with $-q\eta \ll 1$, i.e. superhorizon modes. In this limit the two functions become 
\begin{equation}
\begin{split}
&g_{+}(q ; \eta)\simeq \left(\frac{-q\eta}{2}\right)^{-\nu} \frac{\Gamma(\nu)}{\pi}\sqrt{\frac{-\pi \eta}{2}} \xrightarrow[\nu \to 3/2]{} \frac{1}{q^{3 / 2} \eta}\\
&g_{-}(q ; \eta)\simeq \left(\frac{-q\eta}{2}\right)^{\nu} \frac{1}{\Gamma(\nu+1)}\sqrt{\frac{-\pi \eta}{2}}\xrightarrow[\nu \to 3/2]{} \frac{1}{3} q^{3 / 2} \eta^2\\
\label{eq:growingdecayingexpansion}
\end{split}
\end{equation}
The decaying mode becomes more and more negligible as the number of e-folds after horizon crossing grows. The intrinsic decoherence papers~\cite{polarski_semiclassicality_1996} used to neglect it, as mentioned in the introduction. Instead, the ``importance'' of considering the decaying mode in the context of decoherence has been underlined many times in the recent epoch, see e.g. section 4.2 of~\cite{colas_benchmarking_2022}. As we are going to see, also Boyanovsky~\cite{boyanovsky_effective_2015}, even empolying the quantum master equation formalism, neglects the decaying mode: this however prevents from computing decoherence and 
non-unitary corrections.

\subsection{The interactions}
\label{subsec:interaction}
The interactions we will consider for decoherence are derived by expanding the single-field inflation action rewritten in terms of the ADM formalism variables
~\footnote{See e.g. eq. (A:4) in~\cite{burgess_minimal_2023}.} at higher orders in the number of field fluctuations beyond the free (linear) ones. As explained well in the introduction of~\cite{burgess_minimal_2023}, 
the interactions involving three fields are less suppressed in a $1/M_{\rm Pl}$ than the ones with four fields, so we are only going to consider trilinear  (i.e. cubic) interactions. Differently from~\cite{burgess_minimal_2023}, since we are considering a deep subhorizon environment, there is no reason to discard the interactions with time derivatives, which in this case contribute just as the spatial derivatives one. We want to consider the interactions between a subhorizon tensor environment and a superhorizon scalar system, which we can call ``mixed'', by only using the interactions derived by General Relativity.  As computed in~\cite{maldacena_non-gaussian_2003}, the trilinear mixed lagrangian is:
\begin{equation}
    S=\frac{\varepsilon M_{\rm Pl}^2}{8} \int d t d^3 x\left(a^3 \zeta \dot{h}_{i j} \dot{h}_{i j}+a \zeta \partial_l h_{i j} \partial_l h_{i j}-2 a^3 \dot{h}_{i j} \partial_l h_{i j} \partial_l\left(\nabla^2\right)^{-1} \dot{\zeta}\right).
\end{equation}
We want to write the corresponding action for the canonically normalized variables $\theta_{ij}$~eq.(\ref{thetaij})  and $v$~eq.{ (\ref{v})}. This turns out ot be
\begin{equation*}
\begin{aligned}
&S= \frac{\varepsilon M_{\rm Pl}^2}{8} \frac{4}{M_{\rm Pl}^2} \frac{1}{\sqrt{2 \varepsilon} M_{\rm Pl}} \int d^3 x d \eta\left(a^4 \frac{1}{a} v \frac{1}{a}\left(\frac{\theta_{i j}}{a}\right)^{\prime} \frac{1}{a}\left(\frac{\theta_{i j}}{a}\right)^{\prime}+a^2 \frac{v}{a} \partial_l\left(\frac{\theta_{i j}}{a}\right) \partial_l \left(\frac{\theta_{i j}}{a}\right)-\right. \\
& \left.-2 \frac{a^4}{a}\left(\frac{\theta_{i j}}{a}\right)^{\prime} \partial_l \frac{\theta_{i j}}{a} \frac{1}{a} \partial_l\left(\nabla^2\right)^{-1}\left(\frac{v}{a}\right)^{\prime}\right)=
\end{aligned}
\end{equation*}
\begin{equation}
=\frac{\sqrt{\varepsilon}}{2 \sqrt{2} M_{\rm Pl}} \int d^3 x d \eta\left(2 a H^2 \theta_{i j} \theta_{i j} v+\frac{1}{a} v \partial_l \theta_{i j} \partial_l \theta_{i j}+\frac{2}{a} \theta_{i j}^{\prime \prime} \partial_l \theta_{i j} \partial_l\left(\nabla^2\right)^{-1} v\right),
\label{eq:interaction}
\end{equation}
where we have integrated by parts, both in conformal time and in configurational space, discarding the boundary terms. Notice, however, that as investigated in~\cite{ning_decoherence_2023}, also time boundary terms may have an effect on decoherence. Thus, we leave for future investigations to compute their effect. 

In order to obtain the previous result, we have assumed to be at lowest order in slow-roll expansion. Since the interactions are already proportional to the first slow roll parameter $\varepsilon$, the slow roll corrections to the mode functions would be of higher order in slow roll parameters. Thus we will consider the mode functions in de Sitter space time. In the end we are left with 3 interactions:
\begin{equation}
    \begin{aligned}
H_{I N T 1} & =-\frac{\sqrt{\varepsilon}}{ \sqrt{2} M_{\rm Pl}} a H^2 \theta_{i j} \theta_{i j} v=\frac{\sqrt{\varepsilon}}{ \sqrt{2} M_{\rm Pl}} \frac{H}{\eta} \theta_{i j} \theta_{i j} v, \\
H_{I N T 2} & =-\frac{\sqrt{\varepsilon}}{2 \sqrt{2} M_{\rm Pl}} \frac{1}{a} v \partial_l \theta_{i j} \partial_l \theta_{i j}=\frac{\sqrt{\varepsilon}}{2 \sqrt{2} M_{\rm Pl}} H \eta v \partial_l \theta_{i j} \partial_l \theta_{i j}, \\
H_{I N T 3} & =- \frac{\sqrt{\varepsilon}}{ \sqrt{2} M_{\rm Pl}} \frac{1}{a} \theta_{i j}^{\prime \prime} \partial_l \theta_{i j} \partial_l\left(\nabla^2\right)^{-1} v= \frac{\sqrt{\varepsilon}}{\sqrt{2} M_{\rm Pl}} H \eta \theta_{i j}^{\prime \prime} \partial_l \theta_{i j} \partial_l\left(\nabla^2\right)^{-1} v.
\label{eq:interactions}
\end{aligned}
\end{equation}
We will call the first one derivativeless interaction; it will prove to be the most important in our analysis. This interaction is used in many models in the literature~\cite{boyanovsky_effective_2015, Burgess:2024eng,Brahma:2024yor,Hollowood:2017bil}, involving  a conformal scalar field as a proxy for a subhorizon environemnt, so it may be important to analyze this interaction in itself to have a direct comparison. Instead, in the papers considering an environment of short wavelength modes, this interaction, to our knowledge, has never been considered before. Notice that we have directly translated into interaction Hamiltonian the Lagrangian interaction using $H_{int}=-L_{int}$ since no derivative terms as $v'$ is present in the various considered interactions. 
By contrast, we will call derivative interactions the ones labelled by $H_{INT2}$ and $H_{INT3}$. Apart from the presence of derivatives, notice that this latter group of interactions is different from the derivativeless interaction for the power of conformal time which stays in front of it. We will be interested in the end of inflation, e.g. as $\eta \rightarrow 0$. In particular, as $\eta \rightarrow 0$, we can say that the derivativeless interactions are relevant, as the prefactor of the interactions grow, while the derivative interactions are irrelevant, i.e. the prefactor of the interaction becomes zero. Going back in time, instead, the derivative interactions coupling becomes bigger. As we will see, this has an important effect on how to properly take the Markovian approximation.

In this way, the total Hamiltonian reads
\begin{equation}
    \mathcal{H}(\eta)=\mathcal{H}_S(\eta) \otimes \mathcal{I}_E+\mathcal{I}_S \otimes \mathcal{H}_E(\eta)+\mathcal{H}_{\text {int }}(\eta),
    \label{total Hamiltonian}
\end{equation}
where we have called $\mathcal{H}_S(\eta)$ the free Hamiltonian of the system, $\mathcal{H}_E(\eta)$ the free Hamiltonian of the environment, and the interaction Hamiltonian as $\mathcal{H}_{\text {int }}(\eta)$. Notice that the free Hamiltonians live each in either the Hilbert space of the system or the Hilbert space of the environment. The interaction Hamiltonian, instead, does not live in either the Hilbert space of the system or of the environment, lying instead in the cross product of the two. For this reason, it can create entanglement from initially factorized states. 
In general, fields belonging to the system Hilbert space can be written as: 
\begin{equation}
    v(\eta, x):=\int \frac{\mathrm{d}^3 \boldsymbol{k}}{(2 \pi)^{3 / 2}} \Theta\left(a H-k\right) v_{\boldsymbol{k}}(\eta) e^{i \boldsymbol{k} \cdot \boldsymbol{x}},
\end{equation}
and $H_{S}$  is given by~eq.(\ref{eq:free field Hamiltonian}) for this type of degrees of freedom. On the other hand the field belonging to the environment can be written as
\begin{equation}
    \theta_{ij}(\eta, x):=\int \frac{\mathrm{d}^3 \boldsymbol{k}}{(2 \pi)^{3 / 2}} \Theta\left(k-10 a H\right) \theta_{\boldsymbol{k} i j}(\eta) e^{i \boldsymbol{k} \cdot \boldsymbol{x}},
    \label{eq:environmental fields decompostiion}
\end{equation}
where the choice of having modes with $k>10 aH$ is completely arbitrary and will be explained better in the section\ref{subsection:env}.

\subsection{Quantum master equation: microphysical derivation}
\label{sec:qme}
In this section we present the microphysical derivation of the quantum master equation. This approach is different from deriving the same quantum master equation out of the formal Nakajima-Zwanzig equation, which is possible to find, e.g., in~\cite{colas_benchmarking_2022} and~\cite{burgess_minimal_2023}. 

It is known that the information about the state of the system must be encoded in a density matrix when computing decoherence. This happens because the quantum correlations are lost and the original wavefunction is transformed, after the computations, in a statistical ensemble. In Appendix A, we review what decoherence is, and clarify its meaning.
 The idea is simply starting from the customary Heisenberg equations and then to trace over an unobservable environment. This is not the only approach to decoherence in the early universe: for example, in~\cite{nelson_quantum_2016,ning_decoherence_2023}, the authors directly trace over the wavefunctional of the primordial perturbations in the Schrodinger representation in order to obtain the decoherence time. Also, a path integral representation of the open quantum system dynamics, named the Feynman Vernon integral, has been widely used in the literature.  These approaches are equivalent (see e.g.~\cite{Boyanovsky:2015xoa,Colas:2023wxa} ). 
 
 As we are going to explain below, the application of the QME to cosmology is not straightforward, and presents many differences with respect to the usual applications  in condensed matter physics. 
 After reviewing the microphysical derivation, we will review the Markovian approximation, stressing the points which create some difficulties in translating it to a cosmological background.

\subsubsection{Microphysical derivation}
\label{subsec:Microphysical derivation}
The equation of motion for the total density matrix in the Schrodinger representation is the Von Neumann equation:
\begin{equation}
    \rho'(\eta)=-i[H,\rho(\eta)],
\end{equation}
where $H$ is the total Hamiltonian in~eq.(\ref{total Hamiltonian}). It is customary to consider the interaction representation, in which instead the density matrix evolves only with the interaction Hamiltonian~eq.(\ref{eq:interactions})~\footnote{Even though it is an abuse of notation, but we will sometimes omit any distinctive symbol on the operators according to the representation in which they are written. In general, if not specified, all the operators will be written in the interaction picture.}
\begin{equation}
    \rho_{I}'(\eta)=-i g(\eta) [H_{int},\rho_{I}(\eta)],
\end{equation}
where for simplicity we have called $g$ the coupling constant and taken it out of the interaction Hamiltonian for keeping track of its powers. By formally solving it we obtain
\begin{equation}
\rho_I(\eta)=\rho_I\left(\eta_0\right)-i \int_{\eta_0}^\eta g(\eta')\left[H_{int}\left(\eta^{\prime}\right), \rho_I\left(\eta^{\prime}\right)\right] d \eta^{\prime}.
\end{equation}
The first assumption that we make is the perturbativity of the coupling between the system and the environment. The aforementioned important constraints on  primordial non-Gaussianities by {\it Planck} \cite{akrami_planck_2020} sustain this approximation scheme in inflation. Sometimes, this is also called the Born approximation. By substituting back $\rho_{I}(\eta)$ inside the integral and truncating the series since it would be higher order in the coupling one gets
\begin{equation}
\rho_{I}^{\prime}(\eta)=-i g(\eta) \left[H_{int}(\eta), \rho_{I}\left(\eta_0\right)\right]-\int_{\eta_0}^\eta g(\eta) g(\eta')\left[H_{int}(\eta),\left[H_{int}\left(\eta^{\prime}\right), \rho_I\left(\eta^{\prime}\right)\right]\right] d \eta^{\prime}+O(g^3),
\label{eq:Bornapprox}
\end{equation}
where we stress that the perturbativity is still controlled by the coupling. Eq.~(\ref{eq:Bornapprox}) can be also integrated to obtain the equivalent equation
\begin{equation}
\begin{split}
&\rho_{I}(\eta)=-i\int_{\eta_0}^\eta d \eta' g(\eta') \left[H_{int}(\eta'), \rho_{I}\left(\eta_0\right)\right]-\\
&-\int_{\eta_0}^{\eta}  d \eta^{\prime} \int_{\eta_0}^{\eta'} d \eta^{\prime \prime} g(\eta') g(\eta'')\left[H_{int}(\eta'),\left[H_{int}\left(\eta^{\prime\prime}\right), \rho_I\left(\eta^{\prime\prime}\right)\right]\right]
+O(g^3).\\
\label{eq:Bornapproxintegrated}
\end{split}
\end{equation}
The initial density matrix is usually specified to be factorizable at the beginning of the dynamics. Notice that this is a \textit{second approximation} since in fact we are assuming that, before the dynamics becomes effective, the system and the environment were independent and unentangled, and the entanglement rises only because of the interactions  
\begin{equation}
    \rho_{I}(\eta_{0})=\rho_{E}(\eta_{0}) \otimes \rho_{S}(\eta_{0}) \quad {\rm where}\quad \rho_{E}=Tr_{S} \rho_{I} \quad {\rm and} \quad \rho_{S}=Tr_{E} \rho_{I},
    \label{eq:secondapprox}
\end{equation}
where remember that all the $\rho$ are written in  theinteraction picture, and in the following we will usually call $\rho_{S}=\rho_{r}$, i.e. the reduced density matrix, as defined in eq.(\ref{eq:reduceddensitymatrixexample}) (after tracing over the environment).

One of the key features of the environment should be that it is not affected by the interactions with the system: it should be in the same state $\rho_{E}(\eta)=\rho_{E}(\eta_0)$(in interaction picture) at every time as in the initial time. In condensed matter settings, the environment is considered to be a thermal, large reservoir, so any perturbative interaction is easily lost in the reservoir without affecting it. The latter approximation in these situations is thus naturally verified.  For inflationary perturbations there is probably no exact correspondence. Usually, the environment, in the interaction picture, is assumed to be at the beginning in a Bunch-Davies vacuum, the corresponding vacuum in flat spacetime. Then, in analogy, we should require that it stays in this state for the whole evolution described by the quantum master equation. Of course, the validity of this approximation is challenged if the environment states are easily excitable. Conversely, for states which are heavily energetic, as particles with a huge mass ($M \gg H$) or subhorizon modes of a field, we expect that the approximation is more appropriate. In fact, excitations are unlikely; or, equivalently, since the lifetime of an excitation is expected to be inversely proportional to the energy of the modes, the excitations are of very small duration, so the validity of the approximation is always verified.

The \textit{third approximation} we made can be stated as:
\begin{equation}
    \rho_{E}=\prod_{k \in E}|0\rangle_{k} \,\, _{k} \langle0|
    ,
    \label{eq:thirdapprox}
\end{equation}
where $\rho_{E}$ is always written in the interaction representation. The trace over the environment is so reduced to a vacuum expectation value.

Since the final goal is to write an equation for the reduced density matrix of the system, we have to trace~eq.(\ref{eq:Bornapproxintegrated}) over the environment. The first term on the RHS of~eq.(\ref{eq:Bornapprox}) 
\begin{equation}
\operatorname{Tr}_{E} [H_{int},\rho],
\end{equation}
is usually assumed to have zero environmental expectation value.  If this condition is not already satisfied by the Hamiltonian, it can be usually obtained by a simple field redefinition. This \textit{fourth approximation} can so be easily relaxed.
The second term on the RHS of~eq.(\ref{eq:Bornapprox}) instead features the non-unitary contributions. This is the leading term for decoherence, and it is of second order in the coupling interactions. At this point our quantum master equation looks like:
\begin{equation}
   \rho_{\text {r }}(\eta)=-\operatorname{Tr}_{E(\eta)} \int_{\eta_{0}}^\eta \mathrm{~d} \eta^{\prime} g(\eta') \int_{\eta_0}^{\eta'} \mathrm{~d} \eta^{\prime \prime} g(\eta'')\left[{H}_{\text {int }}(\eta'),\left[{H}_{\text {int }}\left(\eta^{\prime \prime}\right), \rho\left(\eta^{\prime \prime}\right)\right]\right],
    \label{eq:integratednonmarkqme}
\end{equation}
 and we have to derive it with respect to time to obtain an equation for the reduced density matrix.  In particular, notice that for a time dependent environment, for each instant of time, we have a different environment in~eq.(\ref{eq:integratednonmarkqme}), and so we have to trace over a different subspace of the Hilbert space. When applying the derivative, we should thus vary independently the environment and the reduced density matrix. We will apply this modification in section \ref{subsec:timedependentenvironment}. 
 
 As we will show, the effects of the time dependent environment (under some conditions and approximations we will apply) can approximately be neglected; so we will for now stick with the usual derivation. By applying the time derivative one gets
\begin{equation}
    \frac{\mathrm{d} \rho_{\text {r }}}{\mathrm{d} \eta}=-g(\eta) \int_{\eta_0}^\eta \mathrm{~d} \eta^{\prime} g(\eta')\operatorname{Tr}_{E}\left[{H}_{\text {int }}(\eta),\left[{H}_{\text{int }}\left(\eta^{\prime}\right), \rho\left(\eta^{\prime}\right)\right]\right].
    \label{eq:nonmarkqme}
\end{equation}

We assume that the interaction Hamiltonian can be written as
 \begin{equation}
 H_{int}(\eta)=\int \mathrm{d}^3x g(\eta) O_{S}(\eta,\boldsymbol{x})\otimes O_{E}(\eta,\boldsymbol{x}),
 \end{equation}
i.e. it can be factorized in an operator $O_{S}$ acting on the system Hilbert space and an operator $O_{E}$ acting on the environmental Hilbert space.
We can thus write the quantum master equation in the following form
\begin{equation}
\begin{split}
 \frac{\mathrm{d} \rho_{\mathrm{r}}}{\mathrm{d} \eta}=&-g(\eta) \int_{\eta_0}^{\eta} \mathrm{~d} \eta^{\prime} g(\eta')\int \mathrm{d}^3 \boldsymbol{x} \int \mathrm{d}^3 \boldsymbol{y}\left\{ O_{\mathcal{S}}(x) {O}_{\mathcal{S}}(y) \rho_{r}(\eta')-{O}_{\mathcal{S}}(y) {\rho}_{r}
(\eta') {O}_{\mathcal{S}}(x)\right] G^{>}(x, y)\\
&\left.-\left[{O}_{\mathcal{S}}(x) {\rho}_{r}(\eta') {O}_{\mathcal{S}}(y)-{\rho}_{r}(\eta'){O}_{\mathcal{S}}(y) {O}_{\mathcal{S}}(x)\right]\left[G^{>}(x, y)\right]^*\right\},\\
\label{eq:QMEposition}
\end{split}
\end{equation}
where notice that all the variables now refer to the system, and the information about the environment has been isolated into the correlation function and its conjugate:
\begin{equation}
G^{>}(x, y) \equiv \operatorname{Tr}_{E}\left[{O}_{E}(\boldsymbol{x},\eta) O_{E}(\boldsymbol{y},\eta') {\rho}_{E}\right]=_{\epsilon}\left\langle 0|{O}_{E}(\boldsymbol{x},\eta) {O}_{E}(\boldsymbol{y},\eta')|0\right\rangle_{\epsilon}.
\end{equation}
while the $G^{*}(x,y)$ is the time reversal correlation function.  Notice that  in the second equality we have considered the vacuum as the state of the environment, as customary in a cosmological setting. In this paper, we will work with the Fourier transform of the quantum master equation eq.(\ref{eq:QMEposition})  which reads
\begin{equation}
\begin{split}
&\rho_r^{\prime}(\eta)=-g(\eta) \int_{\eta_0}^\eta {\mathrm{d} \eta^{\prime}} g(\eta') \sum_{\boldsymbol{k}}\left[O_{S,\boldsymbol{k}}(\eta) O_{S,-\boldsymbol{k}}\left(\eta^{\prime}\right) \rho_r(\eta') K\left(k, \eta, \eta^{\prime}\right)+\rho_r(\eta') O_{S,-\boldsymbol{k}}\left(\eta^{\prime}\right) O_{S,\boldsymbol{k}}(\eta)\right. \\
&\left. K^*\left(k, \eta, \eta^{\prime}\right)-O_{S,\boldsymbol{k}}(\eta) \rho_r(\eta') O_{S,-\boldsymbol{k}}\left(\eta^{\prime}\right) K^*\left(k, \eta, \eta^{\prime}\right)-O_{S,-\boldsymbol{k}}\left(\eta^{\prime}\right) \rho_r(\eta') O_{S,\boldsymbol{k}}(\eta) K\left(k, \eta, \eta^{\prime}\right)\right],\\
\label{eq:QMEmomentum}
\end{split}
\end{equation}
where we have introduced the Fourier transform of the environmental correlation functions $K(k,\eta,\eta')$:
\begin{equation}
    \begin{aligned}
G^{>}\left(\boldsymbol{x}-\boldsymbol{y}, \eta, \eta^{\prime}\right) & =\frac{1}{V} \sum_{\boldsymbol{k}} K\left(k, \eta, \eta^{\prime}\right) e^{-i \boldsymbol{k} \cdot(\boldsymbol{x}-\boldsymbol{y})} \\
G^{<}\left(\boldsymbol{x}-\boldsymbol{y}, \eta, \eta^{\prime}\right) & =\frac{1}{V} \sum_ {\boldsymbol{k} }K^*\left(k, \eta, \eta^{\prime}\right) e^{-i \boldsymbol{k} \cdot(\boldsymbol{x}-\boldsymbol{y})} .
\label{eq:correlationfunctionsdef}
\end{aligned}
\end{equation}
For future convenience, we have also considered a finite volume V of the universe and discrete momenta. However, when the environmental momenta are far bigger than the system ones, e.g. in the case where there is a separation of scales between the two (as in our case), the environmental momenta can be still considered in a good approximation a continuum, if the box is sufficiently big. So, the sum is replaced by an integral. When passing from discrete to continuum momenta we can employ the substitution:
\begin{equation}
    \int \frac{d^3k} {(2\pi)^3} = \frac{1}{V}\sum_{\boldsymbol{k}} ,
    \label{eq:continuumtodiscretemomenta}
\end{equation}
and, taking into account that also the operators must change their normalization when considered as operators in a finite universe, we can also say that (see, e.g, eq.(D.3) of~\cite{burgess_minimal_2023})
\begin{equation}
    O^{\textit{discrete}}=O^{\textit{continuum}} \sqrt{\frac{(2\pi)^3}{V}}.
    \label{eq:discretecontinop}
\end{equation}
In the end the two contributions in eq.(\ref{eq:discretecontinop}) and eq.(\ref{eq:continuumtodiscretemomenta}) simplify inside eq.(\ref{eq:QMEmomentum}) and the final expression of the quantum master equation has no volume factor inside, even in the case of discrete operators. In all the situations we will consider, after a Fourier transform, as a system operator the Sasaki-Mukhanov variable 
\begin{equation}
    O_{S}(\boldsymbol{k},\eta)=v_{\boldsymbol{k}}(\eta).
    \label{eq:sysoperator}
\end{equation}
As written in eq.(\ref{eq:QMEmomentum}), the quantum master equation is unsolvable. This is because the evolution of the system state at a certain instant of time $\rho_{r}(\eta)$ depends on the whole previous history $\rho_{r}(\eta')$. The only way out of the problem is to consider some more approximations, which, for their importance, we will discuss in a separate paragraph.

\subsubsection{The Markovian approximation}
\label{subsec:the Markovian approximation}
Here it comes maybe the most discussed point on the applications of the quantum master equation: the Markovian approximation. Markovianity is a semantically ambiguous word which stands for many connected (but different) concepts in the quantum master equation literature. To have an overview on this point, see, e.g.,~\cite{Colas:2023wxa}.

The logic behind the Markovian approximations involves a hierarchy between the time scales of the environment, the system and the interactions between the system and the environment. In particular, the Markovian approximation is justified if the environmental correlation function is non-zero for a time, $\tau_{E} \ll \tau_{r}$, where $\tau_{r}$ represents instead the typical time of evolution of the system due to interactions. This time depends roughly by the inverse of the coupling in the interaction. In the interaction picture, this is the time of evolution of the density matrix of the system. When we derive the quantum master equation, the time evolution can be considered coarse grained on time scales $ \Delta \eta $, chosen in such a way that $\tau_{E} \ll \Delta \eta \ll \tau_{r}$. Therefore, under this condition, we may consider the density matrix in the memory integral on the RHS of the quantum master equation eq.(\ref{eq:QMEmomentum}) as approximately constant for all the time in which the integral is non zero (i.e. the environmental correlation functions are non zero). We can thus develop in series $\rho(\eta')$:
\begin{equation}
    \rho_r(\eta') \simeq \rho_r(\eta) + (\eta'- \eta) \rho_r'(\eta)+\dots\simeq \rho_r(\eta),
    \label{eq:markovianapprox}
\end{equation}
inside the integral of the eq.(\ref{eq:QMEmomentum}), and truncate this series at the zero order. Notice that this is equivalent to say that the term $\rho'(\eta)$ is subdominant and negligible. In this sense, we could justify this approximation also if we think that $\rho'(\eta)$ is of order $g^2$, so the second term in the series, if substituted in the RHS of eq.(\ref{eq:QMEmomentum}), would be of order $g^4$, and thus subdominant in a perturbative expansion in the coupling. This underlines how the Markovian approximation is tightly coupled to the Born approximation. Sometimes, for this reason, the approximation is called the Born-Markov approximation. The quantum master equation that we obtain is:
\begin{equation}
\begin{split}
&\rho_r^{\prime}(\eta)=-g(\eta) \int_{\eta_0}^\eta {\mathrm{d} \eta^{\prime}} g(\eta') \sum_{\boldsymbol{k}}\left[v_{\boldsymbol{k}}(\eta) v_{-\boldsymbol{k}}\left(\eta^{\prime}\right) \rho_r(\eta) K\left(k, \eta, \eta^{\prime}\right)+\rho_r(\eta) v_{-\boldsymbol{k}}\left(\eta^{\prime}\right) v_{\boldsymbol{k}}(\eta)\right.\\
&\left. K^*\left(k, \eta, \eta^{\prime}\right)\right. 
\left.-v_{\boldsymbol{k}}(\eta) \rho_r(\eta) v_{-\boldsymbol{k}}\left(\eta^{\prime}\right) K^*\left(k, \eta, \eta^{\prime}\right)-v_{-\boldsymbol{k}}\left(\eta^{\prime}\right) \rho_r(\eta) v_{\boldsymbol{k}}(\eta) K\left(k, \eta, \eta^{\prime}\right)\right],\\
\label{eq:blochredfield}
\end{split}
\end{equation}
where we have we have approximated $\rho_{r}(\eta') \simeq \rho_{r}(\eta)$ and considered as system operator the Sasaki-Mukhanov variable as in eq.(\ref{eq:sysoperator}). 
\\
\\
\textbf{Connection of the Markovian approximation with the other approximations}
We want to underline that the Markovian approximation is connected very tightly, not only to the Born approximation, but also to all the other approximations we already did in section (\ref{subsec:Microphysical derivation}), in order to derive the quantum master equation eq.(\ref{eq:QMEmomentum}).  Actually, also what we have called the third approximation, eq.(\ref{eq:thirdapprox}), can be redirected to the Markovian approximation. It is not actually necessary to require that the environment is always in a fixed state in order to apply eq.(\ref{eq:thirdapprox}). It is only sufficient that the possible excitations of the environmental degrees of freedom relax in a typical time much smaller both of the aforementioned coarse graining scale, and of the typical time of evolution of the system due to interactions. In this way, even if the state of the environment is excited, it will relax to the Bunch-Davies vacuum before the system actually realizes it. In other words, from the point of view of the system, the excitation was so rapid that the environmental state can be considered as it was always the Bunch-Davies vacuum~\cite{Breuer:2007juk}. Following a similar logic, also the second approximation to obtain the quantum master equation, regarding initial correlations between system and environment, eq.(\ref{eq:secondapprox}) can be connected to the Markovian approximation. Suppose that there was a previous dynamics which actually could have entangled the system and the environment already at the initial time of the dynamics $\eta_{0}$. In an inflationary setting, we can imagine, for example, a preinflationary stage in which interactions between system and environment were strong enough to create an entanglement in the initial state of inflationary curvature pertubations. However, when inflation starts this strong interaction stops, also because during inflation the interactions of the curvature perturbations are observed to be weak. This leaves just an initial state of perturbations which is a mixed state between the inflationary perturbations and the environment, akin to the one analyzed, although in a unitary setting, in ~\cite{albrecht_cosmological_2014}. If the previous interactions, however, stop being effective, how long would the quantum correlations between system and environment that they have built stay active? If the environment has a really short ``memory'', the free evolution of the environment makes it readily loose this quantum correlations, so actually returning to the unentangled state eq.(\ref{eq:secondapprox})~\cite{Tasaki_2007}, before this can have any effect on the state of the system. 
\\
\\
\textbf{Bloch Redfield equation vs. Markovian master equation}
The equation eq.(\ref{eq:blochredfield}) is called Bloch-Redfield equation. The microphysical derivation, as we have seen, is very clear and allows to develop some physical intuition, besides to verify the approximations while doing them. Notice that we have applied a Born-Markov approximation in order to derive eq.(\ref{eq:blochredfield}), but the Bloch-Redfield equation \textit{is not} a Markovian master equation. This semantical ambiguity usually generates confusion, which we will try to clarify in the next lines. In fact, eq.(\ref{eq:blochredfield}) is local in time, since the dynamics of the density matrix, at each time $\eta$ depends only on the density matrix at the present time $\eta$ in the RHS. However, in the Bloch Redfield equation there is still some memory left, which corresponds to the time integration  on the RHS. We define a Markovian master equation, instead, as an equation where there is no memory at all, i.e. no integration on time on the RHS.

 In a more rigorous way~\cite{Breuer:2007juk}, we say that a master equation is Markovian if it considers the evolution of the density matrix as a semigroup map: 
\begin{equation}
    {\rho}_r(t) \rightarrow {\rho}_r\left(t^{\prime}\right)=\mathcal{V}_{t \rightarrow t^{\prime}} \rho_r(t)=e^{\mathcal{L}(t'-t)} {\rho}_r(t) .
    \label{markovian}
\end{equation}
where the Lindblad super-operator $\mathcal{L}$ is defined by $\rho_r'(t)=\mathcal{L}\rho_r(t)$.  The Lindblad theorem states that there is only one Markovian master equation that can describe the dynamics while maintaining positivity, hermiticity and trace unit of the density matrix, called Lindblad or GKSL (Gorini-Kossakowski-Sudarshan-Lindblad) equation~\cite{Lindblad:1975ef,Kossakowski:1972sbn,Gorini:1975nb}:
\begin{equation}
    \dot{\rho}_r=-\mathrm{i}[H, \rho_r]+\sum_j \gamma_j\left(\hat{L}_j \rho_r \hat{L}_j^{\dagger}-\frac{1}{2}\left\{\hat{L}_j^{\dagger} \hat{L}_j, \rho_r\right\}\right)
    \label{eq:lindblad}
\end{equation}
where, we stress, the $\gamma_k$ coefficient are here time independent and must be positive to guarantee a the positivity of the density matrix. Usually, in condensed matter settings, the Lindblad super-operator $\mathcal{L}$ is taken time independent. For this reason, the Lindblad theorem  is often expressed for time independent Lindblad operators $L$.
In cosmological setting, general covariance prescribes the presence of time dependent Jacobians depedending on some powers of the scale factor $a(\eta)$, which can be seen effectively as a time dependent coupling constant $g(\eta)$. Thus, even in the case of a memoryless, Markovian dynamics, the operators $L$, and thus the super-operator $\mathcal{L}$, should depend on time.
Notice that, if the super-operator  $\mathcal{L}$ depends on time, the definition with a semigroup property in eq.(\ref{markovian}) is not suitable. A possible rigorous extension to time dependent Markovian processes is instead presented in \cite{Rivas_2014}, based on divisibility. Intuitively, we argue that in order for a time dependent property to satisfy a composition law just like:
\begin{equation}
      \mathcal{V}_{t \rightarrow t^{\prime}}=\mathcal{V}_{t \rightarrow t^{\prime\prime}} \mathcal{V}_{t^{\prime \prime} \rightarrow t' }, \quad \textit{where} \quad e^{\int_{t}^{t'} ds \mathcal{L(s)}}=\mathcal{V}_{t \rightarrow t^{\prime}}
\end{equation}
the generators of the transformation must satisfy the commutator relation:
\begin{equation}
[\mathcal{L}(t),\mathcal{L}(s)]= 0\, \, \, {\rm for}\,\,\, t \neq s,
\label{eq:commmarkovianity}
\end{equation}
 otherwise the exponential of linear operators is known in general not to commute. This can be seen as a statement on the dynamics: if the commutator is zero, there must be no memory and thus there is no overlap between the dynamics at instant $t$ of time and the one at another instant $s$ of time~\cite{Lidar:2019qog}. In other words, we can neglect the correlation of the dynamics between two different instants of time. Thus, if the property in eq.(\ref{eq:commmarkovianity}) is valid,  we could consider ``time-dependent Lindblad equation'' (as in~\cite{martin_observational_2018,Burgess:2024eng}). 

 Eq. (\ref{eq:blochredfield}) is far from eq. (\ref{eq:lindblad}),  because of the aforementioned memory integration on the RHS of eq. (\ref{eq:blochredfield}). In particular, the integration on the RHS of eq. (\ref{eq:blochredfield}) features the system operator $v_{k}(\eta')$, the coupling $g(\eta')$ and the environmental correlation function $K(\eta,\eta')$. If $K(\eta, \eta') \propto \delta(\eta-\eta')$, however, eq.(\ref{eq:blochredfield}) would reduce to a time dependent Lindblad equation; in fact, the RHS now would feature only $v_{\boldsymbol{k}}(\eta)$.  The Markovianity of the Master equation depends thus strongly on the properties of the environment, and, in particular, on the memory of the environment.  The connection between Markovianity and the properties of the environemnt opens the way to another aspect of Markovianity. In a system described by a Markovian master equation, the information can only flow in one direction: from the system to the environment, as the environment readily dissipates every information received from the system. 

However, the case in which the environment correlation function is just represented by a Dirac Delta is, of course, an ideal case which is not represented exactly by any concrete physical situation. This happens also in cosmology. For example, in many works~\cite{boyanovsky_effective_2015,Burgess:2024eng}, a conformal masselss scalar field  was considered as an environment. The environmental correlation function is in the form $1/(\eta-\eta')$. Thus, even if the correlation function is peaked at $\eta' \simeq \eta$, there are some non negligible tails. As a consequence, when there is some information transferred from the system to the environment, this may be not readily forgotten by the environment, but ``stored'' in these long time tails of the environmental correlation function. This information may possibly return some time later to the system and so render the dynamics of the system dependent on its past values. As we are going to see, this will be the case also for the environment we will consider. 

A Markovian quantum master equation seems not ideal to properly describe these systems, and we will have to consider a Bloch-Redfield equation instead.
\\
\\
\textbf{Positivity problems and the ``Strong Markovian Approximation''}
Whenever we can extract a meaningful result by solving the Bloch-Redfield equation, we should believe it. However, there are some issues with this type of equations. It is known that they do not guarantee that during its evolution the density matrix is always positive definite. This would mean that the result of the evolution may in some specific situation be unphysical. This problem was also observed in a cosmological setting~\cite{colas_quantum_2023, burgess_minimal_2023}; analogously, we will observe in our calculations a similar problem, as shown in eq.(\ref{eq:Dnegativemixedresult}). We will talk diffusely about this problem and how to diagnostic it in section \ref{sec:canonicalform}. These problems may arise because some approximations, in particular the Born-Markov one, are actually not well justified in some specific settings. The memory is thus not properly modelled by the quantum master equation. So, what to do in these cases?

In condensed matter settings, the ``rotating wave approximation''\footnote{For more details on this, see the paragraph before eq.(3.136) in~\cite{Breuer:2007juk}. } is a general presctiption that consents to extract a healthy evolution out of the Bloch-Redfield master equation. In other words, a further approximation is necessary, in order to have a meaningful evolution for the quantum master equation. However, the same type of approximation cannot usually be considered in an inflationary setting, because of the time dependent background, as explained e.g. in~\cite{Colas:2023wxa}.    

So, a possible attempt could be to find another approximation which plays the same role. A possible prescription was suggested by~\cite{burgess_minimal_2023}. Instead of just developing in series the density matrix as we did in eq.(\ref{eq:markovianapprox}) the authors suggested to also develop in series all the system variables, i.e. also the system operators $v_{k}(\eta')$. By explicitly expanding:
\begin{equation}
    \left[{v}_{\boldsymbol{k}}(\eta), {v}_{\boldsymbol{k}}\left(\eta^{\prime}\right) \rho_r\left(\eta^{\prime}\right)\right] \simeq\left[{v}_{\boldsymbol{k}}(\eta), {v}_{\boldsymbol{k}}(\eta) \rho_r(\eta)\right]+\left(\eta^{\prime}-\eta\right)\left[{v}_{\boldsymbol{k}}(\eta),\left[{v}_{\boldsymbol{k}}(\eta) \partial_\eta \rho_r(\eta)+{p}_{\boldsymbol{k}}(\eta) \rho_r(\eta)\right]\right]+\cdots
    \label{eq:smar1}
\end{equation}
and then truncating at the zero order. In the following, we will call this approximation the ``Strong Markovian Approximation''.

A possible justification for this approximation could be that it is not really consistent to just neglect the change in the density matrix, without neglecting the change in the other system variables in a cosmological setting. This approximation relies on the fact that the evolution of the environment correlation function, in a typical time $\tau_{\epsilon}$, is much more rapid than the evolution of the density matrix, in a typical time of the interaction $\tau_{r}$, but also than the evolution of the typical time of the system operators $\tau_{S}\gg\tau_{\epsilon}$. Then, of course, the approximation appears more justified as far as the system operator variables evolve slower than the correlation function of the environment, or in general of what is left inside the integration. Notice that in the interaction picture the system operators evolve with the typical time of free evolution, $\tau_S$, which for superhorizon modes is of order $1/H$.

In cosmology also the coupling constant are time dependent. What we have called derivativeless interaction in eq.(\ref{eq:interactions}), $H_{int1}$, have a coupling constant which grows as $\eta \rightarrow 0$. This means that they are maximum at the endpoint of the memory integration on the RHS of eq.(\ref{eq:blochredfield}), and suppressed before: the relevant interactions in the $\eta \to 0$ expansion thus tend to be Markovian. Instead, the derivative interactions in eq.(\ref{eq:interactions}) decrease as time proceeds, being irrelevant as $\eta \to 0$, but they increase as time goes back: memory effects are thus enhanced. When considering these interactions, we will have problems in considering the correct Markovian approximation. Thus, beyond the environment, another element which should be important in analyzing the Markovianity of a quantum master equation is the relevance in a $1/\eta$ expansion, for $\eta \to 0$ of the interactions: derivativeless interactions are relevant, and thus we will not have problems in applying the machinery of the open quanutm systems. Instead, derivative interactions are irrelevant, and then they are more problematic. This probably requires a proper approximation, to arrive to a physical result. Our approach will thus be to benchmark the Strong Markovian approximation; we will explicitly verify that also in our case it will provide a meaningful result in the end.

\subsection{The environment}
\label{subsection:env}
\subsubsection{Our working hypothesis}
One of the main original points of this work is the environment. We will consider a subhorizon, time dependent, environment of tensor degrees of freedom (i.e. of primordial gravitational waves).~\footnote{As shown in~\cite{burgess_minimal_2023}, there is only a factor 2 that differentiates a scalar environment from a tensorial one. Thus, many of the conclusions we will elaborate in this case, may directly be translated to a scalar environment. For this reason, in the following, when we will talk about scalar environments, e.g. to compare to other models already in the literature, a direct connection can be driven. \label{footnotetensscalar} } In principle, we should consider the entire form of the mode function (eq.(\ref{eq:fullmodefunction})), for tensor degrees of freedom. However, in order to simplify calculations, we choose to focus during this work only on the ``deep subhorizon modes'', which we define to be the ones with $k>10\, aH$. The choice of the constant 10 is totally arbitrary, provided it satisfies sufficiently well the conditions that we are considering for the environment and which we explain below. 

Consider the expression eq.(\ref{eq:fullmodefunction}) for the mode functions. Since for deep subhorizon modes $-k\eta>10$, the $-i/k\eta$ part is suppressed of at least an order of magnitude with respect to the other part of the mode function. 
Therefore, as a work hypotheses, we will neglect the $-i/k\eta$ in eq.(\ref{eq:fullmodefunction}).
Our environmental mode functions are thus:
\begin{equation}
    \theta_k=\frac{e^{-i k \eta}}{\sqrt{2k}} \quad \quad k>10\, aH.
    \label{eq:approxflatenvironment}
\end{equation}
However, as discussed below, this perspective is not complete, and this is why we refer to a work hypothesis and not a justified approximation. 

By neglecting this part of the mode functions, we will find some couples of interactions which give no contribution to either decoherence or observables. 
This happens for the irrelevance of the derivative interaction operators $H_{\rm int 2-3}$ in eq.(\ref{eq:interactions}) in a $1/\eta$ expansion. Instead, derivativeless interactions, $H_{int1}$, are relevant in a $1/\eta$ expansion and will give some contributions to decoherence and correlation functions, also in this approximation, as we show in section \ref{sec:derivativelesscanonical}.  However, when reinstating the $-i/k\eta$ inside the mode functions, the irrelevance of the derivative interaction operators $H_{int2-3}$ would be compensated and these interactions would also probably give some contributions, which by considering just eq.(\ref{eq:approxflatenvironment}) we miss.
Thus, we cannot claim that our results are definitive also for deep subhorizon modes of minimally coupled fields. 
It is possible to consider this work still as an intermediate step towards analyzing the full effect of subhorizon modes on superhorizon modes in inflation. We think that, in this sense, it is more realistic than the models considering an environment of all the modes of a conformal coupled scalar field (see footnote \ref{footnotetensscalar}) as a proxy for subhorizon modes~\cite{boyanovsky_effective_2015}.

By the working hypothesis~eq.(\ref{eq:approxflatenvironment}), we reduce noticeably the number of terms to compute, so simplifying the analysis on this point and allowing us to focus on other original points. These novelties are well underlined in the introduction (see~\ref{sec:bulletpoints}). On the other hand, the technique we employ for the computations can be easily generalized, as the terms generated by reinstating the $-i/k\eta$ contribution in the mode functions produces calculations of the same form as the ones we compute here from the various interactions.  
One of our next goals is to extend them to $-k\eta \simeq1$, i.e  considering all the sub-horizon environmental modes, also the ones crossing the horizon by employing the technique we establish in this work. 

Actually, notice that the working hypothesis that we do is equivalent to considering Minkowski mode functions (i.e. oscillating functions) for the subhorizon environment. Our approximation is thus equivalent to neglecting the effect of the Hubble horizon during inflation on the environment. By comparing our results to the ones we will obtain in a future work featuring also the horizon crossing modes in the environment, we may point out the importance of the effect of being in inflation
not only for the superhorizon modes, which notoriously are affected by a huge particle creation, but also for 
subhorizon environmental modes. The importance of the presence of the horizon for decoherence has been underlined also in other contexts in cosmology~\cite{Danielson:2022sga}.

Nevertheless, we think our work is a first important step in the analysis of time dependent environments in open quantum field theory during inflation which has still to be developed and whose results still miss a full benchmarking, in particular by testing the validity of the numerous cited approximations in the derivation of the quantum master equation (and, above all, the Markovian approximation). 

\subsubsection{Deep subhorizon environment and the validity of the approximations}

Indeed, the choice eq.(\ref{eq:approxflatenvironment}) should go into the direction of strengthening the validity of the Markovian approximation, since more energetic modes are intuitively ``faster''. It is in fact intuitive to think to the characteristic correlation time of the full subhorizon environmental functions, considering also the $-i/k\eta$ terms, as $\tau\simeq 1/H$. In the case of deep subhorizon modes, instead, intuitively, we should have a shorter correlation time, $\tau_{\epsilon} \simeq 1/10H$. 
The typical time of the free evolution of the system $\tau_{S}$ is still of order an Hubble time. This is important, in particular, for the validity of the Strong Markovian approximation, which we will apply in the end to obtain our final results for decoherence and quantum corrections. As we discussed  in section \ref{subsec:the Markovian approximation}, in particular in the paragraph following eq.(\ref{eq:smar1}), the Strong Markovian approximation requires that the typical time of the evolution of the system operators (i.e. the time of free evolution) is much longer than the environmental correlation time. In our case, we have a clear separation of scales between the typical time of free evolution of the system and of the environment, and not only between the typical time of interactions and the environment. This justifies and strengthens the validity of this Strong Markovian approximation in our case.

For the aforementioned intersection between the various approximations of the QME, it is reasonable to expect that also the other approximations that we made in order to derive it should be strenghtened by the choice of a deep subhorion environment. 
 
The \textit{third approximation} stated that the state of the environment, in the interaction picture, is fixed in the Bunch Davies vacuum state. States near to the horizon may instead be excited more easily than ``deep'' subhorizon states. By giving an estimate, we can consider the ``thermal'' interpretation of de Sitter: the typical populations of an excited state is obtained by comparing the ``temperature'', which is essentially $T=\frac{H}{2\pi}$, with the energy $E$, $e^{-E/T}$; in our case the maximum energy of the environmental vacuum state is further suppressed by a factor 10 in the negative exponent; this is a huge suppression.

\medskip
\subsubsection{A time dependent environment: comparison with the literature}
\label{subsubsec:timedependentenvironment}
\begin{figure}
    \centering
    \includegraphics[scale=0.8]{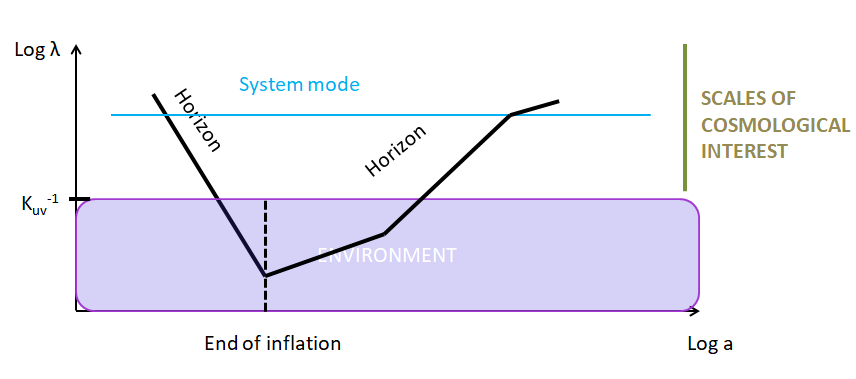}
    \caption{In this figure we show the environment/system setting that was analyzed in~\cite{burgess_minimal_2023}. In this case, the authors considered a fixed separation between environment and system at a wavelength $k_{UV}^{-1}$. Notice that the results obtained in~\cite{burgess_minimal_2023} about decoherence consider the interval of time since when at least part of the environment becomes superhorizon. }
    \label{fig:fixedenvironment}
\end{figure}

 The other aspect which is innovative about our choice of the environment is that it is a {\it time dependent environment}. In particular, the cutoff for environmental modes is time dependent, and thus some modes that initially are part of the environment, as inflation goes by, can leave the environment. This goes in a different direction with respect to other models  where the environment is kept fixed at a certain comoving scale, as, for example,~\cite{burgess_minimal_2023}. The framework is sketched in fig.(\ref{fig:fixedenvironment}). In~\cite{burgess_minimal_2023}, the cutoff scale is fixed around 2500 times smaller than the largest scales observable in cosmology, i.e. it is chosen to correspond to the smallest observable cosmological  scale today. Consequently, the scale of the cutoff will cross the horizon at a certain point during inflation. Therefore, naturally, also some of the modes in the environment will become superhorizon. 
 
 In~\cite{burgess_minimal_2023}, the results are computed just taking into account the superhorizon part of the environment, since for the interaction considered (corresponding to $H_{int2}$ in our eq.(\ref{eq:interactions})), the results are dominant in that region.\footnote{As we will prove in section \ref{sec:derivativeinteractions}, our results are coherent with this claim, since for a deep subhorizon environment we have negligible contributions to both decoherence and corrections to the power-spectrum from interactions of this kind, which we call derivative.} Notice that the fixed cutoff of the environment first crosses the horizon around 8 e-folds after the CMB scales do. In a minimal scenario, only with gravitational interactions,decoherence, for the smallness of gravitational couplings, happens at least around 13 e-folds after horizon crossing of the system mode. As a consequence, a fixed cutoff as the one defined before has already crossed the horizon. However, in situations where couplings between short wavelength and long wavelength modes are stronger, decoherence may be efficient before 8 e-folds. In those cases, a fixed cutoff is still very deeply subhorizon, and so all the modes of the environment would be even more deeply subhorizon. At the same time, there will be some subhorizon modes which are would not be included in the environment with the fixed cutoff. Instead, an environment featuring all the subhorizon modes would, in this situation, include more modes in the environment and can be expected to be more efficient in decoherence. 
 
 In this sense, we believe our work is complementary to the one of~\cite{burgess_minimal_2023}, and that an analysis on a subhorizon environment could be helpful in order to generalize to stronger couplings our works.

In order to have always only subhorizon modes in the environment, we necessarily have to consider a time dependent environment. In the open quantum field theory applications to cosmology, this has been done by many papers in the past:~\cite{brahma_universal_2022},~\cite{brahma_quantum_2022},~\cite{gong_quantum_2019}.
We stress we are going to consider different interactions with respect to these works. Also, differently from these works, we will also be the first to actually justify the choice of a time dependent environment, and compute the corrections to the results due to this choice in section \ref{subsec:timedependentenvironment}.

\begin{figure}
    \centering
    \includegraphics[scale=0.8]{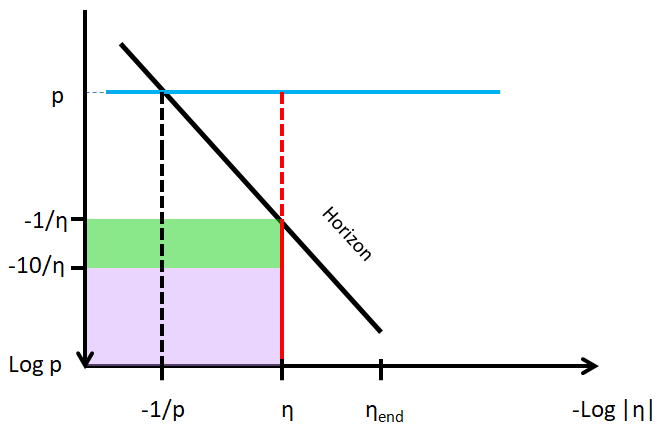}
    \caption{In this graph we try to give an image of the modes we are considering and the ones we are missing in our approach. On the $x$ axis we have conformal time, in a logarithmic scale, which starts from $\eta_{0}$ and arrives to the end of inflation $\eta_{f}$. On the $y$ axis we have the comoving wavenumbers. $p$ is the wavenumber of the system. Notice that the arrow of the $y$ axis goes downside, because we plot, as customary in this type of diagrams, the modes with larger wavelength (e.g smaller wavenumber) in the upper part of the diagram.  The mode $p$ goes out of the horizon and begins to be considered at that moment; then, at a generic moment $\eta$ we only consider as environment the modes which have always been deeply subhorizon (i.e. $k>10 aH$), from the beginning until $\eta$, indicated in purple. We have omitted the green part, by not considering the modes near to horizon crossing, left to further studies, but which can be straightforwardly included in our framework.}
    \label{fig:whatmissing}
\end{figure}

We have sketched the modes we are considering and those we are missing in our sub-horizon environment in Fig.\ref{fig:whatmissing}. In particular, by considering a certain system-mode with momentum $p$, after it first crosses the horizon, at every instant of time $\eta$ we consider as an environment all the modes with momentum $k>-10/\eta$. Consequently, as already stressed, we are missing in our picture the modes which are actually crossing the horizon, and it would be interesting to explore in a future work how important  they are as an environment.

\subsubsection{Computing the correlation function}
The effect of the environment is encoded in the correlation functions, $K(p,\eta,\eta')$, which we defined in eq.(\ref{eq:correlationfunctionsdef}). 

For simplicity, we begin by computing the environmental correlation function just for the derivativeless couple of interaction $H_{int1}-H_{int1}$, as defined in eq.(\ref{eq:interaction}). 

\begin{figure}
\begin{minipage}{.49\textwidth}
   \includegraphics[width=.9\linewidth]{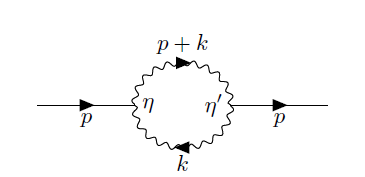}
\end{minipage}
\begin{minipage}{.49\textwidth}
    \includegraphics[width=.9\linewidth]{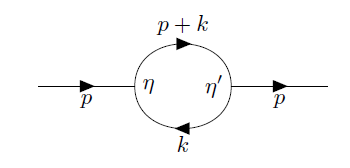}
\end{minipage}
\caption{Feynman diagrams for tensor and scalar environment. The loop represents the effect of the environment acting on the system (the two external legs) thanks to two trilinear interactions, i.e. vertices; notice that the vertices refer to interactions happening, in principle, in different instants of time $\eta$ and $\eta'$. }
\label{fig:1feyn}
\end{figure}

The others will be straightforward generalizations. Since the environment is imposed to be fixed in Bunch-Davies vacuum state, the correlation functions that we have to compute are:
\begin{equation}
    { }_\epsilon\left\langle 0\left|\theta_{i j}(\eta) \theta_{i j}(\eta) \theta_{a b}\left(\eta^{\prime}\right) \theta_{a b}\left(\eta^{\prime}\right)\right| 0\right\rangle_\epsilon,
    \label{eq:environmentalcorrelationfunctionfull}
\end{equation}
where $\theta_{ij}$ are the normalized tensor perturbation fields. By using the Wick Theorem this reduces to compute all the possible couples of correlation; for the present moment, we ignore the tadpoles disconnected contribution. As explained in detail in e.g.~\cite{Frob:2012ui,burgess_minimal_2023} the  correlation function can be expressed as:
\begin{equation}
    \left\langle 0\left|\theta^{a b}(\eta, \boldsymbol{q}) \theta^{c d}\left(\eta^{\prime},\boldsymbol{k} \right)\right| 0\right\rangle=\delta^3(\boldsymbol{k}+\boldsymbol{q}) \frac{1}{2 k} e^{-i k\left(\eta-\eta^{\prime}\right)} P^{\text {abcd }}(\boldsymbol{k}) .
    \label{eq:envcorrelationfunctionsinglecouple}
\end{equation}
where $P^{\text {abcd }}$ is the polarization term:
\begin{equation}
    P^{\text {abcd }}(\boldsymbol{k})=\frac{1}{2}(P^{a d}(\boldsymbol{k}) P^{b c}(\boldsymbol{k})+P^{a c}(\boldsymbol{k}) P^{b d}(\boldsymbol{k})-P^{a b}(\boldsymbol{k}) P^{c d}(\boldsymbol{k})),
\end{equation}
and the tensors $P^{i j}(\boldsymbol{k})=\delta_i^j-\frac{k^i k^j}{k^2}$ are essentially projectors. We can summarize the environmental correlation functions in a diagram in spirit similar to a Feynman diagram, see Fig.~\ref{fig:1feyn}.   
The amplitude we obtain for the correlation function is:
\begin{equation}
    K\left(p, \eta, \eta^{\prime}\right)=\int_{10 a H}^{+\infty} \frac{d^3 k}{(2 \pi)^3} \frac{1}{|k|} \frac{1}{2|p+k|} e^{\left.-i k\left(\eta-\eta^{\prime}\right)\right)} e^{\left.-i|\boldsymbol{p}+\boldsymbol{k}|\left(\eta-\eta^{\prime}\right)\right)} P^{i j k l}(\boldsymbol{p}+\boldsymbol{k}) P_{i j k l}(\boldsymbol{k}).
    \label{eq:corrfunctionexplicit}
\end{equation}
Notice the factor 2 which is due to the two possible  unequal time contractions in the environmental correlation function.

Now we have to make an assumption that greatly simplifies the calculations. For how we considered our environment, the deep subhorizon modes momenta are always at least 10 times bigger than the superhorizon momenta; besides, we will be interested mostly in the deep superhorizon evolution, e.g. after the system modes have crossed the horizon since a couple e-folds at least. As a consequence, we can state $p\ll aH\ll k$, with the system and environment momenta $p$ and $k$. It is possible to explicitly check\footnote{see, e.g., equation (E.31) of~\cite{burgess_minimal_2023}.} that this approximation enters quadratically in the expansion $p/k$, so that the corrections to the correlation functions are at least order 0.01 for $p/k \simeq 0.1 $; as a consequence, since other approximations we will do are probably more influent, we can simply send $p \rightarrow 0$ in our following computations. This strikingly simplifies the correlation function, making it dependent only on time and not on the superhorizon momenta:
\begin{equation}
K\left(\eta, \eta^{\prime}\right)=\int_{10 a H}^{+\infty} \frac{d^3 k}{(2 \pi)^3} \frac{2}{2|k|} \frac{1}{2|k|} e^{-i k\left(\eta-\eta^{\prime}\right)} e^{-i|\boldsymbol{k}|\left(\eta-\eta^{\prime}\right)} P^{i j k l}(\boldsymbol{k}) P_{i j k l}(\boldsymbol{k})\, .
\end{equation}
The contraction between the two polarization tensors can then be straightforwardly evaluated, by using the well known properties of the projector $P^{i j}$:
\begin{equation}
    P^{a b} P_{a b}=\left(\eta^{a b}-\frac{p^{a} p^{b}}{p^{2}}\right)\left(\eta_{a b}-\frac{p_{a} p_{b}}{p^{2}}\right)=3+1-2=2 \quad\quad P^{i a} P_{a b}=P^a_b
\end{equation}
\begin{equation}
\begin{split}
   &P^{i j k l}(\boldsymbol{k}) P_{i j k l}(\boldsymbol{k})=\frac{1}{4}\left(P^{i l} P^{j k}+P^{i k} P^{j l}-P^{i j} P^{k l}\right)(\boldsymbol{k})\left(P_{i l} P_{j k}+P_{i k} P_{j l}-P_{i j} P_{k l}\right)(\boldsymbol{k})=\\
   &\frac{1}{4}(4+4+4+P_{k}^{l} P_{l}^{k}-P_{j}^{l} P_{l}^{j}+P_{k}^{l} P_{l}^{k}-P_{j}^{k} P_{k}^{j} -P_{j}^{l} P_{l}^{j}-P_{j}^{k} P_{k}^{j})=\frac{1}{4}(4+4+4-2-2)=2\, .
   \label{eq:polarizationssummed}
\end{split}
\end{equation}
The large simplification arises also in evaluating now the angular part of the integral in the environmental functions:
\begin{equation}
    K\left(\eta, \eta^{\prime}\right)=\frac{1}{8 \pi^2} \int_{10 a H}^{\infty} d k e^{-2 i k\left(\eta-\eta^{\prime}\right)} 4. 
\end{equation}

The upper extremum of the integration over the momenta tends to infinity. Thus,
the integrand will contain highly oscillating functions in the high momenta region. The upshot is that we won't be able to assign a numerical value to it. As costumary in the literature, and following in particular the method adopted by~\cite{boyanovsky_effective_2015,Burgess:2024eng}, we can employ the trick to rotate $\eta \rightarrow \eta-i \epsilon$, with $\epsilon >0$. In this way, $\epsilon$ becomes also an UV cutoff, regularizing the coincidence limit $\eta \rightarrow \eta'$. By performing the integration, we obtain

    \begin{equation}
\begin{split}
   & K(\eta,\eta')=\frac{1}{8 \pi^2} \int_{10aH}^{\infty}dk e^{-2ik(\eta-\eta'-i\epsilon)} \cdot 4 =\frac{4}{8 \pi^2 } \frac{1}{-2i}\left[\frac{e^{-2ik(\eta-\eta'-i\epsilon)}}{(\eta-\eta'-i\epsilon)}\right]_{10aH}^{+\infty}=\\
   &=\frac{1}{-i4\pi^2}\left(\frac{e^{-\epsilon \infty}- e^{-20iaH(\eta-\eta')}}{(\eta-\eta'-i\epsilon)}\right)=\frac{1}{i4\pi^2}\frac{e^{20i\frac{(\eta-\eta')}{\eta}}}{\eta-\eta'-i\epsilon} \,\,  , 
   \label{1Kdef}
   \end{split}
\end{equation} 
where we have used the de Sitter approximation eq.(\ref{scalefactor}). Notice that, while similar to the expression for the correlation function in the conformal field environment used by Boyanovsky (see, e.g., eq.(5.3) in~\cite{boyanovsky_effective_2015}), there are some differences. Boyanovsky considers all the modes of an environment of a conformal scalar field, from infinity down to the environmental momentum $k \rightarrow 0$. Unlike this, in the integration in eq.(\ref{1Kdef}), we have an infrared cutoff at $10 aH$. The integrand is the same. In the Boyanovsky case, after computing the integral, the value for the environmental momentum zero explicitly shows the momentum of the system; in the end the author applies the limit of the system momentum going to zero $p \rightarrow 0$. This kills the exponent, and thus the correlation functions presents no oscillating function in it. In our case, instead, the exponent is never null, and is exactly determined by the infrared cutoff at $10aH$. The result is that we have an oscillating function as in the last equality of eq.(\ref{1Kdef}). This oscillation will be found also when considering an environment constituted by all the subhorizon modes, and not just the deep, and in this sense it represents a step towards more complete calculations.  Coming back to the form of eq. (\ref{1Kdef}) turns out useful to rewrite it by considering the dimensionless variable $x=\frac{\eta'}{\eta}$:
\begin{equation}
    K(x,\eta)=\frac{1}{i4\pi^2\eta}\frac{e^{-20i(x-1)}}{1-x+i\epsilon}\, .
        \label{eq:correlationfunctionK}
\end{equation}
Notice the abuse of notation by writing again $\epsilon$ even though we have divided by $\eta$; being the conformal time negative in inflation, we have changed sign in front of  $\epsilon$ to keep it positive. 

In order to deal with the coincident limit, we consider the Sokhotsi-Plemelj theorem:
\begin{equation}
\frac{1}{x-1 \pm i \varepsilon}=\mp i \pi \delta(x-1)+P.V. \left(\frac{1}{x-1}\right)=\mp i \pi \delta(x-1)+\lim_{\epsilon \to 0} \frac{x-1}{(x-1)^2+\epsilon^2},
\end{equation}
where $P.V.$ denotes the Principal Value.
By substituting this expression inside the previous equation  we get 
\begin{equation}
    K\left(\eta, x\right)=\frac{1}{4 \pi^2 i\eta} e^{20 i(1-x)} \operatorname{P.V.} \left(\frac{1}{1-x}\right)-\frac{1}{4 \pi \eta} \delta(1-x)\, .
\end{equation}
For later purposes, we can divide the correlation function in the Real and Imaginary parts as 
\begin{equation}
   \operatorname{Im} K=-\frac{1}{4 \pi^2 \eta} \cos(20(x-1)) \operatorname{P.V.}\left(\frac{1}{1-x}\right), \quad \quad \operatorname{Re} K=-\frac{1}{4 \pi \eta} \delta(1-x) +\frac{\operatorname{sin}(20(1-x))}{4\pi^2 \eta} \operatorname{P.V.}\left(\frac{1}{1-x}\right)\, .
   \label{eq:correfunctionrealnonreal}
\end{equation}
We will show in detail how the Real part is the only one contributing to decoherence and amplitude corrections; the imaginary part instead contributes to the spectral index corrections.

In the real part we have a Markovian Dirac delta contribution and a non-local in time contribution. We will comment diffusely on these later, but the non-local in time contribution will necessarily call for a non-Markovian master equation in order to be properly accounted for. The corrections to the spectral index, sourced by the modification of the Hamiltonian that we called Lamb Shift in the introduction, instead are only sourced by the non-local imaginary part, as already found by~\cite{boyanovsky_effective_2015}. This is the part of the ``loop'' diagrams we have shown in fig.\ref{fig:1feyn} which will  be authomatically resummed by the Quantum master equation.

Keeping in mind the limit $p \rightarrow 0$, the correlation functions for the mixed couple of interactions (i.e. the couples which feature a derivative and a derivativeless interactions in the two vertices of the loop in \ref{fig:1feyn})
and the derivative interactions will be of the type:
\begin{equation}
    K\left(\eta, \eta^{\prime}\right)\propto  \int_{10 a H}^{\infty} d k e^{-2 i k\left(\eta-\eta^{\prime}\right)} k^{2n} \quad \quad n=1,2,
    \label{eq:dercorrform}
\end{equation}
with a coefficient in front determined by the angular part. The trick we will employ is to derive $2n$ times with respect to $\eta'$:
\begin{equation}
     K\left(\eta, \eta^{\prime}\right)\propto  \int_{10 a H}^{\infty} d k\frac{1}{(2i)^n} \frac{d^{2n}}{d\eta'^{2n}}e^{-2 i k\left(\eta-\eta^{\prime}\right)}.
     \label{eq:dercorreq}
\end{equation}
For $n=1$, this will add a minus sign in front; for $n=2$ no change in sign instead. This observation will also prove to be important later.
We underline that the difference in all the couples of interactions that we are considering from now on relies only on the form of the correlator and the power of the coupling.

\subsection{Canonical form of the quantum master equation}
\label{sec:canonicalform}
We have already seen that when the dynamics is Markovian, with time independent map, the Lindblad theorem guarantees that there is a unique equation that can describe the dynamics while maintaining positivity, hermiticity and trace unit of the density matrix. We stress that, as customary in condensed matter settings, in the eq.(\ref{eq:lindblad}) the $\gamma_{k}$ are time independent.
In eq.(\ref{eq:lindblad})
we have to impose,  however, that $\gamma_{k}$ are all non negative; if not, the theorem states that the density matrix may evolve in an unphysical way. In particular, it may happen that the density matrix does not preserve the positivity of its eigenvalues. 

However, we know that in our case $\gamma_{k}(\eta)$ are time dependent. This happens for two reasons: because the coupling constant are time-dependent (see discussion after eq.(\ref{eq:lindblad})), and because we may have non-Markovian effects.

In the open quantum system literature non-Markovianity is a widely explored topic, and still an active field of research (see e.g.~\cite{Breuer_2016,Rivas_2014} for reviews). A way to analyze non-Markovian effects in a master equation is writing the equation in a canonical form akin to eq.(\ref{eq:lindblad}) (see, e.g.~\cite{Hall_2014}, in the open quantum system literature, or in cosmology, 
~\cite{colas_benchmarking_2022, burgess_minimal_2023, colas_quantum_2023}\dots):
\begin{equation}
     \dot{\rho}=-\mathrm{i}[H+H_{ls}, \rho]+\sum_k \gamma_k(t)\left(\hat{L}_k \rho \hat{L}_k^{\dagger}-\frac{1}{2}\left\{\hat{L}_k^{\dagger} \hat{L}_k, \rho\right\}\right),
     \label{eq:noncanonicallind}
\end{equation}
\footnote{Notice that with respect to the previous paragraph \ref{subsection:env} we have employed different label to denote the system momentum: instead of using $p$, as also indicated in fig. \ref{fig:whatmissing}, we will use, from now on, $k$, always to indicate the system momentum. }where the main differences, with respect to the Lindblad equation, are due to:
\newline

1)~The Lamb Shift Hamiltonian; the unitary evolution is dressed by the interaction with the environment and consequently there is an Hamiltonian piece to add to the system one;
\\
\\
2)~The coefficients $\gamma_{k}$ are now time dependent. Of course, if they are always positive, it is easy to see that it is a sufficient condition in order to guarantee a ``healthy'' evolution of the density matrix; however, if some of them are negative, it means that violations of positivity \textit{may} show up. It is important to underline that there is no direct implication in what we are stating here: there are counterexamples (e.g.~\cite{Whitney_2008}) in which negative coefficients have lead to a physical evolution of the reduced density matrix. The presence of negative coefficients is even often cited as a criterium for non-Markovianity~\cite{Breuer_2016}.
\newline
 The goal of this section is to put the quantum master equation eq.(\ref{eq:blochredfield}) in a form similar to eq.(\ref{eq:noncanonicallind}). As we are going to see, the $\gamma_{k}$ coefficients of eq.(\ref{eq:noncanonicallind}) contain in themselves the integration over the past history of the system that we can see in the quantum master equation eq.(\ref{eq:blochredfield}).

As a first step, we separate the real and imaginary part of the correlation function $K$ as in eq.(\ref{eq:correfunctionrealnonreal}):
\begin{equation}
    \begin{aligned}
&
\frac{\mathrm{d} \rho_{\boldsymbol{r}}}{\mathrm{d} \eta}= -g(\eta) \sum_{\boldsymbol{k}} \int_{\eta_0}^\eta \mathrm{d} \eta^{\prime} g(\eta')(\Re \left[\mathcal{K}\left(\eta, \eta^{\prime}\right)\right] \left[v_{\boldsymbol{k}}(\eta),\left[v_{-\boldsymbol{k}}\left(\eta^{\prime}\right), \rho_{r}(\eta)\right]\right]+ \\
& +i \Im \left[\mathcal{K}\left(\eta, \eta^{\prime}\right)\right]\left[v_{\boldsymbol{k}}(\eta),\left\{v_{-\boldsymbol{k}}\left(\eta^{\prime}\right), \rho_{r}(\eta)\right\}\right]).\\
\label{eq:quantummastereq}
\end{aligned}
\end{equation}
The next step is to express $v_{-\boldsymbol{k}}(\eta')$ in function of operators evaluated at $\eta$. In order to do that consider eq.(\ref{eq:decomposition}) for $v_{\boldsymbol{k}}(\eta)$ and $v_{\boldsymbol{k}}(\eta')$, and for the conjugate momentum $p_{\boldsymbol{k}}(\eta)=v'_{\boldsymbol{k}}(\eta)$ separately; then, express $c_{\boldsymbol{k}}$ as a function of $p_{\boldsymbol{k}}(\eta)$ and $v_{\boldsymbol{k}}(\eta)$. This can be done by multiplying $v_{\boldsymbol{k}}(\eta)$ by $u'_{k}(\eta)$, and $p_{\boldsymbol{k}}(\eta)$ by $u_{k}(\eta)$, then subtracting side by side, using eq.(\ref{eq:wronskian}), we obtain:
\begin{equation}
    i c_{\boldsymbol{k}}=u'_{k}(\eta)v_{\boldsymbol{k}}(\eta)-u_{k}(\eta)p_{\boldsymbol{k}}(\eta).
\end{equation}
Substituting back in eq. (\ref{eq:decomposition}), the final result reads:
\begin{equation}
\begin{split}
 &v_{\boldsymbol{k}}(\eta')=A(\eta,\eta',k) v_{\boldsymbol{k}}(\eta)+B(\eta,\eta',k) p_{\boldsymbol{k}}(\eta)=\\
&=(i) \left(u^{*}_{k}(\eta')u'_{k}(\eta)-u_{k}(\eta')u'^{*}_{k}(\eta)\right) v_{\boldsymbol{k}}(\eta)+(i) p_{\boldsymbol{k}}(\eta)\left(-u^{*}_{k}(\eta')u_{k}(\eta)+u_{k}(\eta')u^{*}_{k}(\eta)\right),\\
\label{eq:AB}
\end{split}
\end{equation}
where we have introduced for the first time the expressions A and B:
\begin{equation}
    \begin{split}
&A=2 \Im \left(u_{k}(\eta')u'^{*}_{k}(\eta)\right)\\
&B=2 \Im\left(u_{k}(\eta)u^{*}_{k}(\eta')\right).
    \end{split}
    \label{eq:precise definition AB}
\end{equation}
Notice that both A and B are real numbers. These two expressions encode the memory expressed by the system (in this case cosmological perturbations, that we assume in our approximation to be massless minimally coupled fields). In particular, it is important to underline that we consider the modes belonging to the system since when they cross the horizon. Thus, we can identify the initial moment of integration $\eta_{0}$ with $-1/k$, the conformal time at which the mode of interest in the system crosses the horizon. 
If we consider the approximation of a massless scalar field, it is possible to explicitly compute the value of both $A$ and $B$:
\begin{equation}
     \begin{split}
&A=\sin{k(\eta-\eta')}\left(-\frac{1}{k}\left(\frac{1}{\eta}-\frac{1}{\eta'}\right)-\frac{1}{k^3\eta^2\eta'}\right)+\cos{{k(\eta-\eta')}}\left(1-\frac{1}{k^2\eta^2}\left(1-\frac{\eta}{\eta'}\right)\right);\\
&B=\frac{ \sin{k(\eta-\eta')}}{k}\left(1+\frac{1}{k^2\eta \eta'} \right)+\frac{1}{k}\left(\frac{1}{\eta}-\frac{1}{\eta'}\right) \frac{\cos{k(\eta-\eta')}}{k}.
    \end{split}
    \label{eq:explicitdefinition AB}
\end{equation}
These equations can be further simplified by noticing that they depend essentially on two expressions that we introduce here, and which will prove useful for the rest of paper, namely:
\begin{equation}
    a=-\frac{1}{k\eta}\,  \quad\quad x=\frac{\eta'}{\eta}.
    \label{eq:defxa}
\end{equation}
By substituting we finally arrive at:
\begin{equation}
\begin{split}
    & A(x,a)=\cos{\frac{1-x}{a}}(1-a^2(1-1/x))-\sin{\frac{1-x}{a}}\left(a\left(1-\frac{1}{x}\right)+\frac{a^3}{x}\right);\\
    &k \, B(x,a)=\sin{\frac{(1-x)}{a}}\left(1+\frac{a^2}{x}\right)+\cos{\frac{1-x}{a}} a \left(1-\frac{1}{x}\right).\\
    \end{split}
    \label{eq:ABwithxanda}
\end{equation}
Notice that the price to pay to make the equation local in time is to introduce also the variable $p_{\boldsymbol{k}}(\eta)$ in the quantum master equation; consequently, we have to write a 2x2 matrix of the coefficients instead of a single one for each mode k associated to the system. In particular, by substituting $v_{\boldsymbol{k}}(\eta')$ with $v_{\boldsymbol{k}}(\eta)$ and $p_{\boldsymbol{k}}(\eta)$ it is possible to see that the equation (\ref{eq:quantummastereq}) can be written as:
\begin{equation}
    \begin{aligned}
& \frac{d \rho_r}{d \eta}=g(\eta) \sum_{\boldsymbol{k}}\int_{-\frac{1}{k}}^\eta d \eta^{\prime} g\left(\eta^{\prime}\right) 
[2 \Re K(\eta,\eta') A(k,\eta,\eta') \left(v_{\boldsymbol{k}}(\eta) \rho_r(\eta) v_{-\boldsymbol{k}}(\eta)-\frac{1}{2}\left\{v_{\boldsymbol{k}} v_{-\boldsymbol{k}}(\eta), \rho_r(\eta)\right\}\right)+\\
&+B(k,\eta,\eta') K^{*}(\eta,\eta')  \left(v_{\boldsymbol{k}}(\eta) \rho_r(\eta) p_{-\boldsymbol{k}}(\eta)-\frac{1}{2}\left\{p_{-\boldsymbol{k}}(\eta) v_{\boldsymbol{k}}(\eta), \rho_r(\eta)\right\}\right)+\\
&+B(k,\eta,\eta') K(\eta,\eta')  \left(p_{-\boldsymbol{k}}(\eta) \rho_r(\eta) v_{\boldsymbol{k}}(\eta)-\frac{1}{2}\left\{ v_{\boldsymbol{k}}(\eta) p_{-\boldsymbol{k}}(\eta), \rho_r(\eta)\right\}\right)]-\\
&-i g(\eta) \int_{-\frac{1}{k}}^\eta d \eta^{\prime} g\left(\eta^{\prime}\right) \Im K(\eta, \eta') (A(k,\eta,\eta') [v_{\boldsymbol{k}}(\eta) v_{-\boldsymbol{k}}(\eta), \rho_r(\eta)]+B(k,\eta,\eta') [v_{\boldsymbol{k}}(\eta) p_{-\boldsymbol{k}}(\eta), \rho_r(\eta)]).
\label{eq:canonicalqme}
\end{aligned}
\end{equation}
Since the interactions (\ref{eq:interactions}) are always linear in the field, it is possible to prove that the density matrix evolution of each momentum $\boldsymbol{k}$ is actually independent, so the density matrix is factorizable:
\begin{equation}
    \rho_r(\eta)=\bigotimes_{k \in S } \rho_{\boldsymbol{k}}(\eta),
\end{equation}
and we can separately consider the equation eq.(\ref{eq:canonicalqme}) for each mode of the sum over momenta on the RHS.

In eq.(\ref{eq:canonicalqme}) we have clearly separated the Lamb shift, unitary contributions corresponding to the dressing from the environment of the unitary dynamics, and the non-unitary ones. We focus first on the non-unitary part of the quantum master equation. It is possible to write the coefficients in a matrix $\mathcal{D}_{k, m n}$ (analogously to eq. (3.26) of~\cite{burgess_minimal_2023}, or to eq. (3.30) in~\cite{colas_benchmarking_2022}), if we rewrite the quantum master equation as
\begin{equation}
     \frac{d \rho_k}{d \eta}=\sum_{n, m=1}^2 \mathcal{D}_{k, n m}\left[O_{k, n} \rho_k(\eta) O_{k, m}^{\dagger}-\frac{1}{2}\left\{O_{k, m}^{\dagger} O_{k, n}, \rho_k(\eta)\right\}\right],
\end{equation}
where where in particular $O_{k 1}=v_{\boldsymbol{k}}$ and $O_{k 2}=p_{\boldsymbol{k}}$; the matrix $\mathcal{D}_{k, n m}$ is defined as:
\begin{equation}
    \left(\begin{array}{ll}
g(\eta) \int_{-\frac{1}{k}}^\eta d \eta^{\prime} g\left(\eta^{\prime}\right) 
2 \Re K(\eta,\eta') A(k,\eta,\eta') & \quad g(\eta) \int_{-\frac{1}{k}}^\eta d \eta^{\prime} g\left(\eta^{\prime}\right) B(k,\eta,\eta') K^{*}(\eta,\eta')  \\
 g(\eta) \int_{-\frac{1}{k}}^\eta d \eta^{\prime} g\left(\eta^{\prime}\right) B(k,\eta,\eta') K(\eta,\eta') &    
 \quad\quad\quad\quad\quad0
\end{array}\right).
\label{eq:mathcalDmatrix}
\end{equation}
The $\mathcal{D}_{k,mn}$ matrix is called the dissipation matrix.
However, in order to consider the non-unitary features of the evolution, we will see that it is useful to define also a matrix whose coefficients are only dependent from the \textit{real} part of the correlation functions. In this way it is possible to introduce the matrix $D_{k,m n}$ as:
\begin{equation}
    \left(\begin{array}{ll}
g(\eta) \int_{-\frac{1}{k}}^\eta d \eta^{\prime} g\left(\eta^{\prime}\right) 
2 \Re K(\eta,\eta') A(k,\eta,\eta') & \quad g(\eta) \int_{-\frac{1}{k}}^\eta d \eta^{\prime} g\left(\eta^{\prime}\right) B(k,\eta,\eta') \Re K(\eta,\eta')  \\
 g(\eta) \int_{-\frac{1}{k}}^\eta d \eta^{\prime} g\left(\eta^{\prime}\right) B(k,\eta,\eta') \Re K(\eta,\eta') &    
 \quad\quad\quad\quad\quad0
\end{array}\right).
\label{eq:Dmatrix}
\end{equation}
The $D_{k,m n}$ matrix is sometimes referred to as the diffusion matrix, and it will prove to be useful in the following.

As observed in~\cite{burgess_minimal_2023}, the eigenvalues of the dissipation matrix $\mathcal{D}_{k,mn}$ correspond to the $\gamma_{k}(\eta)$ in eq.(\ref{eq:noncanonicallind}); so, a sufficient condition for the dynamics to be completely positive can be stated as the fact that both the eigenvalues are non negative. However, what we are going to show is that the off diagonal elements of the matrix, featuring $B(k,\eta,\eta')$ are, for all the cases at our use, negligible. Effectively, both the $D$ and $\mathcal{D}$ matrix can be approximated by a diagonal matrices without any problem, with eigenvalues 0 and the $D_{1 1}$. In order to show it, in the section \ref{par:themixedterm}  we are going to perform the calculations for the derivativeless case (which is also the one giving the most important results); but for now, we invite the reader to a ``leap of faith''. Essentially, what we are saying is that the only non zero term, for both $D$ and $\mathcal{D}$ matrix, is $D_{1 1}$, and its sign determines if we may apply the sufficient condition expressed by the Lindblad theorem, which would ensure us that the dynamics is physical for sure. 

The Lamb shift Hamiltonian can in the same way be written by means of a matrix $\Delta_{ij}$, as:
\begin{equation}
H_{LS}(\eta)= \frac{1}{2}\left(\begin{array}{ll}
v & p
\end{array}\right) \Delta \left(\begin{array}{l}
v \\
p
\end{array}\right), 
\label{eq:delta11def}
\end{equation}
where $\Delta_{ij}$ corresponds to:
\begin{equation}
\Delta=    \left(\begin{array}{ll}
2 g(\eta) \int_{-\frac{1}{k}}^\eta d \eta^{\prime} g\left(\eta^{\prime}\right) 
 \Im K(\eta,\eta') A(k,\eta,\eta') & \quad  g(\eta) \int_{-\frac{1}{k}}^\eta d \eta^{\prime} g\left(\eta^{\prime}\right) B(k,\eta,\eta') \Im K(\eta,\eta')  \\
 g(\eta) \int_{-\frac{1}{k}}^\eta d \eta^{\prime} g\left(\eta^{\prime}\right) B(k,\eta,\eta') \Im K(\eta,\eta') &    
 \quad\quad\quad\quad\quad0
\end{array}\right)
\label{eq:deltamatrix}
\end{equation}
The by far dominant term in the matrix is $\Delta_{1 1}$, analogously to what has been said for $D$ matrix, and consequently we can infer that the net effect of the $H_{L S}=\frac{1}{2} \Delta_{1 1} v_{\boldsymbol{k}}(\eta) v_{-\boldsymbol{k}}(\eta)$ is just to renormalize the mass. 

In order to take into account this effect, we can come back to the Heisenberg representation and reintroduce the free evolution of system perturbations in our equations. By writing down the Heisenberg equation for the field, equivalent to eq.(\ref{eq:freeeom}), we have simply to change the value of $\nu=3/2$ for a massless minimally coupled field by inserting a perturbatively small mass to 
\begin{equation}
\nu_{LS}=\frac{3}{2}-\frac{1}{3}\left(\frac{m^2}{H^2}+ \Delta_{1 1} \eta^2\right).
\label{eq:numassren}
\end{equation}.
The mode function solutions to the new equations of motion are now Hankel functions, as indicated in the first equality of eq.(\ref{eq:fullmodefunction}), with the appropriate $\nu$. Of course, the regime in which we consider these solutions is the superhorizon one, in the late time limit. By the decomposition of the perturbations field in growing and decaying modes exposed in eq.(\ref{eq:growingdecayingexpansion}), it is clear that the decaying modes go to zero rapidly. When computing the power-spectrum, it is thus a very good approximation in slow roll inflation to consider only the dependence on the growing mode, e.g. $g_{+}(k,\eta)$, squared. The power-spectrum can so be expressed as:
\begin{equation}
    P_{v v}(k, \eta) \propto (-k\eta)^{(-3+\frac{2 \Delta_{11}\eta^2}{3})},
    \label{eq:approxcorretopowerspectrum}
\end{equation}
and the final result is effectively a resummation of the loop correction $\Delta_{1 1}$ and it is just a modification of the spectral index. 

\subsubsection{Transport equations}

In the following, we will obtain again the result of eq.(\ref{eq:approxcorretopowerspectrum}) by following~\cite{colas_benchmarking_2022}, writing and solving the transport equation, that we introduce below. Despite the fact that, in this paragraph, we will show that we arrive at the same result, such a  computation gives us the opportunity to derive the non-unitary contributions to the power-spectrum, analogous to the ones found e.g. in~\cite{martin_observational_2018}. 
The transport equations are the equations of motion for the power spectra of $v$ and $p$, and the mixed one; all these variables could be encoded in the covariance matrix, defined as
\begin{equation}
    \Sigma(\eta)=\left(\begin{array}{cc}
P_{vv}(\eta) & P_{vp}(\eta)\\
P_{vp}(\eta) & P_{pp}(\eta) \\
\end{array}\right),
\end{equation}
where, as usual, $P_{zz}$ indicates the power-spectrum of the related quantity $z=v$ or $z=p$, i.e. the expectation value of each couple of operators. 
In our case, the covariant matrix is an important object, as it contains all the physical information about the quantum state of perturbations. This happens because the system density matrix has still a Gaussian form, being the system operator linear in the interactions we consider inside the quantum master equation. We obtain the covariant matrix entries by remembering that
\begin{equation}
    \langle O(\eta) \rangle= \operatorname{Tr} (O\rho(\eta)),
    \label{eq:operatorexpvalue}
\end{equation}
and substituting, for the appropriate power-spectrum, the quantum master equation solution for the density matrix $\rho(\eta)$.  If instead we work in the Schrodinger representation, all the time dependence is encoded in the density matrix itself and the operator are time independent. In that case, to write the transport equations, it is possible to directly differentiate eq.(\ref{eq:operatorexpvalue}) w.r.t time in order to replace the derivative of the density matrix with the quantum master equation. This has been done, e.g., in~\cite{colas_benchmarking_2022}, and, with different notations and method, previously, in~\cite{martin_observational_2018}.

The Transport equation obtained reads 
\begin{equation}
\frac{\mathrm{d} \Sigma}{\mathrm{d} \eta}=  \omega\left(H^{free,sys}+\Delta\right) \Sigma-\Sigma\left(H^{free.sys}+{\Delta}\right) \omega  -\omega D \omega-2 {\Delta}_{12} \Sigma,
\end{equation}
where $\omega=\left(\begin{array}{cc}
0 & 1 \\
-1 & 0
\end{array}\right)$. By applying to the specific case in which only $D_{1 1}$ is different from zero, however, things simplify. The solution to the transport equation is:
\begin{equation}
\Sigma(\eta)=  g_{\mathrm{LS}}\left(\eta, \eta_0\right) \Sigma\left(\eta_0\right) g_{\mathrm{LS}}^{\mathrm{T}}\left(\eta, \eta_0\right) -\int_{\eta_0}^\eta \mathrm{d} \eta^{\prime}g_{\mathrm{LS}}\left(\eta, \eta^{\prime}\right)  \left[\omega {D}\left(\eta^{\prime}\right) \omega\right] g_{\mathrm{LS}}^{\mathrm{T}}\left(\eta, \eta^{\prime}\right) .
\label{eq:transporteqsolution}
\end{equation}
where we have introduced the Green function for the unitary evolution $g_{L S}$. It is simply a matrix of the unitary propagators which describe the Bogoliubov transformation for the perturbation fields:
\begin{equation}
 g_{\mathrm{LS}}\left(\eta, \eta^{\prime}\right)=2\left(\begin{array}{l}
\Im \mathrm{m}\left[u_{\mathrm{LS}}(\eta) u'^*_{\mathrm{LS}}\left(\eta^{\prime}\right)\right]\quad-\Im \mathrm{m}\left[u_{\mathrm{LS}}(\eta) u^*_{\mathrm{LS}}\left(\eta^{\prime}\right)\right] \\
\Im \mathrm{m}\left[u'_{\mathrm{LS}}(\eta) u'^*_{\mathrm{LS}}\left(\eta^{\prime}\right)\right]\quad-\Im \mathrm{m}\left[u'_{\mathrm{LS}}(\eta) u^*_{\mathrm{LS}}\left(\eta^{\prime}\right)\right].
\end{array}\right)
\label{eq:bogoliubovtransf}
\end{equation}
Notice that we introduced the subscripts ``LS'' to denote the mode functions calculated taking into account the Lamb Shift corrections to the Hamiltonian\footnote{In order to favour the comparison with~\cite{colas_benchmarking_2022}, we underline that in the notation of Benchmarking the mode funcions $u_{LS}$ are indicated with $v_{LS}$ (for the field),  and the mode functions of the momentum $u'_{LS}$ are indicated with $p_{LS}$. } 
Notice the correspondence between the second column of eq.(\ref{eq:bogoliubovtransf}) and eq.(\ref{eq:AB}), eq.(\ref{eq:precise definition AB}). The first term in eq.(\ref{eq:transporteqsolution}) is simply the unitary evolution modified for the Lamb Shift, and its influence on $\Sigma_{1 1}$, i.e. the $v$ power-spectrum, is as already stated before. However, we also notice that another term is introduced, directly related to the $D$ matrix and which is the non-unitary part of eq.(\ref{eq:transporteqsolution}). Since, for now, we are interested just in $\Sigma_{1 1}$, i.e. $P_{v v}$, we work out the equations for it:
\begin{equation}
\begin{split}
    \Sigma_{1 1}(\eta)&=4\left(\frac{[\Im(u_{L S}(\eta) u'^{*}_{L S}(\eta_{0}))]^2}{2k} +\frac{[\Im(u_{L S}(\eta) u^{*}_{L S}(\eta_{0}))]^2k}{2}\right)+4\int_{\eta_0}^{\eta} d\eta'[\Im(u_{k}(\eta) u_{k}^{*}(\eta'))]^2 \times \\ 
    &\times D_{1 1}(\eta, \eta',k).
    \label{eq:powerspectrumcorr}
    \end{split}
\end{equation}

It is straightforward to see that from the first two terms of the RHS of eq.(\ref{eq:powerspectrumcorr}) we just obtain the unitary evolution, corrected by the Lamb Shift effect on the $\nu_{LS}$ . Let's focus only on them. Actually, the imaginary part of the mode function $u_{k}(\eta)$ in $v_{\boldsymbol{k}}(\eta)$ corresponds to the growing mode in eq.(\ref{eq:growing decaying definitions}), and it dominates over the decaying mode, which is instead the real part. Since we are evaluating the power-spectrum well after the horizon crossing, it is legitimate to neglect the decaying mode. Taking into account this observation, and instead substituting the full mode function expression in  eq.(\ref{eq:fullmodefunction}) when considering the mode function at $\eta_{0}$,
we can write
\begin{equation}
    \Sigma_{1 1}(k,\eta)= \frac{1}{2}\Im( u_{L S}(\eta))^2=\frac{1}{2 k^3 \eta^2} (-k\eta)^{-2 \delta\nu_{LS}}=\frac{1}{2 k^3 \eta^2} (-k\eta)^{\frac{2}{3} \Delta_{11} \eta^2}\, , 
\end{equation}
where we have used the deep superhorizon limit for the expression of the growing mode, as seen in eq. \ref{eq:growingdecayingexpansion}.
We have obtained again the result of eq.(\ref{eq:approxcorretopowerspectrum}).
So, notice that a positive $\Delta_{11}$ corresponds to a blue tilt of the spectral index of the inflationary power-spectrum:
\begin{equation}
    \Delta n_{S}= \frac{2}{3} \Delta_{11} \eta^2.
\end{equation}
We should evaluate then this expression at the end of inflation.

Notice how the last term in the RHS of eq.(\ref{eq:powerspectrumcorr}) 
corresponds to the real part correlation function corrections to the power-spectrum found e.g. in~\cite{martin_observational_2018},~\cite{burgess_minimal_2023}. 
We have considered the free, not Lamb Shift corrected, mode functions $u_{k}(\eta)$ in this second term because considering this correction would be only an higher order effect in the perturbative coupling expansion. The connection to the real part of environmental correlation functions comes from $D_{1 1}$, while $\Delta_{1 1}$ is connected to the imaginary part of the environmental correlation function.
By this approach, we have understood the two parts of the corrections to the power-spectrum, but this is not the end of it. 

As we have showed in eq.(\ref{eq:purityforagaussianstate}), we can also compute the purity out of the determinant of the covariant matrix. In particular, it is possible to write a differential equation for the determinant of the covariant matrix, from which it is possibble to compute purity, by the quantum master equation, in Schrodinger picture, as worked out in~\cite{colas_benchmarking_2022}:
\begin{equation}
    \frac{\mathrm{d} \operatorname{det}\left({\Sigma}\right)}{\mathrm{d} \eta}= {D}_{11} {\Sigma}_{ 11}+2 {D}_{12} {\Sigma}_{ 12} -4 {\Delta}_{12} \operatorname{det}\left({\Sigma}\right)\simeq  {D}_{11} {\Sigma}_{ 11},
    \label{eq:determinantfromtcl2}
\end{equation}
where we considered that, being in our case the diffusion matrix highly hierarchical in its elements, we can neglect $\Delta_{1 2}$ and $D_{1 2}$ with respect to $D_{1 1}$. 

In the following, we are going to apply all the results stated here explicitly to the interactions we considered from eq.(\ref{eq:interactions}). Since decoherence is a second order phenomenon in the coupling, it interests couples of interactions. We will thus consider couples of the interactions in eq.(\ref{eq:interactions}), and separately treat the cases in which we consider only derivativeless interactions, only derivative interactions, and mixed derivativeless derivative interactions. In doing this, we will see how, on the one side, divergences, and on the other side negative $D_{11}$ coefficients, will show up. We will apply various methods, already explained in the literature, in order to compute purity and corrections to the power-spectrum.

\section{Derivativeless interactions}
\label{sec:derivativelesscanonical}
\subsection{Quantum master equations coefficients and Lamb Shift}
We consider the couple $H_{int1}-H_{int1}$ from the trilinear interactions lagrangian eq.(\ref{eq:interactions}), where
\begin{equation}
    H_{int1}=\frac{\sqrt{\varepsilon}}{ \sqrt{2} M_{\rm Pl}} \frac{H}{\eta} \theta_{i j} \theta_{i j} v \quad \quad \textnormal{so} \quad \quad g(\eta)=\frac{\sqrt{\varepsilon}H}{ \sqrt{2} M_{\rm Pl}\eta},
\end{equation}
$v$ is the system, and $\theta_{ij}$, environmental operator, represents the canonically normalized operator for tensor perturbations. It is noticeable that in the context of a division of short and long modes such an interaction was never considered before in the literature, while being analyzed in the context of spectator fields \cite{boyanovsky_effective_2015,Brahma:2024yor,Hollowood:2017bil}. 
We will compute the values of the elements of the diffusion matrix one by one. Consider first the value for $D_{1 1}$, whose expression is explicited in eq.(\ref{eq:Dmatrix})
\begin{equation}
\begin{split}
   D_{11}^{int1\, int1} &=\frac{\varepsilon H^2}{2 M_{\rm Pl}  \eta} \int_{-\frac{1}{p}}^\eta \frac{ d \eta^{\prime} }{\eta'} 
2 \left(-\frac{1}{4 \pi \eta} \delta(1-x) +\frac{\operatorname{sin}(20(1-x))}{4\pi^2 \eta}\right) A(k,\eta,\eta')\\
&=\frac{\varepsilon H^2}{4 M_{\rm Pl}  \pi^2 \eta}\int_{1}^{a}\frac{ d x }{x} \left(\pi \delta(1-x) -\frac{\operatorname{sin}(20(1-x))}{1-x}\right) A(x,a),\\
\end{split}
\label{eq:integralDint1int1}
\end{equation}
where $x,a$ were defined in eq.(\ref{eq:defxa}) and $A(x,a)$ in eq.(\ref{eq:ABwithxanda}), and we have considered the real part of the environmental correlation function as in eq.(\ref{eq:correfunctionrealnonreal}). $A(x,a)$ expresses the memory of the system. Notice, in particular, that $x=1$ represents the present moment, while $x=a$ represents the moment of horizon crossing. However, because of the Born-Markov approximation eq.(\ref{eq:markovianapprox}) we made,
it is not immediate to trust the whole region of the integral going from 1 to $a$, in particular because for CMB modes, towards the end of inflation, $a\simeq e^{50-60}$.  Also, we would expect that the value of the integral does not depend on the details of the oscillatory functions which are present into eq.(\ref{eq:ABwithxanda}), since they are important near $x=a$. The Integrand is peaked around $x=1$, so it makes sense to consider the regime where $x \ll a, a\gg 1$ and consequently expand the oscillatory functions in $A$, neglecting the inverse powers of $a$ 
\begin{equation}
A(x,a)\simeq \frac{x^2}{3}+\frac{2}{3x}.
\label{eq:Aapprox}
\end{equation}
Notice the presence of a term which grows as $x$ grows, i.e. as we go back in time in the memory integral. This is connected to the decaying mode, as can be seen if the same calculation for $A$ is performed in the basis of growing and decaying modes:
\begin{equation}
\begin{split}
    A(k,\eta,\eta')&=- \Im[(g_{+}+ ig_{-})(\eta')(g'_{+}-ig'_{-}(\eta)]=-(g_{-}(\eta')g'_{+}(\eta)-g_{+}(\eta')g'_{-}(\eta))=\\
    &= \frac{x^2}{3}+\frac{2}{3x}\,, 
    \label{eq:decayingmodecontributiontoA}
\end{split}
\end{equation}
where in the last equality the first term comes from the decaying term $g_{-}(\eta')$, where $\eta'$ is the conformal time which should be integrated. Thus, we explicitly see in eq.(\ref{eq:decayingmodecontributiontoA}) that the decaying mode is responsible to source a memory effect in the quantum master equation. 

By explicitly solving eq.(\ref{eq:integralDint1int1}) we also notice that the Markovian term, the first one in the integration, is positive, while the second term gives a negative non-Markovian effect. This seems to fit the description of the $D_{1 1}$ coefficients we mentioned in the discussion after eq.(\ref{eq:noncanonicallind}), 
where negative contributions are associated to the non-Markovian effects. By following the prescription for integrating the delta function, e.g. as in~\cite{Burgess:2024eng}, we can use 
\begin{equation}
    \int_{1}^{a} dx \delta(x-1)=\frac{1}{2},
\end{equation}
so that the final result is 
\begin{equation}
\begin{split}
&D_{1 1}^{int1\,int1}= \frac{\varepsilon H^2}{4 \pi^2 M_{\rm Pl}^2 \eta^2}\left(\frac{\pi}{2}-\frac{1}{60} \left(1-40 \left(\text{Ci}(20 a) (20 \cos (20)-\sin (20))+\text{Si}(20-20 a)\right. \right.\right.\\
&\left.\left. \left.+20 \sin (20) \text{Si}(20 a)+\cos (20) \text{Si}(20 a)+\text{Ci}(20) \sin (20)-20 \text{Ci}(20) \cos (20)-20 \text{Si}(20)\right. \right. \right.\\
&\left.\left.\left. \sin (20)-\text{Si}(20) \cos (20)\right)-20 \text{Si}(20-20 a)-\cos (20 (a-1))+
\frac{\sin (20-20 a)}{a}\right)\right) =\\
&\simeq  \frac{\varepsilon H^2}{4 \pi^2 M_{\rm Pl} ^2 \eta^2} \left(\frac{\pi}{2}-1.52 \right)\\
& \simeq  \frac{\varepsilon H^2}{4 \pi^2 M_{\rm Pl} ^2 \eta^2} 0.05, \\
\label{eq:derivativelessresult}
\end{split}
\end{equation}
which is positive and reassure us about the fact that Markovian effects are still dominant on non-Markovian ones. This result was also obtained by solving the integration numerically. In particular, by varying the numerical value of the upper extreme of the integral $a$ it was found that the result was largely insensitive to its value, apart from the last two terms in the sum in eq.(\ref{eq:derivativelessresult}), which are oscillatory terms in $a$. Their value is of order $0.01\frac{\varepsilon H^2}{4 \pi^2 M_{\rm Pl} ^2 \eta^2}$; as we are going to see in eq.(\ref{eq:averagingout}), we can average these values to zero because the period of these oscillations is much shorter than the typical time scale of the interaction. Also, the period of oscillations is much shorter than the time in which the quantum master equation we obtain is coarse grained. Thus, we can safely ignore these oscillatory terms. 

The fact that the value of this coefficient is insensitive to the precise time at which the integration begins (in conformal time $\eta$)  is a very important sanity check of the all procedure. It is only in this way, as discussed e.g. in~\cite{burgess_gravity_2023}, that the quantum master equation solutions can be trusted at long times. Actually, if the quantum master equation is independent from the initial value of the integration, the only quantities which are relevant are the time intervals on which the coarse graining is operated, and how they are related to the typical time of evolution of system and environment. 

The Lamb Shift corrections to the power-spectrum of the Sasaki Mukhanov variable can instead be computed by $\Delta_{1 1}$, as explicit in eq.(\ref{eq:deltamatrix}) :
\begin{equation}
    \Delta_{1 1}^{int1\, int1}=-\frac{2 \varepsilon H^2}{8 M_{\rm Pl} ^2 \pi^2 \eta}\int_{1}^{a}\frac{ d x }{x} \cos{[20(1-x)]}\operatorname{P.V.}\left(\frac{1}{x-1}\right) \left(\frac{x^2}{3}+\frac{2}{3x}\right).
 \label{eq:delta11derivativeless}
\end{equation}
We notice that this integral, as it is written, diverges in $x=1$, and the divergence is regulated by the Principal Value. In order to treat this properly, we follow the lead of Boyanovsky in~\cite{boyanovsky_effective_2015} and rewrite
\begin{equation}
    P.V.\left(\frac{1}{x-1}\right)=\frac{x-1}{(x-1)^2+\epsilon^2}=\frac{1}{2} \frac{d}{d x} \operatorname{log} \left((x-1)^2+\epsilon^2\right)\, ,
    \label{eq:regularizedivprincipalvalue}
\end{equation}
where $\epsilon$ regulates the divergence and eventually we will take $\epsilon \to 0$, and then we integrate by parts inside eq.(\ref{eq:delta11derivativeless})
\begin{equation}
\begin{split}
   & \Delta_{1 1}^{int1\,int1}=-\frac{\varepsilon H^2}{4 M_{\rm Pl} ^2 \pi^2 \eta^2}\left( \left[\left(\frac{x}{3}+\frac{2}{3x^2} \right) \cos (20(x-1)) \ln \left((x-1)^2+\epsilon^2\right)\right]_1^{-\frac{1}{k \eta}}\right.\\
   &\left.-\int_{1}^{a}  d x \frac{d}{d x}\left(\cos (20(x-1)) \left(\frac{x}{3}+\frac{2}{3x^2} \right)\right) \ln (x-1)^2\right).\\
    \end{split}
    \label{eq:delta11derivativeless2}
\end{equation}
The lower limit of the boundary term (i.e. $x=1$) produces a divergence, which can be reabsorbed by an infinite renormalization term for the mass:
\begin{equation}
   \Delta{M^2}=-\frac{\varepsilon H^4}{8 M_{\rm Pl} ^2 \pi^2 }\ln \epsilon.
   \label{eq:renormalizationofthemass}
\end{equation}
as it is evident from eq.(\ref{eq:numassren}), where the relation between the coefficient $\Delta_{11}$ and the mass is highlighted.

What we are interested in are instead the finite corrections. However, in this case, we see that both the upper limit of the boundary term, and the value of the integration produced by $a \rightarrow \infty$ in eq.(\ref{eq:delta11derivativeless2}), become leading terms in inverse powers of $\eta$ with respect to all other terms. This is in opposition to what we saw before for $D_{1 1}$, and would mean that the result would be very sensitive to the value of $a$, which, we remember, represents the beginning of the integration in terms of the conformal time, and physically the time of horizon exit for the mode system. Besides, we cannot trust, as explained multiple times, the memory of the integral, for we are not sure of the validity of the Born-Markov approximation eq.(\ref{eq:markovianapprox}), that we did when we derived the equation, on such a long interval of time.

If, instead we consider again eq.(\ref{eq:delta11derivativeless}), we realize that as $x\to a$ inside the integration there should arise only an oscillatory behavior and not a divergent one. 
The aforementioned divergent behavior we observe in eq.(\ref{eq:delta11derivativeless2}) is caused by the logarithm, which was introduced by out trick to cure the divergence problems at $x=1$.

This artificial behavior, caused by the presence of the logarithm, connected to the principal value of the environmental function, in the region near $x=a$, is also artificially enhanced by the part of $A(x,a)$  in eq.(\ref{eq:Aapprox}) which grows, going back in time, because of memory effects (as it is connected to the decaying mode). 
One idea could so be to cut out this artificial behavior by considering an approximated form for $A(x,a)$, which coincides with it near $x=1$ but does not grow further. If we actually expand $A$ in $t=x-1$ it is simple to see that we obtain
\begin{equation}
    A(t)\simeq 1+t^2+\dots
    \label{eq:smardelta}
\end{equation}
so, we may approximate $A\simeq 1$. Notice that this approximation is equivalent to  what we called in this paper the ``strong Markovian approximation'' used  in~\cite{burgess_minimal_2023} which we explained in  relation to eq.(\ref{eq:smar1}).

Taking only the first term in the expansion in eq.(\ref{eq:smar1}) is actually equivalent to substitute the $v(\eta')$ inside the integral with $v(\eta)$; but, as can be seen by the definition eq.(\ref{eq:AB}), this is equivalent to take $A=1$. As underlined after eq.(\ref{eq:smar1}) the choice of a deep subhorizon environment strenghtens the validity of this approximation, since the evolution time of $v(\eta')$ is of order $\simeq 1/aH$, while the environmental correlation function time is of order $\simeq 1/10aH$; so, in the memory integral in the quantum master equation, we can assume that $v(\eta')$ is almost constant, since it varies more slowly, and take it out of the integration, assuming the value $v(\eta)$.

In this case, we apply this prescription in order to erase the memory contribution in eq.(\ref{eq:delta11derivativeless2}) which were artificially introduced by the trick in eq.(\ref{eq:regularizedivprincipalvalue}). 

Performing the integration in eq.(\ref{eq:delta11derivativeless2}) using the strong Markovian approximation in eq.(\ref{eq:smardelta}) we obtain
\begin{equation}
\begin{split}
   \Delta_{1 1}^{int1int1}&=\frac{\varepsilon H^2}{4 M_{\rm Pl} ^2 \pi^2 \eta^2} \int_{1}^{a}  d x  \left(\left(\frac{1}{x^2}\right) \cos[20(-1+x)]+\left(\frac{20}{x}\right) \operatorname{Sin}[20(-1+x)]\right) \log [(x-1)] \\
   &\simeq \frac{\varepsilon H^2}{4 M_{\rm Pl} ^2 \pi^2 \eta^2} 3.6,\\
   \end{split}
   \label{eq:delta11derivativelesssmar}
\end{equation}
corresponding to a variation of the spectral index:
\begin{equation}
    \delta n_{S}= \frac{2}{3} \eta^2 \Delta_{1 1}^{int1int1}\simeq \frac{\varepsilon H^2}{4 M_{\rm Pl} ^2 \pi^2} 2.4,
    \label{eq:derivativelessspectralindexcorr}
\end{equation}
 
This is a blue correction to the spectral index of the power-spectrum. 
Notice that the temporal dependence drops out of the expression. However, differently e.g. from~\cite{boyanovsky_effective_2015}, there is no logarithmic running, proportional to $\ln k$, in the spectral index; the correction to the spectral index is scale invariant. This may be a feature of an environment of deep subhorizon modes which do not emerge in a model with a conformally coupled scalar field, which is just a proxy for this type of calculations, as~\cite{boyanovsky_effective_2015}. A possible explanation could be that the infrared limitation of our environment, differently from a conformal field not infrared limited in the case of~\cite{boyanovsky_effective_2015}, may avoid the presence of a logarithmic running.

The precision of the present data on the spectral index of the scalar primordial power-spectrum at CMB scales is around $10^{-3}$; instead, by the newest limits on the scale of inflation~\cite{akrami_planck_2020}, the order of magnitude of $\varepsilon \frac{H^2}{M_{\rm Pl} ^2}$ is, at most, $\simeq (10^{-5})^2 \times 10^{-3}\simeq 10^{-13}$, way out of the present measurements.  The goal of this paper, however, is mostly to consider the form of the correction; and without the huge suppression of the GR interactions, e.g. considering a stronger coupling, it is simple to see that we can reach a bigger correction thanks to possible higher non-Gaussianities. We plan to develop more on this in a future work. 

Apart from the Lamb Shift corrections, we can also consider the non-unitary corrections to the power-spectrum. These are derived from the real part of the correlation function in eq.(\ref{eq:powerspectrumcorr}) 
\begin{equation}
4\int_{\frac{-1}{p}}^{\eta} d\eta'[\Im(u_k(\eta) u_k^{*}(\eta'))]^2 D_{1 1}( \eta',p)=-\eta \int_{1}^{a} d x [B(x,a)]^2 D_{1 1 }(\eta',p),
\label{eq:nonunitcorrect}
\end{equation}
where the initial time of integration is imposed to be the horizon crossing $\eta=-1/k$. We substitute the function $B(x,a)$ from eq.(\ref{eq:ABwithxanda}) in eq.(\ref{eq:nonunitcorrect}). As we will consider the system mode as deeply superhorizon, we can take an expansion of the final expression around the limit $a \to \infty$:
\begin{equation}
\begin{split}
&\Delta P_{vv}(\eta)=-\frac{\varepsilon H^2}{4 \pi^2 M_{\rm Pl} ^2 \eta} \int_{1}^{a} \frac{d x}{x^2} B(x,a)^2 \left(\frac{\pi}{2}-1.52\right) \simeq \\
&\simeq \frac{\varepsilon H^2}{4 \pi^2 M_{\rm Pl} ^2 k^2}\frac{1}{\eta} \frac{a}{3} (2-\cos{2}-\sin{2}-\operatorname{Si}{2}) \left(\frac{\pi}{2}-1.52\right)\simeq  \frac{1}{2 k^3\eta^2} \frac{\varepsilon H^2}{6 \pi^2 M_{\rm Pl} ^2} 0.1 \left(\frac{\pi}{2}-1.52\right).\\
\label{eq:nonunitcorrections}
\end{split}
\end{equation}
The result of eq:(\ref{eq:nonunitcorrections}) is a fractional shift of the amplitude of the Primordial power-spectrum, analogous to the one found by \cite{burgess_minimal_2023} (notwithstanding the very different environment employed).
From these results, we can already infer that the corrections act at the same time on the amplitude of the power-spectrum and on the spectral index; this is an interesting feature also from the phenomenological point of view, as a contemporary correlated shift in the two may be a distinctive feature of the decoherence process to look for in the observations.

 It is interesting to obtain the same result as the one in eq.(\ref{eq:nonunitcorrections}) by considering instead an approximate expression for the function $B(x,a)$ , as already done for $A(x,a)$ in the regime where $x \ll a$ and $a\gg 1$, 
 \begin{equation}
   k  B(x,a)\simeq \frac{1-x}{a}+\frac{1}{3}\frac{(1-x)^3}{ax}\simeq \frac{1-x^3}{3ax},
    \label{eq:Bapprox}
\end{equation}
which will prove to be useful also in the following.
By substituting eq.(\ref{eq:Bapprox}) into eq.(\ref{eq:nonunitcorrect}) we have:
\begin{equation}
\begin{split}
&\Delta P_{vv}(\eta)=\frac{\varepsilon H^2}{4 \pi^2 M_{\rm Pl} ^2}\left(\frac{\pi}{2}-1.52\right) \int_{a}^1 dx \frac{a^2}{x^2}\frac{(x^3-1)^2}{9a^2x^2}=\\
&\simeq \eta \int_{a}^{1}\frac{\varepsilon H^2}{16 \pi^2 M_{\rm Pl} ^2} \frac{x^2}{9}\left(\frac{\pi}{2}-1.52\right)=\\
&=\left(\frac{\pi}{2}-1.52\right)\frac{\varepsilon H^2}{54 \pi^2 M_{\rm Pl} ^2} \frac{1}{2k^3 \eta^2},\\
\end{split}
\end{equation}
which indeed is approximately equal to the value in eq.(\ref{eq:nonunitcorrections}).

\subsection{The mixed term coefficients}
\label{par:themixedterm}
 We have worked out our calculations until now following the hypothesis that the  matrices of coefficients of the quantum master equation written in eq.(\ref{eq:Dmatrix}) and eq.(\ref{eq:deltamatrix}) are diagonal. Equivalently, the coefficients of the second or third lines of eq.(\ref{eq:canonicalqme}), which feature both the operators $v$ and $p$, should be zero. Here we verify this hypothesis by computing explicitly these coefficients for the derivativeless interactions. We employ again, for the function $B(x,a)$ the approximation in  eq.(\ref{eq:Bapprox}). This is justified as we expect that the results should not be sensitive to the precise value at which the mode crossed the horizon. The coefficient of the third line of eq.(\ref{eq:canonicalqme}) can be written as
\begin{equation}
    D^{int11}_{12}=\frac{\varepsilon H^2}{8 i \pi^2 M_{\rm Pl} ^2 \eta^2 k} \int_{1}^{a} \frac{dx}{x}\left[\frac{1-x^3}{3ax}\right] \frac{e^{i20(x-1)}}{(1-x+i\epsilon)},
\end{equation}
where $k$ is the momentum of the system. Notice that, since 
\begin{equation}
    B(x=1,a)=0,
\end{equation}
 the Markovian contribution to $D_{12}$ always goes to zero. Also, differently from $A$, in eq.(\ref{eq:Bapprox}), $B$ is at least suppressed by a factor $1/a$ so in the limit when $a\to \infty$ the $B$ terms go to zero.
 We compute the real and imaginary parts of the integral. The analytic expression e.g. for the real part is not really insightful, but we report it fully:
\begin{equation}
    \begin{aligned}
& \frac{\varepsilon H^2}{8 \pi^2 M_{\rm Pl} ^2 \eta^2 k} \frac{1}{3 a}\left(1-\frac{\cos(20 (-1+a))}{a}-5 \operatorname{Ci}(20 a)(\cos(20)-4 \sin(20))-20 \operatorname{Ci}(20) \sin(20)+5 \sin(20)\right.\\
&\left. \operatorname{Si}(20)-\frac{1}{20}  \sin(20-20 a)+  5 \cos(20)(\operatorname{Ci}(20)+4 \operatorname{Si}(20))-5 (4 \cos(20)+\sin(20)) \operatorname{Si}(20 a)\right).
\label{eq:Bderivativelessrealpart}
\end{aligned}
\end{equation}
where $Ci$ and $Si$ are cosine integral and sine integral.Notice that the result in eq.(\ref{eq:Bderivativelessrealpart}) inherits from $B(x,a)$ the proportionality to $1/a$. This is equivalent to say that it has a power of $\eta$ more in the numerator, which makes the result subdominant with respect to the other elements in the D matrix at late times $\eta \to 0$ as eq.(\ref{eq:derivativelessresult}). A similar result, also subdominant at late times as $\eta \to 0$ with respect to eq.(\ref{eq:delta11derivativeless}) is obtained for the imaginary part. We can so safely neglect them and consider the $D$ matrix as diagonal with only the first entry being non vanishing. A similar conclusion holds also for the $\Delta$ matrix.

This may be seen as a spontaneous driving of the system towards Markovianity as the time increases, since the effect makes the $D$ matrix diagonal. Off diagonal coefficients of the matrix eq.(\ref{eq:Dmatrix}) unavoidably create negative eigenvalues, as can be easily checked.  Other works also observed that as $\eta \to 0 $, in a similar setting, the system under study naturally tends to become Markovian~\cite{burgess_minimal_2023}.  

When employing the Strong Markovian approximation, we are extracting $B$ out of the $D_{12}$ integral, and computing it at $x=1$; thus, we effectively obtain that we impose $D_{12}=0$. The fact that $D_{12}$ is already negligible by direct computation shows that, considering this aspect, the validity of the approximation is greatly verified by our system.

\subsection{Time dependent environment}
\label{subsec:timedependentenvironment}
One of the main points on which we focus in this work is considering a time dependent environment. By time dependent we mean that some of the modes of the fields which, at a certain instant of time, belong to the environment, will later cross the horizon and so leave the environment. This type of environment has already been considered in the literature before~\cite{brahma_quantum_2022, brahma_universal_2022, gong_quantum_2019}; however, as pointed out e.g. by~\cite{burgess_gravity_2023}, there was no estimation in these works of the effects of this choice of a time dependent environment on the dynamics. Other more recent works have instead considered a fixed cutoff~\cite{burgess_minimal_2023}. Here we try to give an estimate of the impact of this choice. 

Let's consider eq.(\ref{eq:Bornapproxintegrated}). We then trace over the environment, arriving to eq.(\ref{eq:integratednonmarkqme}). We should then derive this equation with respect to conformal time. For a fixed environment, the only dependence on the time $\eta$ appears in the upper extreme of the integral of the double commutator part in the RHS of the equation.

However, in our case, with a time dependent cutoff, we know that, after tracing, the environmental correlation function looks (schematically) like: 
\begin{equation}
    \int_{\frac{-10}{\eta}} ^{\infty} dk \langle 0| O_{E}(\eta)O_{E}(\eta') |0 \rangle,
    \label{eq:timedependence}
\end{equation}
where the dependence on the environmental ``loop'' momentum $k$ is hidden inside the operators $O_{E}$, as can be seen e.g. in eq.(\ref{eq:corrfunctionexplicit}). From eq.(\ref{eq:timedependence}) we deduce that also the environmental correlation function has an explicit dependence on $\eta$. 

In order to compute the derivative in this latter case, by definition, we could independently vary the $\eta$ in the two places, for a finite increment of time $\Delta \eta$, and then divide and take the limit as $\Delta \eta \to 0$. Notice that, from the derivation of the quantum master equation, implicitly we have to coarse grain the increments in conformal time. In particular, we have to find an increment which is bigger than the typical environmental correlation time but smaller than the typical time of evolution of the system. If we use the hypothesis that $|\eta/10| \ll |\Delta \eta| \ll |\eta|$, where $\eta$ is the typical time of evolution of the system and $\eta/10$ is the typical time of the environment, we may find an appropriate $\Delta \eta$ for all the approximations we are doing to be valid.  As $\eta$ decreases in absolute value, notice that $\Delta \eta$ would have to decrease accordingly. 

We can obtain $\rho_{r,I}(\eta+\Delta \eta)$ by varying independently  the time dependent environment and expanding:
\begin{equation}
\begin{split}
   &\rho_{r,I}(\eta+\Delta\eta)=-\int_{\eta_{0}}^{\eta+\Delta\eta} d\eta_{1} g(\eta_{1})\int_{\eta_{0}}^{\eta_{1}} d\eta_{2} g(\eta_{2}) \operatorname{Tr}_{E(\eta)} [H_{int} (\eta_{1}),[H_{int}(\eta_{2}),\rho(\eta_{2})]]\\
   &-\int_{\eta_{0}}^{\eta} d\eta_{1} g(\eta_{1})\int_{\eta_{0}}^{\eta_{1}} d\eta_{2} g(\eta_{2}) \operatorname{Tr}_{E(\eta+\Delta \eta)-E(\eta)} [H_{int} (\eta_{1}),[H_{int}(\eta_{2}),\rho(\eta_{2})]].\\
   \end{split}
\end{equation}
By subtracting then eq.(\ref{eq:integratednonmarkqme}):
\begin{equation}
\begin{split}
&\rho_{r,I}(\eta+\Delta\eta)- \rho_{r,I}(\eta)=-\Delta\eta  g(\eta) \int_{\eta_{0}}^{\eta} d\eta_{2} g(\eta_{2}) \operatorname{Tr}_{E(\eta)} [H_{int} (\eta),[H_{int}(\eta_{2}),\rho(\eta_{2})]]\\
&- \int_{\eta_{0}} ^ {\eta} d\eta_{1} g(\eta_{1}) \int_{\eta_{0}} ^{\eta_{1}} d\eta_{2}
g(\eta_{2}) \operatorname{Tr}_{E(\eta+\Delta \eta)-E(\eta)} [H_{int} (\eta_{1}),[H_{int}(\eta_{2}),\rho(\eta_{2})].\\
\end{split}
\label{eq:timedependentqme}
\end{equation}
 A priori, we are not even sure that this equation that we obtain is a quantum master equation, because it is not guaranteed that by tracing over different environments at each time we obtain the same structure as in eq.(\ref{eq:quantummastereq}). The first term on the RHS of eq.(\ref{eq:timedependentqme}) is well known, and we have already worked out its consequences, while the second one is completely new. We focus on the latter in this paragraph, ignoring the first.

We now explicitly use our environmental correlation function, in the specific case of derivativeless interactions, by taking the limit
\begin{equation}
\begin{split}
    &\rho'_{r,I}(\eta)=-\lim_{\Delta \eta \to 0} \frac{\varepsilon H^2}{ 2 M_{\rm Pl}^2}  \int_{\eta_{0}} ^ {\eta} \frac{d\eta_{1}}{\eta_{1}}   \int_{\eta_{0}} ^{\eta_{1}} \frac{ d\eta_{2}}{\eta_{2}} \left(\Re \frac{K_{\eta+\Delta\eta}(\eta_{1},\eta_{2})-K_{\eta}(\eta_{1},\eta_{2})}{\Delta\eta} \right. \\
    &\left.[v (\eta_{1}),[v (\eta_{2}),\rho(\eta_{2})]]+i
  \Im \frac{K_{\eta+\Delta\eta}(\eta_{1},\eta_{2})-K_{\eta}(\eta_{1},\eta_{2})}{\Delta\eta} A(\eta_{1},\eta) A(\eta_{2},\eta) [v (\eta_{1}),\{v (\eta_{2}),\rho(\eta_{2})\}]\right),
 \end{split}
 \label{eq:qmewithtimedepednentenv}
\end{equation}
where $K_{\eta}(\eta_{1},\eta_{2})$ is the environmental correlation function written in eq.(\ref{eq:correlationfunctionK}), computed between $\eta_{1}$ and $\eta_{2}$ by considering the environment at $\eta$.  We focus on the environmental function variation:
\begin{equation}
    \frac{K_{\eta+\Delta\eta}(\eta_{1},\eta_{2})-K_{\eta}(\eta_{1},\eta_{2})}{\Delta\eta}=\frac{e^ {\frac{20 i (\eta_{1}-\eta_{2})}{\eta+\Delta\eta}}-e^{\frac{20 i (\eta_{1}-\eta_{2})}{\eta}} } {4 \pi^2 i (\eta_{1}-\eta_{2})}=\frac{e^{\frac{20 i (\eta_{1}-\eta_{2})}{\eta} }}{4 \pi^2 i(\eta_{1}-\eta_{2})} \frac {(e^{\frac{20i(\eta_{1}-\eta_{2})}{\eta}\frac{-\Delta\eta}{\eta+\Delta \eta}}-1)}{\Delta \eta}.
    \label{eq:differenceofkernels}
\end{equation}
Notice that in the last equation we have factorized the environmental correlation function. We can only take appropriately the limit $\Delta \eta \rightarrow 0$ if we can develop in series the exponential in the last equality; we know that $|\Delta \eta|\ll|\eta|$, but we also have to say that $|\eta_{1}-\eta_{2}| \lesssim |\eta|$. A priori this is not true, since we could consider every $\eta_{1}$ and $\eta_{2}$, but for the presence of the $(\eta_{1}-\eta_{2})$ at the denominator we know that the whole expression is suppressed for $|\eta_{1}-\eta_{2}|>|\eta|$.
By keeping in mind that we are working in a regime where $|\eta_{1}-\eta_{2}|$ is small, we develop in series the term in the parenthesis in the last equality of eq.(\ref{eq:differenceofkernels}) and obtain
    \begin{equation}
    \frac {(e^{\frac{20i(\eta_{1}-\eta_{2})}{\eta}\frac{-\Delta\eta}{\eta}})-1}{\Delta \eta}\simeq-\frac{20 i }  {\eta^2} (\eta_{1}-\eta_{2}).
    \label{eq:timedependentdevelopinseries}
\end{equation}
Notice that the factor 20 in the numerator is the velocity of the change of the cutoff. This is due to our choice of considering environmental modes with wavelength ten times smaller than the horizon. This factor brings an enhancement because we are considering a deep subhorizon environment. Sholud we consider a superhorizon environment, we would have had a suppression .We now susbtitute back eq.(\ref{eq:timedependentdevelopinseries}) into eq.(\ref{eq:qmewithtimedepednentenv}) and compute the limit $\Delta \eta\to 0$:
\begin{equation}
\begin{split}
   & \rho'_{r}(\eta)= \frac{H^2 \varepsilon}{2 M_{\rm Pl} ^2 \eta^2} \int_{a}^{1}\frac{d x_{1}} { x_{1}}\int_{a}^{x_1}\frac{d x_{2}} { x_{2}} \frac{5}{ \pi^2} \left(\cos{(20  (x_{1}-x_{2}))} [v(\eta_{1}),[v(\eta_{2}),\rho_{r}(\eta_2)]]-\right.\\
   &\left.-i \sin{20  (x_{1}-x_{2})} [v(\eta_{1}),\{v(\eta_{2}),\rho_r(\eta_{2})\}]\right).\\
    \label{eq:qmetimedependentseparatingrealimag}
\end{split}
\end{equation}
The equation (\ref{eq:qmetimedependentseparatingrealimag}) is an integrodifferential equation, which is difficult to handle analitically. Therefore, we adopt the Born-Markov approximation eq.(\ref{eq:markovianapprox}), which is justified since the difference between $\rho(\eta_{2})$ and $\rho(\eta)$ is higher order in the coupling constant. 

We focus first on the real part contribution. We can rewrite it in the canonical form:
\begin{equation}
    \rho'_{r}(\eta)=\frac{H^2 \varepsilon}{2 M_{\rm Pl} ^2 \eta^2} \int_{a}^{1}\frac{d x_{1}} { x_{1}}\int_{a}^{x_1}\frac{d x_{2}} { x_{2}} \frac{5}{\pi^2} \cos{(20(x_{1}-x_{2}))} A(x_{1}) A(x_{2}) [v(\eta),[v(\eta),\rho_{r}(\eta)]].
    \label{eq:qmewithtimedepenvrealpart}
\end{equation}
Notice that in eq.(\ref{eq:qmewithtimedepenvrealpart}) the form of the equation is far more familiar, comparing for example to the first row of eq.(\ref{eq:canonicalqme}). In particular, if the limit exists, the RHS of eq.(\ref{eq:qmewithtimedepenvrealpart}) looks like simply a correction to the coefficient of the first row of eq.(\ref{eq:canonicalqme}),i.e. to $D_{11}$, the first entry of eq.(\ref{eq:Dmatrix}).
Notice that we have neglected the modifications in the other entries of the D matrix apart from $D_{11}$, since, we are assuming they are subdominant (as explained in detail in the section \ref{par:themixedterm} they are subdominant in the late time $\eta$ expansion by at least one power). Analyzing these other entries could shed light on possible corrections to the related coefficients $D_{12}$ and $\Delta_{12}$, which we leave to analyze in future works. By using the expression eq.(\ref{eq:Aapprox}), it is evident that the memory part in the double integral in eq.(\ref{eq:qmewithtimedepenvrealpart}) would peak in the far past. This would spoil the validity of the Markovian approximations. Moreover, in eq.(\ref{eq:qmewithtimedepenvrealpart}) the integrand form is actually valid only for $x_{1}-x_{2} \ll1$, because of the hypotheses we made before writing eq.(\ref{eq:timedependentdevelopinseries}), so memory effects would not be compatible even with our derivation. 

All these effects reveal a non trivial interplay between the corrections due to a time dependent environment and memory effects (due for example to non-Markovianity). In this work, we limit ourselves to consider the Markovian effects of the time dependent environment. In order to take the Markovian limit, we consider an approximation similar to what we called ``Strong Markovian approximation''. Thus, we consider the approximations $A(x_{1})=1$  and $A(x_{2})=1$, i.e. akin to what is done below eq.(\ref{eq:smardelta}). We are therefore supposing that the region $(x_1,x_2)$ which gives the most important results in the  integral in  eq.(\ref{eq:qmewithtimedepenvrealpart}) is around $x_1 \simeq x_2 \simeq 1$. Substituting these values into eq.(\ref{eq:qmewithtimedepenvrealpart}) and, also, sending the limit of integration on $x_{1}$ from $a=-1/k\eta$ to $\infty$, since we numerically verified it has no influence:
\begin{equation}
\begin{split}
     &\delta D_{11}=- \frac{H^2 \varepsilon}{2 M_{\rm Pl} ^2 \eta^2} \int_{\infty}^{1}\frac{d x_{1}} { x_{1}}\int_{\infty}^{x_1}\frac{d x_{2}} { x_{2}} \frac{10}{ \pi^2}  \cos (20  (x_{1}-x_{2}))=\\
     &=- \frac{10 H^2 \varepsilon}{2 M_{\rm Pl} ^2 \pi^2 \eta^2} \left(\frac{1}{8} \left(4 \text{Ci}(20)^2+(\pi -2 \text{Si}(20))^2\right)\right) \simeq- \frac{ H^2 \varepsilon}{4 M_{\rm Pl} ^2 \pi^2 \eta^2} 0.02,\\
     \label{eq:markoviantimedependentmasterequation}
     \end{split}
\end{equation}
which is of the same order of magnitude of our result for the coefficient $D_{11}$ (eq.(\ref{eq:derivativelessresult})), still not able to change the sign of the result. This means that we are still in the case in which the coefficient $D_{11}$ in front of the Lindblad-like part of the quantum master equation is positive, and therefore we are guaranteed that the evolution is physical for the derivativeless interaction  (see the discussion below eq.(\ref{eq:noncanonicallind})).

Notice that having applied the ``Strong Markovian approximation'', we have also to consider $B=0$, so $D_{11}$ would be the only coefficient varied in the $D$ matrix. This is consistent with the assumption we made about the smallness of the off diagonal entries in the $D$ matrix.

We now analogously consider the variation of the $\Delta_{11}$ coefficient for the Lamb Shift Hamiltonian:
\begin{equation}
    \delta \Delta_{11}=-  \frac{5 H^2 \varepsilon}{4 M_{\rm Pl} ^2 \eta^2} \int_{-1/p}^{\eta} \frac{d\eta_{1}}{\eta_{1}} \int_{-1/p}^{\eta_{1}} \frac{d\eta_{2}}{\eta_{2}} \sin{\frac{20(\eta_{1}-\eta_{2})}{\eta}}A(\eta_{1},\eta) A(\eta_{2},\eta).
\end{equation}
We consider again the ``Strong Markovian approximation'', so approximating $A(x_1)=A(x_2)=1$, we have
\begin{equation}
\begin{split}
    &\delta \Delta_{11}=  \frac{5 H^2 \varepsilon}{4 M_{\rm Pl} ^2 \eta^2} \int_{1}^{\infty} \frac{d x_{1}}{x_{1}} \int_{1}^{x_{1}} \frac{d x_{2}}{x_{2}} \sin{20(x_{1}-x_{2})}= -   \frac{5 H^2 \varepsilon}{4 M_{\rm Pl} ^2 \eta^2} \frac{1}{4} \left(\pi ^{3/2} G_{2,4}^{2,1}\left(100\left|
\begin{array}{c}
 \frac{1}{2},1 \\
 \frac{1}{2},\frac{1}{2},0,0 \\
\end{array}
\right.\right)\right.\\
&\left.-2 \pi  \text{Ci}(20)\right) \simeq  -   \frac{H^2 \varepsilon}{8 M_{\rm Pl} ^2 \eta^2} 0.5.
\end{split}
\end{equation}
We explicitly checked, by numerical integrations, that under these approximations varying the upper limit of the integration in $x_{1}$ does not change the result appreciably. In this case, the blue corrections to the power-spectrum of the curvature perturbations are suppressed due to the corrections by a factor of order $\simeq 14\%$. In both cases, the modifications due to the time dependent cutoff do not change appreciably the results and the conclusions we have considered before. We stress that this is due to the choice of considering deep subhorizon perturbations until $10 aH$. Considering ``deeper'' subhorizon perturbations would probably enhance the contribution due to the time dependence of the environment.

\subsection{Decoherence}
\label{subsec:decoherencederivativeless}
We finally consider the decoherence arising from the derivativeless interaction. Since the interaction is linear in the system variable, the reduced density matrix will stay Gaussian throughout the whole evolution. As a consequence, we can apply eq.(\ref{eq:purityforagaussianstate}) for obtaining the purity. The determinant of the matrix can be readily obtained instead by integrating in time eq.(\ref{eq:determinantfromtcl2}):
\begin{equation}
\begin{split}
    \operatorname{det}\left({\Sigma}\right)&= \int_{-1/k}^{\eta} d\eta'  {D}_{11}(\eta') P_{vv}(\eta',k) =\int_{-1/k}^{\eta} d \eta' \frac{0.05 \varepsilon H^2}{M_p^2 4 \pi^2 \eta'^2}  \frac{1}{2\eta'^2 k^3}\simeq - \frac{0.05 \varepsilon H^2}{M_p^2 24 \pi^2 k^3 \eta^3}\\
    &=\frac{0.05 \varepsilon H^2}{M_p^2 24 \pi^2} \left(\frac{aH}{k}\right)^3\\,
    \label{eq:purityderivativelessint}
    \end{split}
\end{equation}
where the contribution to the integral from the lower extreme is negligible. Notice, that just like in other previous works, the decoherence is proportional to the growth of the volume since horizon exit of the mode~\cite{burgess_eft_2015,burgess_minimal_2023,Burgess:2024eng}. This has been related to considering environments with short wavelength; our results seem to confirm this observation. As done e.g. in~\cite{burgess_minimal_2023}, we can write this result using quantities connected to observables, as the curvature primordial power-spectrum
\begin{equation}
    \mathcal{P}_\zeta \simeq \frac{H^2}{8 \pi^2 \varepsilon M_{\mathrm{p}}^2} \simeq 2.2 \times 10^{-9},
\end{equation}
or the tensor-to-scalar pertubation ratio $r=16 \varepsilon$ in single field inflation:

\begin{equation}
\begin{split}
 &4\operatorname{det}\left({\Sigma}\right)=5.6 \times 10^{-19}\left(\frac{\mathcal{P}_\zeta}{2.2 \times 10^{-9}}\right)\left(\frac{r}{10^{-3}}\right)^2 e^{3\left(N-N_*\right)} \\
    &=\frac{1}{20}\left( 1.2 \times 10^{-17}\left(\frac{\mathcal{P}_\zeta}{2.2 \times 10^{-9}}\right)\left(\frac{r}{10^{-3}}\right)^2 e^{3\left(N-N_*\right)}\right)\\,
    \label{eq:decoherencederivativelessefolds}
    \end{split}
\end{equation}

where we put a 4 in front of the LHS in order to directly compare to eq.(4.19) of~\cite{burgess_minimal_2023}, reported in parenthesis after the second equality. This result is a factor 20 larger than our result. Notice that we have considered a correction factor of $2/3$ when reporting the result of~\cite{burgess_minimal_2023}, since we stick to considering only the tensor sub-horizon environment in our calculations, while Burgess et al. also consider the scalar sub-horizon environment (although with a different environment, i.e. with a fixed cutoff). 
Let us now assume that this is the only channel of decoherence. We can say that decoherence is reached when the purity becomes less than one. According to eq.(\ref{eq:purityforagaussianstate}), this happens when $\operatorname{det} \Sigma \gg 1$. So, by considering  $r\simeq 10^{-3}$ into eq.(\ref{eq:decoherencederivativelessefolds}) we have that
\begin{equation}
N-N_{*} \gtrsim 14 \,\, \rm{ e-folds},
\end{equation}
e.g., after 14 e-folds from horizon crossing we have decoherence. Notice that the present upper bound on $r\simeq2.8 \times 10^(-2)$ (\cite{Galloni:2022mok}) is one order of magnitude bigger than the one used in our estimates. In the end, our result is not very different from that of Burgess et al.

However, our result on decoherence is only likely to become faster when considering horizon crossing modes. This should happen because the effect of a time dependent environment we commented on in section \ref{subsec:timedependentenvironment} should decrease, while the increase in the dimension of the environment is only likely to speed up decoherence. 

Let us assume we can do a rough estimate to account for these horizon crossing modes. In this case, for the environmental kernel we must employ the full mode functions eq.(\ref{eq:fullmodefunction}), so our calculations are not straightforwardly translated to this more general case. However, we can imagine that the other terms appearing from eq.(\ref{eq:fullmodefunction}) are not likely to cancel with each other, and will probably be at least of the same order of magnitude. Thus, to estimate this contribution, we repeated the calculations we did in eq.( \ref{eq:integralDint1int1}), pushing the cutoff from $10aH$ to $aH$. The result for $D_{11}$ would be
\begin{equation}
     D_{11, horizon}^{int1int1}=\frac{\varepsilon H^2}{4 M_{\rm Pl}  \pi^2 \eta^2}\int_{1}^{a}\frac{ d x }{x} (\pi \delta(1-x) -\frac{\operatorname{sin}(2(1-x))}{1-x}) A(x,a)\simeq \frac{\varepsilon H^2}{4 M_{\rm Pl}  \pi^2 \eta^2} 0.35,
     \label{eq:derivativelesseqhorcross}
\end{equation}
which is around 7 times bigger than the previous result, i.e. just 3 times smaller than the Burgess et al. result. As an the order of magnitude, we expect this interaction alone to give a similar contribution from subhorizon modes as the ones obtained by the interactions $H_{int2}$ in~\cite{burgess_minimal_2023}. This encourages further studies on the effect of subhorizon modes on decoherence.

\section{Derivative interactions}
\label{sec:derivativeinteractions}
In the previous section we have analyzed the derivativeless interaction from eq.(\ref{eq:interactions}). This is important in itself since, despite its simplicity, they have not been analyzed deeply in the context of the interactions of the curvature and tensor perturbations. Because of our choice of the environmental correlation function, these will also be the most relevant interactions, as we will see. However, the other terms which give a sizeable contribution, as well, are the mixed interaction terms, in which one interaction is the derivativeless one $H_{int1}$ and the other is the sum of $H_{int2}$ and $H_{int3}$. It is also important to consider which of the two interactions is integrated in the variable $\eta'$  or instead evaluated on the present moment $\eta$ in eq.(\ref{eq:blochredfield}). We start by considering the case in which the derivativeless interaction is integrated in $\eta'$. For the sake of clarity, we will call the ``memory interaction'' the interaction integrated in $\eta'$.

\subsection{Mixed terms: derivativeless memory}
\label{sec:mixedderivterms123}
By considering again eq.(\ref{eq:blochredfield}), now with the derivativeless interaction $H_{int1}$ inserted inside the memory integration, and the sum of the derivative interactions $H_{int2}$ and $H_{int3}$ evaluated at the present time $\eta$:
\begin{equation}
\begin{split}
   & \rho'_{r}(\eta)=-\frac{H^2 \varepsilon \eta}{4 M_{\rm Pl} ^2}\int_{\eta_0}^{\eta} \frac{d\eta'}{\eta'} \int d^{3} x \int d^3 y \left[v(\boldsymbol{x},\eta) v(\boldsymbol{y},\eta') \sum_{env}  {}_{\epsilon}\langle 0 | \partial_{\boldsymbol{x}} \theta_{ij} (\boldsymbol{x},\eta)  \partial_{\boldsymbol{x}} \theta_{ij} (\boldsymbol{x},\eta)  \theta_{cd}(\boldsymbol{y},\eta')\right.\\
   &\left. \theta_{cd} (\boldsymbol{y},\eta') |0\rangle_{\epsilon}+2 \partial_{\boldsymbol{x}} \nabla^{-2} v(x,\eta) v(\boldsymbol{y},\eta') \sum_{env} {}_{\epsilon} \langle 0 | \partial_{\boldsymbol{x}} \theta_{ij} (\boldsymbol{x}.\eta)  \theta''_{ij} (\boldsymbol{x}.\eta)  \theta_{cd}(\boldsymbol{y},\eta') \theta_{cd} (\boldsymbol{y},\eta') |0\rangle_{\epsilon}\right] \rho_{r}(\eta) + \\
   & (\leftrightarrow \, \, {\rm permutations}),\\
   \end{split}
 \end{equation}
 where ``permutations''  stands for the various possible permutation of operators due to commutators in the eq.(\ref{eq:blochredfield}). Then, taking the Fourier transformations of each field, we can integrate over $x$ and $y$, in order to obtain the momentum conservation expression at each vertex. Also, the derivatives on the environmental fields are now expressed with the relative momenta in Fourier space:
 \begin{equation}
 \begin{split}
      & \rho'_{r}(\eta)=-\frac{H^2 \varepsilon \eta}{4 M_{\rm Pl} ^2} \int \frac{d^{3} l}{(2\pi)^{9}}  \int d^{3}q \int d^{3}b  \int d^{3} k \int d^{3} p \int d^3 w \int_{-1/w}^{\eta} \frac {d\eta'}{\eta'} (2\pi)^{6} \delta^3(\boldsymbol{l}+\boldsymbol{q}+\boldsymbol{b})\\
     &\delta^3(\boldsymbol{w}+\boldsymbol{k}+\boldsymbol{p} )\left[v_{\boldsymbol{w}}(\eta) v_{\boldsymbol{b}}(\eta')  (- \boldsymbol{k} \cdot \boldsymbol{p}) \sum_{env} {}_{\epsilon} \langle 0 |  \theta_{ij,\boldsymbol{k}} (\eta)  \theta_{ij,\boldsymbol{p}} (\eta)  \theta_{cd,\boldsymbol{l}}(\eta') \theta_{cd,\boldsymbol{q}} (\eta') |0\rangle_{\epsilon}  +2 \frac{-i\boldsymbol{w}}{-w^2} v_{\boldsymbol{w}}(\eta)  \right.\\
   &\left. v_{\boldsymbol{b}}(\eta')\frac{(-i \boldsymbol{k}) (-p^2)+ (-i\boldsymbol{p})(-k^2)}{2} \sum_{env}   {}_{\epsilon} \langle 0 |  \theta_{ij,\boldsymbol{k}} (\eta)  \theta_{ij,\boldsymbol{p}} (\eta)  \theta_{cd,\boldsymbol{l}}(\eta') \theta_{cd,\boldsymbol{q}} (\eta') |0\rangle_{\epsilon}  \right] \rho_{r}(\eta) + (\leftrightarrow {\rm perm.}),\\
   \end{split}
 \end{equation}
where we have symmetrized the second vertex (associated to $H_{int 3}$). Notice that the environmental correlation brackets are reduced to the ones in the derivativeless case before (see eq.(\ref{eq:environmentalcorrelationfunctionfull})).
We can apply the Wick theorem to that, and considering each couple of tensor perturbations, we have a product of delta functions, each one exactly as in eq.(\ref{eq:envcorrelationfunctionsinglecouple}). In the associated ``Feynman diagrams''(see fig. \ref{fig:1feyn}), we could associate these other delta functions to the propagator tensor lines, and the number of ways in which to connect the two tensor legs of each vertex to the propagator gives a factor 2. 
In the end, by integrating over all the delta functions, we arrive at:
\begin{equation}
\begin{split}
   & \rho'_{r}(\eta)=-\frac{H^2 \varepsilon}{2 M_{\rm Pl} ^2} \int d^3 w \int_{-1/w}^{\eta} \frac{d\eta'}{\eta'}\int_{10 aH}^{+\infty} \frac{d^{3} k}{(2\pi)^3}   \left[v_{\boldsymbol{w}}(\eta) v_{\boldsymbol{-w}}(\eta') \left( (- \boldsymbol{k}) \cdot (-\boldsymbol{k}-\boldsymbol{w}) - \frac{(-\boldsymbol{w})\cdot(-\boldsymbol{k})}{w^2} \right.\right.\\
   &\left.\left.|\boldsymbol{k}+\boldsymbol{w}|^2+ \frac{(\boldsymbol{w})\cdot(\boldsymbol{k}+\boldsymbol{w})k^2}{w^2}\right) P_{ijcd}(\boldsymbol{k}) P^{ijcd}(-\boldsymbol{k}-\boldsymbol{w}) \frac{e^{-ik (\eta-\eta')}}{2 |\boldsymbol{k}|} \frac{e^{-i|\boldsymbol{k}+\boldsymbol{w}|(\eta-\eta')}}{2|\boldsymbol{k}+\boldsymbol{w}|}\rho_{r}(\eta)+(\leftrightarrow{\rm perm.})\right],
    \label{eq:lindbladderiv}
    \end{split}
\end{equation}
where here $\boldsymbol{w}$ is the momentum of the system, while we are integrating over the environmental momenta $\boldsymbol{k}$.

The other crucial observation for equation (\ref{eq:lindbladderiv}) regards the $w$ system momentum at the denominator, caused by the presence of $H_{int3}$. This diverges in the infrared limit $w \rightarrow 0 $ and so can potentially cause some issues. Also, in that limit (for which $k\gg w$) it has been shown before in eq.(\ref{eq:polarizationssummed}) that the polarization tensor contractions equals 2. We consider for now just the products of momenta inside the angular integral in $\boldsymbol{k}$, by taking the lowest non null order in  the expansion $w/k$:
\begin{equation}
\begin{split}
 & \int dk k^2 \frac{1}{4k^2} e^{-2 i k (\eta-\eta')} (2\pi) \int_{-1}^{1} d \cos \theta \left(k^2 -\frac{\boldsymbol{w}\cdot \boldsymbol{k}}{w^2} k^2- 2\frac{(\boldsymbol{w}\cdot \boldsymbol{k})^2}{w^2}  +\frac{\boldsymbol{w}\cdot \boldsymbol{k}}{w^2} k^2+k^2\right)=\\
 &=\int dk k^2 \frac{1}{4k^2} e^{-2 i k (\eta-\eta')}   (2\pi) \int_{-1}^{1} d \cos \theta \left(2 k^2- 2\frac{(\boldsymbol{w}\cdot \boldsymbol{k})^2}{w^2}\right)=
 \int dk k^2 \frac{1}{4k^2} e^{-2 i k (\eta-\eta')}   (2\pi) \frac{8}{3} k^2.\\
  \end{split}
\end{equation}
By inserting again this inside the previous eq.(\ref{eq:lindbladderiv}) we arrive at
\begin{equation}
    \rho'_{r}(\eta)=-\frac{H^2 \varepsilon \eta}{2 M_{\rm Pl} ^2} \int d^3 \boldsymbol{w} \int_{-1/w}^{\eta} \frac{d\eta'}{\eta'}\frac{4}{3 (2\pi)^2} \int_{10aH}^{\infty} dk k^2 e^{-2 i k (\eta-\eta')} v_{\boldsymbol{w}}(\eta) v_{\boldsymbol{-w}}(\eta') \rho_{r}(\eta)+(\leftrightarrow {\rm permut.}).
\end{equation}
At this point, we exploit the trick in eq.(\ref{eq:dercorreq}) and write:
\begin{equation}
\begin{split}
        \rho'_{r}(\eta)&=-\frac{H^2 \varepsilon \eta}{2 M_{\rm Pl} ^2}   \int d^3 \boldsymbol{w}\int_{-1/w}^{\eta} \frac{d\eta'}{\eta'}\frac{4}{3 (2\pi)^2}\frac{-1}{4} \frac{d^2}{d\eta'^2} \left(\int_{10aH}^{\infty} dk  e^{-2 i k (\eta-\eta')} \right)v_{\boldsymbol{w}}(\eta) v_{\boldsymbol{-w}}(\eta') \rho_{r}(\eta)+(\leftrightarrow ).
\end{split}
\end{equation}
Now it is simple to solve the integral, being analogous to the derivativeless case, and we have:
\begin{equation}
    \rho'_{r}(\eta)=-\frac{i H^2 \varepsilon \eta}{48 \pi^2 M_{\rm Pl} ^2} \int d^3 \boldsymbol{w}\int_{-1/w}^{\eta} \frac{d\eta'}{\eta'}\frac{d^2}{d\eta'^2}   \frac{e^{20 i \frac{ (\eta-\eta')}{\eta}}}{\eta-\eta'-i\epsilon} v_{\boldsymbol{w}}(\eta) v_{\boldsymbol{-w}}(\eta') \rho_{r}(\eta)+(\leftrightarrow {\rm permutations}).
\label{eq:qmederivativeeq}
\end{equation}
We then notice that the equation is Gaussian in the system variable; this means, as usual, that in the calculations we can separate the contribution to the density matrix for each value of the momenta. Furthermore, as discussed when introducing eq.(\ref{eq:continuumtodiscretemomenta}), we can as well consider discrete momenta:
\begin{equation}
    \rho_{r}= \bigotimes_{\boldsymbol{w}} \rho_{\boldsymbol{w}}.
\end{equation}
The next steps correspond to the ones employed in section \ref{sec:derivativelesscanonical} for the derivativeless interaction: we write the quantum master equation in the canonical form, similarly to the Lindblad form in eq.(\ref{eq:canonicalqme}), by also employing the eq.(\ref{eq:AB}) to express $v_{w}(\eta')$ in terms of $v_{w}(\eta)$ and $p_{w}(\eta)$. We start from considering the part of the equation associated to the real part of the environmental correlation function, in particular the coefficient $D_{11}$ in the $D$ matrix:
\begin{equation}
\begin{split}
    \rho'_{\boldsymbol{w}}(\eta)&=-\frac{i H^2 \varepsilon \eta}{48 \pi^2 M_{\rm Pl} ^2\eta^2} \int_{a}^{1} \frac{dx}{x}\frac{d^2}{dx^2} \left(\frac{ e^{-20 i (1-x) }}{1-x-i\epsilon}-\frac{ e^{20 i (1-x) }}{1-x+i\epsilon}\right) A(x,a)\times\\
   &\times[v_{\boldsymbol{w}}(\eta)\rho_{\boldsymbol{w}}(\eta) v_{\boldsymbol{-w}}(\eta) -\frac{1}{2} \{\rho_{\boldsymbol{w}}(\eta), v_{\boldsymbol{w}}(\eta) v_{\boldsymbol{-w}}(\eta) \}],\\
    \end{split}
    \label{eq:qme1-23beforepartint}
\end{equation}
where we have introduced again the variables $x=\eta'/\eta$ and $a=-1 / (w \eta)$. Inside the eq.(\ref{eq:qme1-23beforepartint}), we integrate by parts the derivative in $\eta'$ twice. Notice that
\begin{equation}
    \frac{d^2}{dx^2} \left(\frac{ e^{-20 i (1-x) }}{1-x-i\epsilon}-\frac{ e^{20 i (1-x) }}{1-x+i\epsilon}\right)= \frac{d^2}{dx^2} \left(\frac{ e^{-20 i (1-x) }(1-x)}{(1-x)^2+(\epsilon)^2}-\frac{ e^{20 i (1-x) }(1-x)}{(1-x)^2+(\epsilon)^2}\right)+ 2i\pi \delta''(x-1),
    \label{eq:D11formixedderivativederivativelesssnumerical}
\end{equation}
where we have used the Solkhotsi-Plemelji theorem. Apart from the local term with the Dirac delta, the other terms produce some boundary terms; we will deal with the latter in section \ref{subsection:boundary terms}, showing that we can ignore their contribution. In that subsection we also deal with the mixed $v(\eta)$ and $p(\eta)$ terms, showing that the coefficients $D_{12}$ and $\Delta_{12}$, as defined in eqs.(\ref{eq:Dmatrix}) and (\ref{eq:deltamatrix}) are negligible also in this case. 

By integrating by parts inside the integral we obtain
\begin{equation}
\begin{split}
&D^{int1-23}_{11}=-\frac{H^2\varepsilon}{24\pi^2M_{\rm Pl} ^2\eta^2} \int_{1}^{a} dx \left(\pi\delta(x-1)-\frac{\sin{20(x-1)}}{x-1}\right) \left(\frac{d^2}{dx^2}\frac{A}{x}\right)=\\
&\simeq \frac{H^2\varepsilon}{24\pi^2M_{\rm Pl} ^2\eta^2} (-2\pi+5.5).\\
\label{eq:D11negativefor 2-3}
\end{split}
\end{equation}
The negativity is due to the number of spatial derivatives into the two couplings, as noticed e.g. in eq.(\ref{eq:dercorreq}), and would have not be there if the two couplings were the same.\footnote{In fact, in that case we would have had 0 or 4 derivatives, corresponding to  a positive coefficient.} Such a result, therefore, is a direct consequence of considering 
 a ``mixed term'', with two different couplings in the vertex. Such a framework has never been analyzed, since many papers usually focused on just one coupling, either for simplicity or because, with an educated guess, they claimed that those vertices were the most important ones in the calculation. Here, we will explicitly show, in section \ref{sec:derivativeinteractions}, that couples of derivative interactions ($H_{int2}$, $H_{int3}$) give negligible results for the quantum master equation coefficients. We have previously shown in section \ref{sec:derivativelesscanonical} that the derivativeless couple gives instead an important contribution to the quantum master equation coefficients. Nevertheless, the terms mixed between derivative and derivativeless give sizeable results. This seems to suggest that considering just a couple of interactions, even though they seem to be the dominant ones, may lead to overestimating the correct result, and it may be worth investigating more the interplay between the various interactions.

In particular, by evaluating the $D_{11}$ coefficient arising from the mixed term provided by  derivativeless memory($H_{int1}$)  and the derivative ones ($H_{int2}$ and $H_{int3}$) numerically:
\begin{equation}
    D^{int1-23}_{11}\simeq -0.13\frac{H^2\varepsilon}{4\pi^2M_{\rm Pl} ^2\eta^2}. 
    \label{eq:Dnegativemixedresult}
\end{equation}
By summing the result of eq.(\ref{eq:Dnegativemixedresult}) with the $D_{11}^{int11}$  results in eq.(\ref{eq:derivativelessresult}) it balances out the positive result of the derivativeless interactions, so the net result is negative. The negativity of the $D_{11}$ coefficient, as we already discussed when introducing the canonical form of the quantum master equation in eq.(\ref{eq:noncanonicallind}), does not imply necessarily a violation of the positivity of the quantum master equation. Nevertheless, we cannot apply anymore the sufficient condition associated to the Lindblad theorem, also discussed after eq.(\ref{eq:noncanonicallind}).

This problem has already been noticed (although, with a different environment) in~\cite{burgess_minimal_2023}. A possible prescription, applied recently in the literature~\cite{burgess_minimal_2023,Burgess:2024eng} to overcome this problem, is what we called the ``Strong Markovian approximation''. It has already been considered in this paper for the derivativeless interaction, although for evaluating the Lamb Shift (see eq.(\ref{eq:delta11derivativeless2})). It consists of taking $A=1$, as explained for eq.(\ref{eq:smardelta}), which is equivalent to considering the substitution $v(\eta')\to v(\eta)$ inside the equation. 

One reason to apply the Strong Markovian approximation is that the memory terms may be not accurately modelled by considering just the Born-Markov approximation \ref{eq:markovianapprox}, in particular since we are in an expanding background. See the discussions in \ref{subsec:the Markovian approximation}.
In this case, the net effect changes from
\begin{equation}
    \frac{d^2}{dx^2} \frac{A(x)}{x}= \frac{d^2}{dx^2}\left(\frac{1}{x} \left(\frac{x^2}{3}+\frac{2}{3x}\right) \right)=\frac{4}{x^4}, 
\end{equation}
to
\begin{equation}
    \frac{d^2}{dx^2} \frac{1}{x}=\frac{2}{x^3}.
\end{equation} 
It essentially halves each contribution, and the new value is:
\begin{equation}
    D_{11}^{int1-23}=\frac{H^2\varepsilon}{24\pi^2M_{\rm Pl} ^2\eta^2}\left(-\pi+\int_{1}^{a}\frac{\sin{20(1-x)}}
{(1-x)}  \frac{d^2}{dx^2} \frac{1}{x}\right)\simeq-\frac{H^2\varepsilon}{4\pi^2M_{\rm Pl} ^2\eta^2}0.049,
\label{eq:D11123aftersmar}
\end{equation}
which is again smaller than zero, but in such a way that if summed with the derivativeless case the result is still positive.
Notice that considering $A=1$ is equivalent to considering $B=0$. In principle, this seems to be a good approximation, as the off diagonal coefficients of the D matrix eq.(\ref{eq:Dmatrix}) involving $B$ (defined in eq.(\ref{eq:AB})) are actually shown to be small in \ref{subsection:boundary terms}. 

The imaginary part instead creates a correction of the spectral index, just as we showed previously:
\begin{equation}
    \delta n_{S} =\frac{2}{3} \Delta_{11}^{int1-23} \eta^2.
\end{equation}
We can compute $ \Delta_{11}^{int1-23}$ by remembering its expression from eq.(\ref{eq:delta11def}); in this case we would have indeed, by taking the imaginary part of the correlation functions in eq.(\ref{eq:qmederivativeeq}):
\begin{equation}
\begin{split}
     \Delta_{11}^{int1-23}&=\frac{2H^2\varepsilon \eta}{48\pi^2M_{\rm Pl} ^2}\int_{-\frac{1}{p}}^{\eta} \frac{d\eta'}{\eta'} \Im \frac{d^2}{d\eta'^2}\left(i \frac{e^{20 i (1-\frac{\eta'}{\eta})}}{\eta-\eta'-i\epsilon}\right)=\\
     =&-\frac{2H^2\varepsilon}{48\pi^2M_{\rm Pl} ^2\eta^2}\int_{1}^a \frac{dx}{x} A(x) \frac{d^2}{dx^2} \frac{\cos{20(1-x)}}{1-x}.
     \label{eq:Delta123}
     \end{split}
\end{equation}
We will solve this integral by integrating by parts. When dealing with boundary terms, we can see that we will obtain an ultraviolet divergence, proportional to $1/\epsilon^2$, which is analogous to the one in eq.(\ref{eq:renormalizationofthemass}), and can be renormalized by a mass counterterm. At the same time no ``dangerous term'' emerges by $x=a$ extremum, differently from eq.(\ref{eq:delta11derivativeless2}). After having dealt with boundary terms, we try to extract a finite result by employing $A(x)$:
\begin{equation}
    \frac{d^2}{dx^2} \left(\frac{2}{3x^2}+\frac{x}{3}\right)=\frac{4}{x^4},
\end{equation}
where notice that what we called the ``memory effect associated to the decaying mode'' referring to eq.(\ref{eq:decayingmodecontributiontoA})) is erased by the derivative. This is also connected to not having any `dangerous term'' by $x=a$ extremum. However, we still need to regularize the integral by adopting eq.(\ref{eq:regularizedivprincipalvalue}) and then integrating again by part, since the $1-x$ at the denominator is divergent:
\begin{equation}
\begin{split}
    \Delta_{11}^{int1-23}&=\frac{H^2\varepsilon}{24\pi^2M_{\rm Pl} ^2\eta^2} \int_{1}^a dx \frac{4 \cos{20(1-x)}}{x^4}  \frac{1}{2} \frac{d}{dx} \left(\log(x-1)^2+\epsilon^2\right)=\\
    &=\frac{H^2\varepsilon}{24\pi^2M_{\rm Pl} ^2\eta^2} \left(\left[\log\left((x-1)^2+\epsilon^2\right)\frac{4}{x^4}\frac{\cos{20(1-x)}}{2}\right]_{1}^a-\right.\\
    &\left.-\int_{1}^a dx \frac{d}{dx}\frac{4 \cos{20(1-x)}}{x^4}  \frac{1}{2}  \left(\log(x-1)^2+\epsilon^2\right)\right).\\  
    \end{split}
    \label{eq:delta11per123}
\end{equation}
We again have new boundary terms for this second integration by parts. The one referred to $x=a$ is obviously negligible, while the boundary term for $x=1$ leads again just to a renormalization of the mass just like eq.(\ref{eq:renormalizationofthemass}) and eq.(\ref{eq:delta11derivativeless2}). It is remarkable that even if starting from equations that feature more complicated operators (e.g. with spatial and temporal derivatives), as in interactions like $H_{int2}$ and $H_{int3}$, we end up just renormalizing the mass. This may be an effect of considering a small wavelength environment compared to the system. The numerical result of the integration turns out to be
\begin{equation}
     \Delta_{11}^{int1-23}\simeq-\frac{H^2\varepsilon}{4\pi^2M_{\rm Pl} ^2\eta^2} 2.4 \quad {\rm which\,\, leads\,\, to} \quad \delta n_{S}\simeq-\frac{H^2\varepsilon}{4\pi^2M_{\rm Pl} ^2} 1.6.
    \end{equation}
The result is a red correction to the spectral index, which is smaller than eq.(\ref{eq:derivativelessspectralindexcorr}), so, still, the resultant correction would be blue. 

Nevertheless, for coherence, since in the end we will apply it to the full quantum master equation, we want to see the impact of the Strong Markovian approximation $A=1$ on this calculation; in that case the only change is
\begin{equation}
    \frac{d^2}{dx^2}\frac{1}{x}=\frac{2}{x^3},
    \label{eq:smarderivderivelss}
\end{equation}
by performing all the calculations the result obtained by applying the Strong markovian approximation is
\begin{equation}
     \Delta_{11}^{int1-23}\simeq\frac{H^2\varepsilon}{4\pi^2M_{\rm Pl} ^2\eta^2} 1.2 \rightarrow \delta n_{S}\simeq-\frac{H^2\varepsilon}{4\pi^2M_{\rm Pl} ^2\eta^2} 0.8,
\label{eq:derivativelessderivativeinteractionspectralindex}
\end{equation}
which is around the half, due to the same factor $2$ that we lost in the $D^{int1-23}_{11}$ computation.

\subsection{Mixed terms: derivative memory}
\label{sec:mixedterms231}
Now we consider the case in which the derivative interactions are integrated inside the memory integral. Notice that, with respect to the inverse case analyzed in \ref{sec:mixedderivterms123}, we don't have any variation of the momenta integral calculations; so we can infer the result by eq. (\ref{eq:qmederivativeeq}), but with the substitutions:
\begin{equation}
    \eta \rightarrow 1/\eta; \quad \quad \frac{1}{\eta'}\rightarrow \eta',
\end{equation}
due to the ``swap'' of the interactions. Therefore the equation reads:
\begin{equation}
    \rho'_{r}(\eta)=- \frac{\varepsilon H^2}{48 i \pi^2 M_{\rm Pl} ^2 \eta} \sum_{\boldsymbol{w}} \int_{-1/p}^{\eta} d\eta' \eta' \frac{d^2}{d\eta'^{2}} \left(\frac{e^{20 i(1-x)}}{\eta-\eta'-i \epsilon}\right) v_{\boldsymbol{w}}(\eta) v_{\boldsymbol{-w}}(\eta')\rho_{r}(\eta) +(\leftrightarrow {\rm perm.}).
\end{equation}
Considering the canonical form of the quantum master equation eq.(\ref{eq:canonicalqme}) we can extract the $D_{11}(\eta)$ and $\Delta_{11}$ coefficients. Consider first the part of the quantum master equation associated to 
the real part of the correlation function:
\begin{equation}
D_{11}^{2-3,1}= \frac{\varepsilon H^2}{24 \pi^2 M_{\rm Pl} ^2 \eta^2} \int_{1}^a dx x A(x,a) \frac{d^2}{dx^{2}} \left(\frac{\sin{20(1-x))} (1-x) } {(1-x)^2+ \epsilon^2} - \pi \delta(x-1) \right). 
\label{eq:2,3-1mixedderivative}
\end{equation}
By integrating by parts, we have again a local part, where we can directly ``move'' derivatives, for the properties of the delta functions, and a non local part, from which we have a couple of boundary terms we treat in section \ref{subsection:boundary terms}. For $A(x,a)$ we use the approximate expression (\ref{eq:Aapprox}), but we numerically verified that the final result does not change, since the integral is highly peaked around $x=1$:
\begin{equation}
   \frac{d^2 ( x A(x,a))}{dx^2} =\frac{d^2}{dx^2} \left(\frac{x^3}{3}+ \frac{2}{3}\right)=2x.
\end{equation}
Notice however that, in this case, the only component of $A(x,a)$ that sources a finite result, because of a cancellation, is the one connected to the ``memory effect of the decaying mode'', i.e. the $x^2$ part in eq.(\ref{eq:Aapprox}). 
The expression for the quantum master equation coefficient is
\begin{equation}
\begin{split}
    D_{11}^{2,3-1}= &\frac{\varepsilon H^2}{24 \pi^2 M_{\rm Pl} ^2 \eta^2} \int_{1}^a dx \left(- 2x \pi \delta(x-1)+ \int_{1}^{a} 2x \frac{\sin{20(x-1)}}{x-1}\right)=\\
    &=\frac{\varepsilon H^2}{24 \pi^2 M_{\rm Pl} ^2 \eta^2}  \left(- \pi+\frac{1}{10}(1- \cos{20(1-a)})+2 Si(20(a-1))\right).\\
    \end{split}
    \label{eq:D11derivativeintegrated}
\end{equation}
We managed, thanks to the approximation for $A(x,a)$, to reach a closed analytycal form. As we checked also numerically, but is evident also from eq.(\ref{eq:D11derivativeintegrated}), the dependence on $a=-1/k \eta$ of the non-Markovian part is very weak. Since $a$ is the ratio between the scale factor when the mode is deep superhorizon and at horizon exit, we can take the limit $a\rightarrow \infty$. Then, $\operatorname{SinIntegral}(20a) \rightarrow \pi/2$, but we have some residual oscillations, expressed in the term proportional to $\cos a$ in eq.(\ref{eq:D11derivativeintegrated}). 

It is difficult to believe that these oscillations are physical in any way, in the same way as we observed for the boundary terms dealt with in appendix \ref{subsection:boundary terms}. The Born-Markov approximation (\ref{eq:markovianapprox}) that we did is hardly compatible with terms with a memory extending over decades of e-folds. 

One possible way out is to consider an averaging of the oscillations. We have already underlined the importance of the coarse graining of the non-unitary dynamics in the quantum master equation over scales $\Delta \eta_{coarsegraining} \lesssim  \eta$, where $\eta$ is the typical time scale of evolution of the system, in the derivation of the quantum master equation in section \ref{sec:qme}(see also ~\cite{Lidar_2001,Lidar:2019qog,Agon:2014uxa} for more details). The time scale of the oscillations are of order $\Delta \eta_{oscill}=p \eta^2 \pi/10$, where $p$ is the system momentum, which is very little for superhorizon scales. We can thus imagine to average the coefficients of the quantum master equation over every time scale which is less than $\Delta \eta_{coarsegraining} \lesssim \eta$, the typical time scale of evolution of the system. Since $\Delta \eta_{oscill}<<\Delta \eta_{coarsegraining}$, averaging the oscillatory part of $D_{11}$ over a period(of a oscillation) gives:
\begin{equation}
    \int_{\eta}^{\eta+p \eta^2 \pi/10} d\eta \frac{1}{\eta^2} \cos(20 a)=\int_{a}^{a+\pi/10} da \cos(20a)=0
    \label{eq:averagingout}
\end{equation}
while the other parts of the quantum master equation coefficient in eq.(\ref{eq:D11derivativeintegrated}) is essentially independent of the averaging, being the averaging period so little in conformal time $\eta$ that it does not induce any change in the coefficients. Thus, we will ignore the oscillatory part of eq.(\ref{eq:2,3-1mixedderivative}) and write the numerical result as
\begin{equation}
    D_{11}^{2-3,1}= \frac{\varepsilon H^2}{24 \pi^2 M_{\rm Pl} ^2 \eta^2} \left(-\pi + \pi +1/10\right)\simeq \frac{\varepsilon H^2}{4 \pi^2 M_{\rm Pl} ^2 \eta^2 }0.017.
    \label{eq:derivativederivativelessD11negativememoty}
\end{equation}

In a similar way we can treat the Lamb Shift Hamiltonian. Let us first consider the coefficient $\Delta_{11}$, obtained again by just exchanging with respect to the previous case the integrated interaction:
\begin{equation}
\Delta_{11}=-\frac{\varepsilon H^2}{24 \pi^2 M_{\rm Pl} ^2 \eta^2}\int_{1}^a d x \frac{d^2}{dx^2}\left(\frac{\cos{20(1-x)}}{1-x}\right)(Ax).
\label{eq:delta231}
\end{equation}
As usual, we integrate by parts three times, twice because of the derivative in the environmental correlation function and the third to regularize the divergence of the integrand in $x=1$ as shown in eq.(\ref{eq:regularizedivprincipalvalue}), analogously to what done for eq.(\ref{eq:delta11per123}) . The boundary terms are treated in subsection \ref{subsection:boundary terms}. We obtain
\begin{equation}
\Delta_{11}=-\frac{\varepsilon H^2}{24 \pi^2 M_{\rm Pl} ^2 \eta^2} \int_{1}^{a} dx \frac{d}{dx}\left(x \cos{20(1-x)} \right) \log((x-1)^2+\epsilon^2). 
\label{eq:d11derivativeper231}
\end{equation}
By solving the integral numerically, it is evident that the integral is peaked around $x=a$ and not $x=1$, and grows indefinitely as $a \to \infty$. This is due to the presence of the $x$ factor in the memory integration, related to the irrelevance of the derivative interactions $2-3$ as $\eta \to 0$. 

Such a huge memory is not compatible with our approximations, as explained many times. Moreover, in this specific case this growth appears to be an unphysical effect similarly to what already discussed for the derivativeless interactions (see discussion before eq.(\ref{eq:delta11derivativeless2})). Actually, the logarithm in eq.(\ref{eq:d11derivativeper231}) was introduced by the trick in eq.(\ref{eq:regularizedivprincipalvalue}) to cure the divergence in $x=1$. In order to deal with it, since we showed that it worked for the derivativeless interactions, we adopt the Strong Markovian Approximation, according to which we take $A=1$. The surprising result is that now we only have boundary terms, as after integrating by parts eq.(\ref{eq:delta231}) we obtain: 
\begin{equation}
    \Delta_{11}=\frac{\varepsilon H^2}{48 \pi^2 M_{\rm Pl} ^2 \eta^2} \left[x \frac{d}{dx} \frac{\cos{20(1-x)}(1-x)}{(1-x)^2+\epsilon^2}-\frac{\cos{20(1-x)}(1-x)}{(1-x)^2+\epsilon^2}\right]_{1}^{a}.
    \label{eq:delta231afterSMAR}
\end{equation}
The boundary terms are treated as prescribed in appendix \ref{subsection:boundary terms}. We will have a renormalization of the mass because of an UV divergence and another term which can be averaged out to zero. However, in this case, we have no finite result. So, we can definitely tell that there are no finite Lamb Shift corrections to the couple of interaction derivativeless/derivative, if the derivative ones are integrated. 

\subsection{Derivative interactions}
\label{subsec:derivative interactions}
In this case, we should sum all the contributions coming from the couples of derivative interactions, i.e.: $H_{int2}-H_{int2}, H_{int3}-H_{int2}, H_{int3}-H_{int3}$. Apart from a coefficient coming from the precise angular part of the integration of the momenta,  which we call $\beta$ for simplicity , the quantum master equation will be
\begin{equation}
\rho'_{\boldsymbol{w}}(\eta)= -\frac{\beta \varepsilon H^2 \eta}{8 M_{\rm Pl} ^2} \int_{-1/p}^{\eta} d\eta' \eta' \int_{10 a H}^{\infty} d k k^4 e^{-2 i k (\eta-\eta')} v_{\boldsymbol{w}}(\eta) v_{\boldsymbol{-w}}(\eta') \rho_{\boldsymbol{w}}(\eta)+ ( {\rm permutations} \leftrightarrow ).
\label{eq:qmederivativesquare}
\end{equation}
Just focusing to begin with on the real part of the coefficient $D_{11}$:
\begin{equation}
\begin{split}
\rho'_{\boldsymbol{w}}(\eta)&= +\frac{\beta \varepsilon H^2 }{8 M_{\rm Pl} ^2 \eta^2} \int_{1}^{a} d x x \frac{1}{32} \frac{d^4}{dx^4} \left(\frac{\sin{20(1-x)} }{1-x} - \pi \delta (x-1)  \right) A(x,a) v_{\boldsymbol{w}}(\eta) v_{\boldsymbol{-w}} (\eta) \rho_{\boldsymbol{w}}(\eta)+\\
&+( {\rm permutations} \leftrightarrow ).
\label{eq:derivderivreal}
\end{split}
\end{equation}
We could try to do what already done in the other cases, i.e. integrating by parts 4 times ``moving'' the derivatives to $x A$. However, it is possible to see that
\begin{equation}
    \frac{d^4}{dx^4} (A(x,a) x)\Bigr|_{x=1}=\frac{1}{a^4}-\frac{4}{a^2},
    \label{eq:eqderiv1}
\end{equation}
so the part associated to the Dirac delta is suppressed by powers of $a$. Computing the same quantity $A(x,a)$ at a generic $x$, but by considering an asymptotic expansion around the limit $a \to \infty$, we see that 
\begin{equation}
    \frac{d^4}{dx^4} (A(x,a) x)=\frac{-4x}{a^2}+O\left(\frac{1}{a^4}\right),
    \label{eq:deriv2}
\end{equation} 
which is suppressed at every $x$ by at least two powers of $a$. So, actually, also when integrating the non local in time part of the coefficient, i.e. the sine in eq.(\ref{eq:derivderivreal}), we would also have such a suppression in $a$, which shows why there are no finite terms coming from the derivative interactions in the quantum master equation coefficients. We underline that in both eq.(\ref{eq:eqderiv1}) and eq.(\ref{eq:deriv2}) we used the full form of $A(x,a)$, as defined in eq.(\ref{eq:ABwithxanda}).

This is actually valid as well for the imaginary part, where we simply have a cosine instead of a sine: 
\begin{equation}
\begin{split}
     &\rho'_{\boldsymbol{w}}(\eta)= +\frac{\beta \varepsilon H^2 }{8 M_{\rm Pl} ^2 \eta^2} \int_{1}^{a} d x x \frac{1}{32} \frac{d^4}{dx^4} \left(\frac{\cos{20(1-x)} }{1-x} \right) A(x,a) v_{\boldsymbol{w}}(\eta) v_{\boldsymbol{-w}} (\eta) \rho_{\boldsymbol{w}}(\eta)\\
    &+ ( {\rm permutations} \leftrightarrow ).\\
    \label{eq:derivderivima}
\end{split}
\end{equation}
The suppression, also in this case, makes all the finite result negligible. The divergence of the $1/(x-1)$ can be cured as in eq.(\ref{eq:delta11per123}, by writing it as the derivative of a logarithm, as in eq.(\ref{eq:regularizedivprincipalvalue}); nevertheless, it is easy to see that the aforementioned $a^2$ suppression in the fourth derivative of $A(x,a)$ eventually drives to zero the final result. 

This could be understood also by considering the approximation of $A(x,a)= 2/3x+ x^2/3+O(1/a)$; the fourth derivative is identically null everywhere. This is even more true if we consider the ``Strong Markovian Approximation'' $A=1$. The physical reason for this to happen can be seen by the fact that the $\eta$ power in the derivative interactions makes the operators irrelevant as $\eta \to 0$, and this suppression is not compensated by any other power of $\eta$. In the case of minimally coupled scalar fields in the environment, there could be a different outcome due to the presence of the neglected terms in eq.(\ref{eq:fullmodefunction}), i.e. the $-i/k\eta$ contribution. So, this null result can be interpreted just as a consequence of our approximations, which were too rough in eliminating the $-i/k\eta$.  In a future project, we will consider also the environmental modes crossing the horizon, and then evaluate the importance of these modes. 

On the other hand, this suppression goes into the direction of confirming that deep subhorizon modes contribution to derivative interaction is suppressed, as checked in~\cite{burgess_minimal_2023}.

It is interesting to analyze also the terms coming from the boundary integration. The calculations are developed in \ref{par:boundarytermsfromderivative}. It is worth pointing out that some boundary terms computed in $x=a$,i.e. corresponding to horizon crossing instant of time, turn out to be dominant in the $1/\eta$ expansion and cannot be averaged out: thus, they are ``dangerous'' for our results (see eq.(\ref{eq:danferouscontr})). If we stick to solving the Bloch-Redfield equation (\ref{eq:blochredfield}) we can consider them as ``spurious terms'', analogously to what done in \cite{colas_benchmarking_2022}, and so erase them by hand. However, in this work we have used the Strong Markovian approximation (\ref{eq:smar1}) in order to infer the final results. Thus, it seems a coherent choice to apply this approximation also in this case. Actually, these dangerous contributions, disappear after the application of the Strong Markovian approximation, as discussed in the appendix (\ref{par:boundarytermsfromderivative}). 

Instead, interestingly, we have the appearance of ultraviolet divergences in the non unitary part coefficient in eq.(\ref{eq:worryinguvdivergence}). These divergences do not disappear after applying the Strong Markovian approximation. Also, they cannot be renormalized as we did with the Hamiltonian UV divergences, e.g. in eq.(\ref{eq:renormalizationofthemass}). To our knowledge, there is no widely accepted method to properly take care of these ultraviolet divergences in the literature of open quantum field theory. Some papers have connected their origin to the renormalization of composite operator~\cite{Agon:2014uxa}. 
Recently, their appearance was discussed also in \cite{Burgess:2024heo}. In that paper, it was argued that, since it is expected to have divergent counterterms hidden in the bare couplings in the lagrangian, in order to reabsorb hamiltonian divergences, we should also expect that they then contribute to terms in purity calculation,  and thus divergences should pop up also in the non-unitary part of the quatnum master equation. In this sense, here we present a first confirmation of this intuition in an inflationary context. However, verifying their connection to the Hamiltonian divergence is beyond the scope of this work. We leave a proper treatment of these divergences to further studies. 

\section{Final results, with and without Strong Markovian Approximation}
\label{subsec:final results SMAR}
\subsection{Decoherence}
\label{subsec:finalresultsdeco}
Summing the three non null contributions to $D_{11}$ from derivativeless interactions eq.(\ref{eq:derivativelessresult}) and mixed derivativeless-derivative interactions eq.(\ref{eq:derivativederivativelessD11negativememoty}) and eq.(\ref{eq:D11negativefor 2-3}), obtained by solving the Bloch-Redfield equation, we find a net final $D_{11}$ negative:
\begin{equation}
    D_{11}=\frac{\varepsilon H^2}{M_{\rm Pl} ^2 4 \pi^2 \eta^2}(+0.017-0.13+0.05) \simeq -0.063 \frac{\varepsilon H^2}{M_{\rm Pl} ^2 4 \pi^2 \eta^2}.
\end{equation}
This negative $D_{11}$ does not necessarily imply that there is a violation of positivity of the reduced density matrix. There actually are counterexamples in the literature~\cite{Whitney_2008}. The appearance of a negative coefficient has often been connected to the non-Markovianity of the system~\cite{Breuer_2016}. So, our is another example of how non-Markovianity and memory effects appear to be ubiquitous in an inflationary setting. However, we cannot apply anymore the sufficient condition associated to the Lindblad theorem applied to time dependent quantum master equations, that we introduced in the discussion after eq.(\ref{eq:noncanonicallind}). Already~\cite{burgess_minimal_2023} faced the same issue, and they proposed a possible way out. The idea is that in a dynamical background the usual Born-Markovian approximation eq.(\ref{eq:Bornapprox}) is not appropriate to the quantum master equation in a cosmological context, since memory effects could be enhanced by the expansion of the universe. Extracting correctly the Markovian limit in inflation is probably not a solved problem in general, and it will be one of the main direction of future research in the next years. However, in~\cite{burgess_minimal_2023}, the authors suggested a possible prescription in order to extract the limit, which we have here called ``Strong Markovian approximation''. In the spirit of testing this prescription, we apply it also to this circumstance, in order to try to compute the decoherence rate. 
By applying the Strong Markovian approximation to the mixed derivative-derivativeless interaction, in the case where the derivative interactions are integrated (thus representing the memory part) we find:
\begin{equation}
    D_{11}^{23-1,{\rm smar}}=\frac{-\varepsilon H^2}{24 \pi^2 M_{\rm Pl} ^2 \eta^2} \int_{1}^{a} dx x \frac{d^2}{dx^2} \left(\frac{\sin{20(1-x)}}{1-x}- \pi \delta(x-1)\right).
\end{equation}
Thus, by integrating by parts twice and moving the two derivatives to the $x$ we have that the memory term is cancelled and no finite results, apart from boundary terms (see eq.(\ref{eq:boundarytermsderivmemorysmar}),  are now there. This happens since the only terms which gave rise to the result in eq.(\ref{eq:D11formixedderivativederivativelesssnumerical}) is the ``memory term'', associated to decaying mode in $A(x,a)$, i.e. $x^2/3$, which is now erased by the Strong Markovian approximation.

By applying the Strong Markovian approximation to the derivativeless interaction we instead have
\begin{equation}
    D_{11}^{int11,smar}=-\frac{\varepsilon H^2}{4 M_{\rm Pl} ^2 \pi^2 \eta^2} \int_{1}^{a} \frac{dx}{x} \left(\pi \delta(1-x)-\frac{\sin{20(1-x)}}{1-x}\right)\simeq 0.0498\frac{\varepsilon H^2}{4 M_{\rm Pl} ^2 \pi^2 \eta^2} .
\end{equation}
The result from the other mixed interaction, where the integrated ones are derivativeless, is also changed, as already shown in eq.(\ref{eq:D11123aftersmar}). Notice that the numerical value of the result in eq.(\ref{eq:D11123aftersmar}) is really close to the derivativeless one eq.(\ref{eq:derivativelessresult}). 

For this coincidence, in order to avoid approximation errors, we decided to express their difference in an analytical form, by subtracting the two contributions part by part
\begin{equation}
    D_{11}^{tot,smar}=\frac{\varepsilon H^2}{4 \pi^2 M_{\rm Pl} ^2} \left(\frac{\pi}{3}-\int_{1}^{a} dx \left( \frac{ \sin{20(1-x)}}{1-x} \left(\frac{1}{x}-\frac{2}{x^3}\right)\right)\right).
    \label{eq:d11totssmar}
\end{equation}
By solving analytically the two integrals we obtain, for the second term inside of the parnthesis of eq.(\ref{eq:d11totssmar}):
\begin{equation}
\begin{split}
 & -\frac{\varepsilon H^2}{4 \pi^2 M_{\rm Pl} ^2} \left(  \frac{-404 a^2 \text{Ci}(20) \sin {20}-40 a^2 \text{Ci}(20) \cos {20)}+4 a^2 \text{Ci}(20 a) (101 \sin {20}+10 \cos {20}-}{6a^2}\right.\\
 &\frac{-4 a^2 \text{Si}(20-20 a)-40 a^2 \text{Si}(20) \sin {20}
 +40 a^2 \sin {20} \text{Si}(20 a)+404 a^2 \text{Si}(20) \cos {20}}{6a^2}\\
 &\left.\frac{-404 a^2 \cos {20} \text{Si}(20 a)+20 a^2+2 a \sin {20-20 a}+\sin {20-20 a}-20 a \cos {20 (a-1)}}{6 a^2}\right)=\\
 &-\frac{\varepsilon H^2}{4 \pi^2 M_{\rm Pl} ^2\eta^2}1.04665,\\
  \end{split}
\end{equation}
while the local in time part is
\begin{equation}
    \frac{\varepsilon H^2}{4 \pi^2 M_{\rm Pl} ^2} \frac{\pi}{3}\simeq 1.0472,
\end{equation}
so that, the net $D_{11}$ coefficient can be estimated numerically as
\begin{equation}
     D_{11}^{tot,smar}=\frac{\varepsilon H^2}{4 \pi^2 M_{\rm Pl} ^2} 5 \times 10^{-4},
\end{equation}
which is  really suppressed compared to the one by Burgess et al.~\cite{burgess_eft_2015} and to eq.(\ref{eq:derivativelessresult}). Nevertheless, after applying the ``strong Markovian approximation'', the coefficient is positive, and the dynamics is now ensured to be safe for the evolution of the density matrix. If we want to compute decoherence we may apply eq.(\ref{eq:purityforagaussianstate}). The purity evolution is related to the determinant of the covariance matrix:
\begin{equation}
4\operatorname{det}\left({\Sigma}\right)= \int_{-1/p}^{\eta} d\eta' 4 {D}_{11}(\eta') P_{vv}(\eta',p) =\int_{-1/p}^{\eta} \frac{5 \times 10^{-4} \varepsilon H^2}{M_p^2 4 \pi^2 \eta'^2}  \frac{1}{2\eta'2 p^3}\simeq \frac{20 \times 10^{-4} \varepsilon H^2}{M_p^2 24 \pi^2} \left(\frac{aH}{p}\right)^3,
\end{equation}
which numerically, as in eq.(\ref{eq:decoherencederivativelessefolds}), reads
\begin{equation}
  4\operatorname{det}\left({\Sigma}\right)\simeq 6 \times 10^{-21}\left(\frac{\mathcal{P}_\zeta}{2.2 \times 10^{-9}}\right)\left(\frac{r}{10^{-3}}\right)^2 e^{3\left(N_{\text {end }}-N_*\right)}.
    \label{eq:decoherencetotalefolds}
\end{equation}
Therefore, by considering $r\simeq 10^{-3}$,  we need around:
\begin{equation}
    (N_{\text {end }}-N_{*})\simeq 17 \,\,{ \rm e-folds},
\end{equation}
in order for the system to undergo decoherence.
So, the effect is suppressed, but in the end the number of e-folds required is similar to the one predicted before. This is due to the fact that most of the suppression of the decoherence rate can be attributed to the coupling $\varepsilon H^2/M_{\rm Pl} ^2 \lesssim 10^{-13}$, and only marginally to the numerical value that we derive by our calculations.

\subsection{Corrections to the power-spectrum}
\label{subsec:lambshiftqme}
In this section we collect and sum the contributions from the Lamb Shift corrections to the power-spectrum. 

In order to obtain a positive $D_{11}$ coefficient, we had to make use of the ``Strong Markovian approximation'' in the previous section.This seems to suggest that the ``Strong Markovian approximation'' could be a prescription that works well in order to extract a proper Markovian limit in the quantum master equation. 

When trying to extract the Lamb Shift Hamiltonian coefficients $\Delta_{11}$, we have found a divergent result 
in the case of the couple of derivativeless interaction and the mixed couple where the derivative interaction were integrated. This led us to employ using the Strong Markovian approximation, for a different reason. As explicitly proved before, this automatically gets rid of divergences and makes us extract a reasonable, finite result.

Indeed, the fact that we need to apply the ``Strong Markovian approximation'' for the real part of the coefficient brings us, for coherence, to apply it also to the Lamb Shift Hamiltonian. 

The two not null result for the Lamb Shift are obtained for
the couple $H_{int1}$-$H_{int1}$, with both derivativeless interactions, and the one with $H_{int1}$-$H_{int23}$ mixed derivative/derivativeless interactions, where the derivativeless interactions are integrated. 
The final result is obtained by summing eq. (\ref{eq:derivativelessspectralindexcorr}) and eq. (\ref{eq:derivativelessderivativeinteractionspectralindex}), thus obtaining:
\begin{equation}
 \delta n_{S}\simeq  1.6\frac{H^2\varepsilon}{4\pi^2M_{\rm Pl} ^2}.
 \label{eq:smartotlambshift}
\end{equation}
So, overall, we have a blue, scale invariant correction to the spectral index of the scalar power-spectrum, without any running in the spectral index proportional to $\ln k$, as found e.g. from~\cite{boyanovsky_effective_2015}. The presence of another factor $\ln^2 (-k\eta)$  in the power- spectrum corrections in ~\cite{boyanovsky_effective_2015} is due to the environment which is a conformally coupled field, i.e. not IR limited by the horizon as for an environment composed only of subhorizon modes. In this respect, a conformally coupled field is not a good proxy for a subhorizon environment. 

As stressed many times, the corrections to the power-spectrum would be too small to be accessible to present observations and thus, unless considering some models producing an higher level of non-Gaussianity, it would be impossible to observe them.

We did not have any sizeable correction from the derivative interactions, or, in general, when the interaction in the integration was of the derivative kind. This happens counterintuitively, since one would naively guess that the dominant interactions could be e.g. $H_{int2}$, being proportional to the environmental momenta, which are really taken to be high in a deep subhorizon environment (i.e. high k for the environment).  Actually, our work underlines that the criteria to consider dominant interactions in a setting of quantum master equations should take into account the relevance of the operator in a time expansion. In particular, the derivativeless interactions had a $1/\eta$ coefficient in the coupling which makes them a relevant operator when $\eta \to 0$. Instead, the derivative interactions are irrelevant as $\eta \to 0$ and thus they seem rather to create memory effects. On the other hand, these memory effects are the most difficult to interpret, and a definitive solution in the cosmological literature has not been given yet, so there is still some road open to properly extract a correct value for them.

\section{Comparison to the Boyanovsky method}
\label{sec:boy}
\subsection{Lamb Shift for power-spectrum}
\label{sec:boyps}
We have seen the effect of an environment of deep subhorizon modes on the power-spectrum of a system of superhorizon curvature perturbations. In particular, we have two possible sources of modification of the value of cosmological correlators: the Lamb Shift corrections and the non-unitary contribution (see eq.(\ref{eq:powerspectrumcorr})). We realized that these corrections give different effects. The Lamb Shift corrections change just the spectral index and the non-unitary contributions insert an entirely new term to the power-spectrum calculation. 

However, in the past, Boyanovsky in~\cite{boyanovsky_effective_2015} was able to extract corrections analogous to the Lamb Shift by a procedure implemented through different approximations, i.e. neglecting, in specific points of the calculations, the decaying mode of the field.  The authors of refs.~\cite{colas_benchmarking_2022,Brahma:2024yor} considered the comparison of the result from the TCL2 equation, which is in a form akin to the Redfield equation, to the Boyanovsky approximation (named the IR method in \cite{colas_benchmarking_2022}) for the effect of a spectator field environment interacting through derivativeless interactions with a system of scalar field perturbations. In this section, by applying the Boyanovsky method to our calculations, we extend this comparison to a single field inflation framework, with a subhorizon time dependent environment, both for a derivativeless trilinear interaction and for two more derivative interactions, the ones indicated in eq.(\ref{eq:interactions}). We also compare the results we obtain to the ones derived before (see section \ref{subsec:lambshiftqme}) by using the Strong Markovian approximation. 

We first consider the derivativeless interaction, namely $H_{int1}$ in eq.(\ref{eq:interactions}). We thus consider eq.(\ref{eq:QMEmomentum}):
\begin{equation}
\begin{split}
\label{eq:boyanovskyqme}
\rho_r^{\prime}(\eta)= & \frac{i \varepsilon H^2}{8 \pi^2  M_{pl}^2 \eta} \int_{\eta_0}^\eta \frac{\mathrm{d} \eta^{\prime}}{\eta^{\prime}} \sum_{\boldsymbol{p}}\left[v_{\boldsymbol{p}}(\eta) v_{-\boldsymbol{p}}\left(\eta^{\prime}\right) \rho_r(\eta) \frac{e^{20 i\left(1-\frac{\eta^{\prime}}{\eta}\right)}}{\eta-\eta^{\prime}-i \epsilon}-\rho_r(\eta) v_{-\boldsymbol{p}}\left(\eta^{\prime}\right) v_{\boldsymbol{p}}(\eta) \frac{e^{-20 i\left(1-\frac{\eta^{\prime}}{\eta}\right)}}{\eta-\eta^{\prime}+i \epsilon}\right. \\
& \left.+v_{\boldsymbol{p}}(\eta) \rho_r(\eta) v_{-\boldsymbol{p}}\left(\eta^{\prime}\right) \frac{e^{-20 i\left(1-\frac{\eta^{\prime}}{\eta}\right)}}{\eta-\eta^{\prime}+i \epsilon}-v_{-\boldsymbol{p}}\left(\eta^{\prime}\right) \rho_r(\eta) v_{\boldsymbol{p}}(\eta) \frac{e^{20 i\left(1-\frac{\eta^{\prime}}{\eta}\right)}}{\eta-\eta^{\prime}-i \epsilon}\right].\\
\end{split}
\end{equation}
Instead of considering the canonical form as developed in  \ref{sec:canonicalform} and the sections that follow, we take another route and consider an adaptation of Boyanovsky method to our case. In order to do that we introduce
\begin{equation}
    X_{-\boldsymbol{p}}(\eta) \equiv \int_{\eta_0}^\eta \frac{d \eta^{\prime}}{\eta^{\prime}} v_{-\boldsymbol{p}}\left(\eta^{\prime}\right) \frac{e^{20 i\left(1-\frac{\eta^{\prime}}{\eta}\right)}}{\eta-\eta^{\prime}-i \varepsilon} \quad \bar{X}_{-\boldsymbol{p}}(\eta) \equiv \int_{\eta_0}^\eta \frac{d \eta^{\prime}}{\eta^{\prime}} v_{-\boldsymbol{p}}\left(\eta^{\prime}\right) \frac{e^{-20 i\left(1-\frac{\eta^{\prime}}{\eta}\right)}}{\eta-\eta^{\prime}+i \varepsilon}.
    \label{eq:XXbar}
\end{equation}
Notice that at this point we slightly deviate from the  Boyanovsky method in \cite{boyanovsky_effective_2015}. We introduce two variables $X$ and $\bar{X}$, associated respectively with the full correlation function $K(\eta,\eta')$ as defined in eq.(\ref{1Kdef}) and its canonical conjugate. Instead, when defining the $X$ and $\bar{X}$ variables, Boyanovsky included in the expression only the non-local contribution of the environmental correlation functions. Also, in his case, when the system mode is deeply superhorizon $X \simeq \bar{X}$. This does not happen in our case, because we only consider deep subhorizon modes in the environment, and the exponentials in $X, \bar{X}$ have a strongly oscillating character also when the system mode is superhorizon. We can now decompose the field $v$ as in eq.(\ref{eq:growdecfieldmom}) and subtitute it in eq.(\ref{eq:XXbar}):
\begin{equation}
X_{-\boldsymbol{p}}(\eta)=Q_{-\boldsymbol{p}} \bar{g}_{+}(p ; \eta)+P_{-\boldsymbol{p}} \bar{g}_{-}(p ; \eta) \quad \bar{X}_{-\boldsymbol{p}}(\eta)=Q_{-\boldsymbol{p}} \bar{g}_{+}^*(p ; \eta)+P_{-\boldsymbol{p}} \bar{g}_{-}^*(p ; \eta),
\end{equation}
where we have introduced
\begin{equation}
\begin{split}
& \bar{g}_{+}(p ; \eta)=\int_{-\frac{1}{p}}^\eta \frac{d \eta^{\prime}}{\eta^{\prime}} \frac{1}{p^{\frac{3}{2}} \eta^{\prime}} \cdot \frac{e^{20 i\left(1-\frac{\eta^{\prime}}{\eta}\right)}}{\eta-\eta^{\prime}-i \varepsilon} \\
& \bar{g}_{+}^*(p ; \eta)=\int_{-\frac{1}{p}}^\eta \frac{d \eta^{\prime}}{\eta^{\prime}} \frac{1}{p^{\frac{3}{2}} \eta^{\prime}} \cdot \frac{e^{-20 i\left(1-\frac{\eta^{\prime}}{\eta}\right)}}{\eta-\eta^{\prime}+i \varepsilon},\\
\end{split}
\end{equation}
where we have already substituted the free form for the growing mode in eq.(\ref{eq:growingdecayingexpansion}), and analogous expressions for the decaying mode $g_{-}$. 
By substituting these expressions into the quantum master equation (\ref{eq:boyanovskyqme}) we obtain:
\begin{equation}
\begin{split}
& \rho_r^{\prime}(\eta)=\frac{i \varepsilon H^2}{8 \pi^2  M_{pl}^2 \eta} \sum_{\boldsymbol{p}}\left\{g_{+}(p , \eta) \bar{g}_{+}(p , \eta)\left[Q_{\boldsymbol{p}} Q_{-\boldsymbol{p}} \rho_r(\eta)-Q_{-\boldsymbol{p}} \rho_r(\eta) Q_{\boldsymbol{p}}\right]\right. \\
&+g_{+}(p , \eta) \bar{g}_{+}^*(p , \eta)\left[Q_{\boldsymbol{p}} \rho_r(\eta) Q_{-\boldsymbol{p}}-\rho_r(\eta) Q_{-\boldsymbol{p}} Q_{\boldsymbol{p}}\right]+ \\
& +g_{+}(p , \eta) \bar{g}_{-}(p , \eta)\left[Q_{\boldsymbol{p}} P_{-\boldsymbol{p}} \rho_r(\eta)-P_{-\boldsymbol{p}} \rho_r(\eta) Q_{\boldsymbol{p}}\right]+ \\
& +g_{+}(p , \eta) \bar{g}_{-}^*(p , \eta)\left[Q_{\boldsymbol{p}} \rho_r(\eta) P_{-\boldsymbol{p}}-\rho_r(\eta) P_{-\boldsymbol{p}} Q_{\boldsymbol{p}}\right]+ \\
& +\bar{g}_{+}(p , \eta) g_{-}(p , \eta)\left[P_{\boldsymbol{p}} Q_{-\boldsymbol{p}} \rho_r(\eta)-Q_{-\boldsymbol{p}} \rho_r(\eta) P_{\boldsymbol{p}}\right]+ \\
& \left.+\bar{g}_{+}^*(p , \eta) g_{-}(p , \eta)\left[P_{\boldsymbol{p}} \rho_r(\eta) Q_{-\boldsymbol{p}}-\rho_r(\eta) Q_{-\boldsymbol{p}} P_{\boldsymbol{p}}\right]\right\},
\label{eq:qmePQboyanovsky}
\end{split}
\end{equation}
where, following Boyanovsky, we have omitted the terms presenting two decaying modes, since as $\eta \to 0$ they are negligible. Then consider the adimensional power-spectrum for the variable $\zeta$
\begin{equation}
\begin{split}
    P_{\zeta}(p,\eta)=& \frac{H^2}{4\pi^2\varepsilon M_{pl}^2} 
(-p \eta)^{\frac{2 M_R^2}{3 H^2}}\left[\left\langle Q_{\boldsymbol{p}}^{\dagger} Q_{\boldsymbol{p}}\right\rangle-\frac{1}{3}\left\langle\left(Q_{\boldsymbol{p}} P_{\boldsymbol{p}}^{\dagger}+P_{\boldsymbol{p}} Q_{\boldsymbol{p}}^{\dagger}\right)\right\rangle(-p\eta)^3(-p \eta)^{-\frac{2 M_R^2}{3 H^2}}+\right.\\
&\left.+\frac{1}{9}\langle P_{\boldsymbol{p}}^{\dagger} P_{\boldsymbol{p}}\rangle(-p \eta)^{6}(-p \eta)^{-\frac{4 M_R^2}{3 H^2}}\right]\\,
\label{eq:powerspectrumboy}
\end{split}
\end{equation}
where we have also included a possible mass $M_{R}$ to reabsorb UV divergences and decomposed the field $\zeta$ in the growing/decaying modes basis, using eq.(\ref{eq:growingdecayingexpansion}). The further approximation that Boyanovsky makes is considering negligible in eq.(\ref{eq:powerspectrumboy}) every term with a positive power of $-p\eta$, i.e. every term containing a decaying mode, which leads to
\begin{equation}
P_{\zeta}(p,\eta)= \frac{H^2}{4\pi^2\varepsilon M_{pl}^2} 
(-p \eta)^{\frac{2 M_2^2}{3 H^2}} Y(p,\eta)\quad \rm{where} \quad Y(p,\eta)=\operatorname{Tr}\left(Q_{\boldsymbol{p}}^{\dagger} Q_{\boldsymbol{p}} \rho_r(\eta)\right),
\label{eq:powerspectrumboyapprox}
\end{equation}
where we have also introduced the auxiliary variable $Y(p,\eta)$.
The idea is that since the mode is analyzed when it is deeply superhorizon, $-p\eta\ll1$, then the other terms in eq.(\ref{eq:powerspectrumboy}) should be negligible.
However, without considering the time evolution of the expectation value of $QP$ in eq.(\ref{eq:boyanovskyqme}) it is impossible to establish the time dependence of the terms dubbed subleading, so the approximation may hide some relevant terms. A more precise investigation is needed. 

In order to find the Lamb Shift power-spectrum correction, we consider the auxiliary variable $Y(p,\eta)$ and derive it with respect to time. Since both $Q$ and $P$ are time independent for construction, in the interaction picture, the derivative applied to $Y$ acts only on the density matrix $\rho$. We know how to express the time derivative of $\rho$ with the quantum master equation (\ref{eq:qmePQboyanovsky}). We can thus multiply both sides of eq.(\ref{eq:qmePQboyanovsky}) with $Q_{\boldsymbol{k}}Q_{-\boldsymbol{k}}$ and then trace over the system degrees of freedom. On the RHS, by applying the commutation relations, many cancellation arise between the terms in each line of eq.(\ref{eq:qmePQboyanovsky}), as shown in~\cite{boyanovsky_effective_2015}. It is possible to verify that the only terms which survive are from the last two lines of  eq.(\ref{eq:qmePQboyanovsky}) giving
    \begin{equation}
Y_{\boldsymbol{p}}^{\prime}(\eta)=\frac{i \varepsilon H^2}{8 \pi^2  M_{pl}^2 \eta} \left(2 i Y_{\boldsymbol{p}}(\eta) \bar{g}_{+}(p, \eta) g_{-}(p, \eta)+2 i Y_{\boldsymbol{p}}(\eta) \bar{g}_{+}^*(p, \eta) g_{-}(p, \eta)\right).
\label{eq:tointegrateboy}
\end{equation}
Notice how the coefficient in equation (\ref{eq:tointegrateboy}) depends on the real part of $\bar{g}_{+} (k,\eta)$, which, for how they are defined \footnote{Notice that there is no prefactor $1/(4 \pi^2 i)$ in the definition of $\bar{g}_{+}$ and $\bar{g}^{*}_{+}$, so that actually  $\bar{g}_{+} \propto K(\eta,\eta') $ and $\bar{g}^*_{+} \propto -K^{*}(\eta,\eta') $. }, corresponds to the imaginary part of the correlation function $K(\eta,\eta')$. Now it is clear that the only contribution we can obtain is the Lamb Shift, since, as we saw before in eq.(\ref{eq:deltamatrix}
), this is the only contribution proportional to the imaginary part of the correlation function $K(\eta,\eta')$. The RHS of eq.(\ref{eq:tointegrateboy}) can be developed as
\begin{equation}
    Y_{\boldsymbol{p}}^{\prime}(\eta)=\frac{\varepsilon H^2}{2 \pi^2 M_{pl}^2 \eta^2 } \frac{\eta^2}{3} \frac{1}{\eta} \int_{1}^{-1/p\eta} \frac{dx}{x} \frac{1}{x} \cos{20(1-x)} \operatorname{P.V.}\frac{1}{1-x}   Y_{\boldsymbol{p}}(\eta).
    \label{eq:boyintegratinggrow}
\end{equation}
This expression should be compared to the expression for $\Delta_{11}$, both without and with the Strong Markovian approximation, respectively in eq.(\ref{eq:delta11derivativeless}) and eq.(\ref{eq:delta11derivativelesssmar}). Looking at the $x$ functional dependence of the integrand in eq.(\ref{eq:boyintegratinggrow}), with respect to eq.(\ref{eq:delta11derivativeless}) it is clear that the dependence on the decaying mode is now lost. However, as we already commented after eq.(\ref{eq:delta11derivativeless}), the decaying mode is mostly responsible for the unphysical oscillations which should be removed because they are not consistent with our approximations. The result in eq.(\ref{eq:boyintegratinggrow}) is also very similar to the one obtained with the Strong Markovian approximation in eq.(\ref{eq:delta11derivativelesssmar}), exception made for the presence of the $1/x$ factor that we see in eq.(\ref{eq:boyintegratinggrow}) due to the growing mode. 
When solving the integration on the RHS of eq.(\ref{eq:boyintegratinggrow}), we treat the divergent integral in $x=1$ in the same way as it was done in eq.(\ref{eq:delta11derivativeless}), by means of the Solkhotsi Plemelji theorem.
Integrating eq.(\ref{eq:tointegrateboy}) in conformal time from horizon exit (at $\eta_{*}=-1/p$) to the considered time $\eta$ brings to
\begin{equation}
    Y(p,\eta)=Y(p,\eta_{*}) e^{\frac{3.6 \varepsilon H^2}{6 \pi^2 M_{pl}^2} \log({-p\eta})},
    \label{eq:boydecayintime}
\end{equation}
where we have actually renormalized in an appropriate mass counterterm an UV divergence analogous to the one found in \ref{sec:derivativelesscanonical} and discarded boundary terms. We don't report all the calculations since they are straightforward, and don't add anything new to what we already did in the previous sections. Notice that as it is written in eq.(\ref{eq:boydecayintime}) the result predicts a decay in time of the power-spectrum after horizon crossing. We actually do the hypothesis that the state of the system, in the interaction picture, is unentangled with the environment and corresponds to the Bunch-Davies vacuum until horizon crossing. This is akin to think that the interactions are actually switched on after the system mode has crossed the horizon. Of course, this is not what happens in reality, especially because we are considering gravity. However, we think that this is a good starting point, and we leave for further studies a more complete analysis. 

As already stressed before, eq.(\ref{eq:boydecayintime}) does not have any square logarithm term in the exponent, unlike the result in the original paper by Boyanovsky~\cite{boyanovsky_effective_2015}. As we commented on in section \ref{sec:derivativelesscanonical}, this is probably due to the different environment. In our case, the infrared limitation of the environment (i.e. $k>10 aH$) prevents secular logarithmic terms (as a square logarithm) from being present.

We can interpret eq.(\ref{eq:boydecayintime}) also in another way. By actually resumming the logarithm in the exponent, and considering the power-spectrum at the end of inflation, we can write:
\begin{equation}
\mathcal{P}(p, \eta_{f})=\mathcal{P}(p,\eta_{*}) (-p \eta_{f})^{\frac{2.4 \varepsilon H^2}{ 4 \pi^2 M_{pl}^2}}=\frac{H_{*}^2}{8 \pi^2 \varepsilon_{*} M_{pl}^2} (-p \eta_{f})^{\frac{2.4 \varepsilon H^2}{ 4 \pi^2 M_{pl}^2}},
\label{eq:boyanovskyderivativelessresult}
\end{equation}
where $H_{*}$ and $\varepsilon_{*}$ are evaluated at horizon crossing, while in the exponent we did not consider any variation of $H$ since it would be higher order in the slow roll parameters. Eq.(\ref{eq:boyanovskyderivativelessresult}) shows that the decay in time of the power-spectrum in eq.(\ref{eq:boydecayintime}) is actually equivalent to a blue modification to the spectral index of the power-spectrum, actually equal to the one we already obtained in eq.(\ref{eq:derivativelessspectralindexcorr}). So, in this case, we can say that Boyanovsky method worked as a very good approximation. This is not in contradiction with the results found in~\cite{colas_benchmarking_2022,Brahma:2024yor}, as the differences found with the result of the $TCL_2$ equation in those cases are also really small. 

Let us now consider the case of derivative interactions. In the case where we have a couple of derivative interactions in the RHS of the quantum master equation (\ref{eq:quantummastereq}), as analyzed in the previous sections, it is possible to verify that there is no finite correction to the power-spectrum due to those interactions also with Boyanovsky method, but only UV divergences which can be renormalized. In general, this is true in any case in which the derivative interaction is integrated in time. We do not go through the details in this case, but we just sketch the calculations. It is sufficient to remember that in this case we have $H_{int2,3} \propto \eta$, so in the modified mode growing mode function $\bar{g}(q,\eta)$ we have:
\begin{equation}
\bar{g}_{+} \alpha \int d \eta^{\prime} \frac{H_{\mathrm{INT}}\left(\eta^{\prime}\right)}{\eta^{\prime}} \frac{d^{(k)}}{d \eta^{\prime(k)}} K\left(\eta, \eta^{\prime}\right),
\end{equation}
where k is the total number of derivatives inside the vertices and $H_{int2,3}(\eta')/\eta' \propto {\rm constant} $. Therefore all we obtain are just  boundary terms. 

Instead, we still have corrections for mixed terms in which the derivativeless interaction is integrated (i.e. it is the memory interaction). In particular, we can start from eq. (\ref{eq:qmederivativeeq}) and decompose in growing and decaying mode the equation as in eq.(\ref{eq:qmePQboyanovsky}). The only non vanishing term, after multiplying for the operators $Q_{\boldsymbol{k}}Q_{-\boldsymbol{k}}$ and tracing, are the same as in eq.(\ref{eq:tointegrateboy}). The only thing which is different is the expression for $\bar{g}_{+}(p,\eta)$, which, as it is straightforward to verify, is given by 
\begin{equation}
    \bar{g}_{+}(p , \eta)+ \bar{g}_{+}^{*}(p , \eta)=2 \int_{-\frac{1}{p}}^{\eta} \frac{d \eta^{\prime}}{\eta^{\prime}} \frac{1}{\eta'} \frac{1}{p^{\frac{3}{2}}} \frac{d^2}{d\eta'^2} \frac{\cos{20(1-\frac{\eta'}{\eta})}}{\eta-\eta^{\prime}}.
\end{equation}
The RHS of the analogous of eq.(\ref{eq:tointegrateboy}) is thus:
\begin{equation}
    -\frac{\varepsilon H^2}{36 \pi^2 M_{pl}^2 \eta} \int_{1}^{a} dx \frac{d^2}{dx^2} \frac{1}{x^2}\frac{\cos{20(1-x)}}{1-x}=-\frac{\varepsilon H^2 3.6 }{6 \pi^2 M_{pl}^2 \eta} + {\rm renormalization\,\,terms}.
    \label{eq:derivboy}
\end{equation}
We can compare this expression with the one obtained by eq. (\ref{eq:Delta123}), without any approximation: we notice a difference in the result because of the presence of the decaying mode. Instead, by comparing the integrand in eq.(\ref{eq:derivboy}) and the one by substituting eq.(\ref{eq:smarderivderivelss}) inside eq.(\ref{eq:Delta123}), we can see that we have a $1/x$ more due to the growing mode as before. This brings, by performing the derivatives, to obtain a factor 6 instead of a factor 2 into the integral, and modifies the result. Thus, the final result for mixed interactions, obtained frome eq.(\ref{eq:derivboy}) for the power-spectrum is:
\begin{equation}
    \mathcal{P}(p, \eta_{f})=\frac{H_{*}^2}{8 \pi^2 \varepsilon_{*} M_{pl}^2} (-p \eta_{f})^{-\frac{2.4 \varepsilon H^2}{ 4 \pi^2 M_{pl}^2}}.
    \label{eq:boyanovskyderivativeresult}
\end{equation}
Notice that the result is similar to eq.(\ref{eq:boyanovskyderivativelessresult}), but opposite in the sign of the exponent.
Now consider the net effect of the contemporary action of the mixed interaction and the derivativeless interaction on the power-spectrum, whose results are reported in eq.(\ref{eq:boyanovskyderivativelessresult}) and eq.(\ref{eq:boyanovskyderivativeresult}). By writing a single quantum master equation for all the interactions, the contributions on the RHS of eq.(\ref{eq:tointegrateboy}) from derivativeless and derivative interactions would cancel. 
This goes in the opposite direction to the net blue correction that we have found from our previous calculations using the Redfield equation and the Strong Markovian approximation (see eq.(\ref{eq:smartotlambshift}). So, the approximation of neglecting the decaying mode, on which the Boyanovsky method is based, does not work equally well in the case of derivative interactions.

In general, derivative interactions give different results according to the approximation we used, unlike the derivativeless one. It is sufficient to compare eq. (\ref{eq:Delta123}) and  eq.(\ref{eq:derivativelessderivativeinteractionspectralindex})to realize that also the Strong Markovian approximation was quite different from the Redfield equation result. So, modeling the correct Markovian approximation is fundamental for derivative interactions in particular. Arguably a definitive answer still lacks in the literature and we expect to see in the few next years a development in these methods. Nevertheless, we think that the fact that, after summing all the interactions, the Strong Markovian approximation results in a positive real part of the quantum master equations coefficient $D_ {11}$, differently from the Redfield equation, makes also the result on the Lamb Shift corrections after applying this prescription more reliable.

\subsection{Lamb Shift for bispectrum}
\label{sec:boybisp}
We have seen in the previous section that Boyanovsky method is reliable in the case of a derivativeless interaction. Thus, here we try to extend the calculation to higher-order correlation functions. The three-point function, and its Fourier transform, namely the bispectrum, represents a fundamental instrument of investigation of early universe physics \cite{akrami_planck_2020}. Some works have already investigated the bispectrum from a perspective of  open quantum system applications to cosmology~\cite{daddi_hammou_cosmic_2023,Salcedo:2024smn,martin_non_2018}. However, to our knowledge, no paper considered also the Lamb Shift correction  to the bispectrum by solving the quantum master equation. Exploiting the reasoning by Boyanovsky, we want to show that there are some corrections to the bispectrum, due to the Lamb Shift Hamiltonian, that will deserve future investigation. 

We consider three modes $\zeta(\boldsymbol{k_{1}})$,$\zeta(\boldsymbol{k_{2}})$,$\zeta(\boldsymbol{k_{3}})$ for the curvature perturbations.
The $\zeta$ bispectrum can also be decomposed in a growing/decaying mode basis:
\begin{equation}
\begin{split}
 &\langle \zeta(\boldsymbol{k_{1}},\eta)\zeta(\boldsymbol{k_{2}},\eta)\zeta(\boldsymbol{k_{3}},\eta) \rangle
=- \frac{H^3 \eta^3} {(\sqrt{2} \varepsilon M_{pl})^3} \left(\langle Q(\boldsymbol{k_{1}})Q(\boldsymbol{k_{2}})Q(\boldsymbol{k_{3}}) \rangle g_{+}(k_{1},\eta) g_{+}(k_{2},\eta) g_{+}(k_{3},\eta)+\right.\\
&+\left. \langle P(\boldsymbol{k_{1}})Q(\boldsymbol{k_{2}})Q(\boldsymbol{k_{3}}) \rangle g_{-}(k_{1},\eta) g_{+}(k_{2},\eta) g_{+}(k_{3},\eta)+(\Leftrightarrow {\rm permutations})+ \right.\\
&\left.+\langle P(\boldsymbol{k_{1}})P(\boldsymbol{k_{2}})Q(\boldsymbol{k_{3}}) \rangle g_{-}(k_{1},\eta) g_{-}(k_{2},\eta) g_{+}(k_{3},\eta)+(\Leftrightarrow {\rm permutations})+\right.\\
&+\left.\langle P(\boldsymbol{k_{1}})P(\boldsymbol{k_{2}})P(\boldsymbol{k_{3}}) \rangle g_{-}(k_{1},\eta) g_{-}(k_{2},\eta) g_{-}(k_3,\eta)\right).\\
\label{eq:bispectrumboygrowdec}
\end{split}
\end{equation}
We would like to consider the evolution of the bispectrum after all the three modes have crossed the horizon. According to the approach of~\cite{boyanovsky_effective_2015} we neglect all the terms with a decaying mode in eq.(\ref{eq:bispectrumboygrowdec}) as they become negligible after all the three modes $k_{1},k_{2},k_{3}$ have crossed the horizon. We also introduce an auxiliary variable $Y_3(\{\boldsymbol{k_i}\},\eta)=\operatorname{Tr}(Q(\boldsymbol{k_{1}})Q(\boldsymbol{k_{2}})Q(\boldsymbol{k_{3}})\rho_{r}(\eta))$.
 By following these leads we can write:
\begin{equation}
 (k_{1}k_{2}k_{3})^{3/2}B_{\zeta}(\boldsymbol{k_{1}},\boldsymbol{k_{2}},\boldsymbol{k_{3}},\eta)\simeq - \frac{H^3} {(\sqrt{2 \varepsilon} M_{pl})^3} Y_{3}(\eta,\boldsymbol{k_{1}},\boldsymbol{k_{2}},\boldsymbol{k_{3}}) (-k_{1} \eta)^{3/2-\nu}(-k_{2} \eta)^{3/2-\nu}(-k_{3} \eta)^{3/2-\nu} ,
  \label{eq:bisprecturmandY}
\end{equation}
where we have expanded the growing mode according to eq.(\ref{eq:growingdecayingexpansion}). Our goal now, thus, is to find the evolution of $Y_{3}(\{\boldsymbol{k}_i\},\eta)$ and then infer the evolution of the bispectrum from it. In order to do that, as before, we can derive $Y_{3}(\{k_i\},\eta)$ with respect to conformal time. The only time dependent quantity, in $Y_{3}(\{k_i\},\eta)$ in the interaction picture, is $\rho_{r}$. By deriving $Y_{3}(\{k_i\},\eta)$ with respect to time we can then substitute the quantum master equation in the form of eq.(\ref{eq:qmePQboyanovsky}). The calculation is analogous to the one for the power-spectrum, just inserting three $Q_{\boldsymbol{k}}$ before tracing. It is easy to see that, as before, the first two lines of the RHS eq.(\ref{eq:qmePQboyanovsky}) vanish, by simply commuting the $Q$ between themselves. When considering the next two lines in eq.(\ref{eq:qmePQboyanovsky}), we obtain zero as well:
\begin{equation}
\begin{split}
   & \sum_{\boldsymbol{p}} (g_{+}(p,\eta) \bar{g}_{-}(p,\eta) \operatorname{Tr}(Q_{\boldsymbol{k_1}}Q_{\boldsymbol{k_2}}Q_{\boldsymbol{k_3}}Q_{\boldsymbol{p}}P_{-\boldsymbol{p}}\rho_{r})-(Q_{\boldsymbol{p}}Q_{\boldsymbol{k_1}}Q_{\boldsymbol{k_2}}Q_{\boldsymbol{k_3}}P_{-\boldsymbol{p}}\rho_{r})+\\
   &+g_{+}(p,\eta) \bar{g}^{*}_{-}(p,\eta) \operatorname{Tr}(P_{-\boldsymbol{p}}Q_{\boldsymbol{k_1}}Q_{\boldsymbol{k_2}}Q_{\boldsymbol{k_3}}Q_{\boldsymbol{p}}\rho_{r})-(P_{-\boldsymbol{p}} Q_{\boldsymbol{p}}Q_{\boldsymbol{k_1}}Q_{\boldsymbol{k_2}}Q_{\boldsymbol{k_3}}\rho_{r}))=0.\\
\end{split}
\end{equation}
Instead, we can consider the last two lines in  eq.(\ref{eq:qmePQboyanovsky}), which give
\begin{equation}
    \begin{split}
   & \sum_{\boldsymbol{p}} (g_{-}(p,\eta) \bar{g}_{+}(p,\eta) \operatorname{Tr}(Q_{\boldsymbol{k_1}}Q_{\boldsymbol{k_2}}Q_{\boldsymbol{k_3}}P_{\boldsymbol{p}}Q_{-\boldsymbol{p}}\rho_{r})-(P_{\boldsymbol{p}}Q_{\boldsymbol{k_1}}Q_{\boldsymbol{k_2}}Q_{\boldsymbol{k_3}}Q_{-\boldsymbol{p}}\rho_{r})+\\
   &+g_{-}(p,\eta) \bar{g}^{*}_{+}(p,\eta) \operatorname{Tr}(Q_{-\boldsymbol{p}}Q_{\boldsymbol{k_1}}Q_{\boldsymbol{k_2}}Q_{\boldsymbol{k_3}}P_{\boldsymbol{p}}\rho_{r})-( Q_{-\boldsymbol{p}}P_{\boldsymbol{p}}Q_{\boldsymbol{k_1}}Q_{\boldsymbol{k_2}}Q_{\boldsymbol{k_3}}\rho_{r})).\\
\end{split}
\end{equation}

In this case, there is no direct cancellation between corresponding terms, and instead we have to repeatedly apply the commutation relations eq.(\ref{eq:commrel}), bringing in the end
\begin{equation}
\begin{split}
    Y_{3}'(\{k_i\},\eta)=&- \frac{\varepsilon H^2}{8 \pi^2 M_{pl}^2 \eta} \left[g_{-}(k_{1},\eta) (\bar{g}_{+}(k_{1},\eta)+\bar{g}^{*}_{+}(k_{1},\eta))+\right.\\
    &\left.+g_{-}(k_{2},\eta) (\bar{g}_{+}(k_{2},\eta)+\bar{g}_{+}^{*}(k_{2},\eta))+\right.\\
    &\left.+g_{-}(k_{3},\eta) (\bar{g}_{+}(k_{3},\eta)+\bar{g}^{*}_{+}(k_{3},\eta))\right] Y_{3}(\{k_i\},\eta).\\
    \label{eq:bispectrumboyqme}
    \end{split}
\end{equation}
In general we can say that
\begin{equation}
    g_{-}(k_{1},\eta) (\bar{g}_{+}(k_{1},\eta)+\bar{g}^{*}_{+}(k_{1},\eta))= 2\frac{\eta^2}{3} \int_{-1/k_{1}}^{\eta} \frac{d\eta'}{\eta'} \frac{1}{\eta'} \frac{\cos{20(1-x)}}{\eta-\eta'}.
    \label{eq:RHSbispectrumboyqme}
\end{equation}
This is analogous to what we had already found in the previous calculation for the power-spectrum, so we can just report the result
\begin{equation}
   g_{-}(k_{1},\eta) (\bar{g}_{+}(k_{1},\eta)+\bar{g}^{*}_{+}(k_{1},\eta))=- \frac{7.16 }{3 } + {\rm renormalization\,\, terms},
\end{equation}
where the renormalization terms can be reabsorbed in the $\nu$ factors in eq.(\ref{eq:bisprecturmandY}).
We integrate eq.(\ref{eq:bispectrumboyqme}) since the time $\eta_*$ when the smallest wavelength mode (assume it is $k_{1}$) leaves the horizon. So, when we integrate the RHS of eq.(\ref{eq:bispectrumboyqme}) from the epoch when the smallest wavelength mode $k_1$ crosses the horizon:
\begin{equation}
\begin{split}
   & \int_{-1/k_{1}}^{\eta} \frac{d Y_{3}}{Y_{3}}=-\frac{\varepsilon H^2}{8 \pi^2 M_{pl}^2 }\left( \int_{-\frac{1}{k_{1}}}^{\eta} \frac{d\eta}{\eta} g_{-}(k_{1},\eta) (\bar{g}_{+}(k_{1},\eta)+\bar{g}^{*}_{+}(k_{1},\eta))+\right.\\
   &\left.\int_{-\frac{1}{k_{1}}}^{\eta} \frac{d\eta}{\eta} g_{-}(k_{2},\eta) (\bar{g}_{+}(k_{2},\eta)+\bar{g}^{*}_{+}(k_{2},\eta))+\int_{-\frac{1}{k_{3}}}^{\eta} \frac{d\eta}{\eta} g_{-}(k_{3},\eta) (\bar{g}_{+}(k_{3},\eta)+\bar{g}^{*}_{+}(k_{3},\eta))\right)\simeq\\
   &\simeq \frac{\varepsilon H^2}{8 \pi^2 M_{pl}^2 } \int_{-\frac{1}{k_{1}}}^{\eta} d\eta' \frac{2.4}{\eta'} \left(3 \theta(\eta'-1/k_1) \right ), 
    \end{split}
    \label{eq:bispectrumqme}
\end{equation}
where after the second equality we have evaluated numerically the integrations of $g_{+}(k)$ terms in the RHS of eq.(\ref{eq:bispectrumqme}). The numerical value  of the $\bar{g}_{+}(k)$ integrations is independent of the momentum the mode $k$ (if we assume, as we do, that the mode has already crossed the horizon before). Evaluating the integral far after each mode has crossed the horizon we obtain
\begin{equation}
    \log{\frac{Y_3(\{k_i\},\eta)}{Y_3(\{k_i\},\eta_{*})}}=2.4 \frac{\varepsilon H^2}{8 \pi^2 M_{pl}^2} \left(3 \log(-k_1 \eta)\right),
\end{equation}
and exponentiating each side we find
\begin{equation}
    Y_3(\{k_i\},\eta)=Y_3(\{k_i\},\eta_{*}) e^{2.4 \frac{\varepsilon H^2}{8 \pi^2 M_{pl}^2} \left(3\log(-k_1 \eta)\right)},
\end{equation}
which can be interpreted as a decay of the bispectrum of the curvature perturbations. Notice that $Y_3(k,\eta_{*})$ is the non vanishing bispectrum generated after the smallest mode crosses the horizon, and thus corresponds to the bispectrum we would observe at the end of inflation neglecting this effect of decay.  This calculation reveals that the effect of the entanglement and Lamb Shift on the power-spectrum of superhorizon modes is generic also to higher order correlation functions, and thus it must be explored more in detail in future works. The different effect on the bispectrum and on the power-spectrum of the entanglement with a subhorizon environment may also lead on possible modifications of the Maldacena Consistency relations in single field inflation. We plan to analyze this in a future work.

\section{Conclusions}
\label{Conclusions} 
In this paper, we have analyzed the problem of decoherence of large super-horizon inflationary curvature perturbations from an environment of deep sub-horizon tensor perturbations, e.g. from perturbation modes with $k > 10aH$. In this sense, we considered, as a work hypothesis, the environmental mode function as an oscillating function.  This is justified by the fact that the environmental modes are always deep inside the horizon. As interactions, we have taken the cubic interactions among curvature and tensor perturbation modes that are predicted by General Relativity in single-field models of inflation, see~\cite{maldacena_non-gaussian_2003}. By rewriting those interactions in terms of canonically normalized fields, and applying some integration by parts, we arrived at the form of eq.(\ref{eq:interactions}). We then recognized two different kinds of interactions, which we named ``derivative'' and ``derivativeless'', according to the presence or not of spatial or time derivatives. We first carefully rederived the quantum master equation, discussing in some detail the Markovian approximation. Since decoherence is a second order phenomenon in the coupling, we solve the quantum master equation (\ref{eq:blochredfield}) considering all the possible couples of interactions in eq.(\ref{eq:interactions}). In particular, we computed for each couple of interactions the real and imaginary parts of the coefficients of the quantum master equation, which are associated respectively to the non-unitary effects (decoherence and quantum contributions to correlators) and the unitary corrections (Lamb shift corrections to the Hamiltonian). Both effects are bigger for the case where we consider a couple of derivativeless interactions. Since this couple of interactions was also analyzed in some detail in some different models~\cite{boyanovsky_effective_2015,Hollowood:2017bil, Burgess:2024eng}, we decided to dedicate to it a thorough analysis in Section (\ref{sec:derivativelesscanonical}).

 A first important result is that a positive coefficient was obtained in the real part of the quantum master equation, thus underlining the Markovian character of this derivativeless interaction $H_{int1}$. This was not surprising, since the operator in $H_{int1}$ is relevant as $\eta \to 0$.  The decoherence associated to this interaction is 20 times suppressed with respect to the one obtained by~\cite{burgess_minimal_2023} 
~(eq.~\ref{eq:decoherencederivativelessefolds}), thus seemingly confirming, in this case, the intuition that subhorizon modes contribution is subleading with respect to superhorizon modes in the environment. We also warn that a fully direct comparison of our results on decoherence with the ones by~\cite{burgess_minimal_2023} is not possible, first, because in~\cite{burgess_minimal_2023} only derivative interactions are considered, and then because the environment employed in that work presents a fixed cutoff, differently from ours. In eq. (\ref{eq:derivativelesseqhorcross}), we did a first rough calculation extending the environment, with the same oscillating mode functions, until the  horizon-crossing; the result for decoherence is of the same order as~\cite{burgess_minimal_2023}, giving a hint that subhorizon modes contribution should be investigated in more details; a full calculation is already in preparation.

We took advantage of this simple interaction $H_{int1}$ to explicitly evaluate the effect of a time dependent environment. We carefully modified the quantum master equation derivation in order to take into account the time dependent cutoff, and surprisingly found a small corrections, which we could neglect in a first calculation.  To our knowledge, this is the first time that this effect is explicitly taken into account, at least in a cosmological setting.

 Next, we moved to complicate the setting by evaluating the effects of the derivative interactions. We found that negligible results for both decoherence and quantum corrections to the (large-scale) curvature perturbation were obtained by considering couples of derivative interactions. This is probably due to the irrelevance of $H_{int23}$ as operators when $\eta \to 0$, which underlines the non-Markovian character of these interactions, while, in order to derive the quantum master equation, we have adopted a Born-Markov approximation eq.(\ref{eq:markovianapprox}). Instead, we realized that non negligible effects on decoherence and quantum corrections were obtained when considering mixed couple of interactions, i.e. a derivativeless and a derivative interaction. In particular, the contribution of this mixed term to the real part of the coefficient of the quantum master equation was negative, in such a way that the sum with the derivativeless couple result was negative too. This result underlines the importance of considering also the interplay between different interactions, even in the case when a couple of interaction, by means of an educated guess, is expected to have a dominant effect with respect to the others.  Also, the sign of the real part of the coefficient of the quantum master equation is known to be crucial for the application of a sufficient condition that ensures the dynamics to be physical (see discussion in section \ref{sec:canonicalform}).
 
 It is known in the literature that the quantum master equation~(\ref{eq:blochredfield}) does not guarantee that the evolution of the density matrix is always physical. This may happen, as explained in detail in section \ref{subsec:the Markovian approximation}, because the Born-Markov approximation eq.(\ref{eq:markovianapprox}), in some situations, does not consistently take into account the memory effects. These unphysical features seem to be exacerbated in a cosmological context, as other papers found as well a negative real part of the coefficient of the quantum master equation \cite{burgess_minimal_2023}. A procedure able to recognize unphysical effects, and properly deal with the memory of the system, has still not been established in cosmology. Thus, we adopted an agnostic approach. When it was possible to extract a physical result by only employing the Born-Markov approximation eq.(\ref{eq:markovianapprox}), as for computing decoherence of derivativeless interactions, we did not do any additional approximations. Instead, for derivative interactions, we resorted to the ``Strong Markovian approximation'', a prescription introduced in \cite{burgess_minimal_2023, Kaplanek:2022xrr}, in order to deal with the memory effects of the total system considered. This prescription seemed to work well also in our case, since by applying it we obtained a positive real part coefficient of the quantum master equation for the full calculations, even though the total rate of decoherence was suppressed. 
 
 We then obtained also the full Lamb Shift correction to the curvature perturbation power-spectrum, by showing that there was a blue correction to the scalar spectral index. Notice that, in order to be consistent with the way we obtained the results for decoherence, we stick to applying the Strong Markovian approximations also to obtain the Lamb Shift corrections to the power-spectrum. Differently from other cases in which a proxy to the subhorizon environment was applied, e.g. a conformal scalar field~\cite{Burgess:2024eng, boyanovsky_effective_2015}, we did not see any  running of the spectral index proportional to $\operatorname{ln} k$, but only a scale invariant correction to the spectral index. We connect this difference to the fact that our environment is infrared limited by the horizon, being composed by subhorizon modes, while an environment of a conformally coupled field features also modes with very large wavelength.

We then compared the Lamb Shift corrections to the power-spectrum obtained by solving the transport equations to the ones obtained, for the same interactions, by an approximated method introduced in \cite{boyanovsky_effective_2015}, in which the decaying mode of the cosmological perturbations was neglected consistently. The results showed how the corrections for derivativeless interaction were equal by employing either of the two methods. In contrast, the corrections computed for mixed derivative-derivativeless interactions presented important differences, thus underlining how neglecting the decaying mode worsens the results for interactions associated to non-Markovian effects.

Since the results for the derivativeless interactions with the method in \cite{boyanovsky_effective_2015} revealed to be solid, we applied this method to compute, for the first time, the Lamb Shift corrections to the bispectrum. The results showed a decay in time of the bispectrum, after the three modes have crossed the horizon, which is akin to the one of the power-spectrum. This opens up interesting scenarios for the analysis of non-Gaussianity in open quantum system applications to inflation.

We believe that the results we obtained in this paper represent another contribution to a variety of studies on the subject that have tried to address various aspects related to decoherence during inflation in an open-quantum system approach.  This work can help in shedding light on many issues, in particular on the importance of considering also a subhorizon mode environment. We also highlighted the importance of accounting for the interplay between different interactions when evaluating decoherence (necessary for a complete and fully consistent computation). We evaluated explicitly the effect of a time dependent environment,  proposing 
a method to fully take into account its effect.

Our calculations can be improved in many ways.
First, we should consider the full mode functions in the environmental modes, so to properly include the effects of the curvature of spacetime also in the environment. As shown in eq.(\ref{eq:derivativelesseqhorcross}), this is likely to increase both the amount of decoherence and the quantum corrections. 
A very important progress would be also to shed light on how to treat non-Markovian effects, which seems ubiquitous in open quantum system applications to inflation. A possible way could be to consider the Time-Convolutionless expansion of the quantum master equation, which is a consistent way to take into account the memory effects in a series expansion. The second order of this expansion, TCL2 has the very same form of the quantum master equation (\ref{eq:blochredfield}) employed in this work. It would be interesting to consider the next order (which would be the TCL4) to actually consistently take into account the non-Markovian features in the quantum master equations.
As a further development, we might consider some specific inflation models, well known for the higher level of primordial non-Gaussianity with respect to the minimal one due to gravity in single-field models of slow-roll inflation~\cite{Acquaviva:2002ud,maldacena_non-gaussian_2003}, which could lead to stronger decoherence, and see if we can obtain a bound on the interaction strength from the quantum corrections to the $n$-point correlation functions of inflationary perturbations (see, for a first attempt in this direction,~\cite{daddi_hammou_cosmic_2023,martin_non_2018}). A very ambitious, but remarkable development, would be to identify a possible distinguishable signature of the quantum-to-classical transition, which would indirectly prove the quantum nature of the primordial fluctuations. 

With the application of the open quantum system theory to cosmology, many recent investigations are trying to address a missing piece to the history of inflation. This opens a new perspective on the theory of inflation, which has always been recognized as a great theoretical laboratory of quantum field theory and gravity, but may also show its potential as a scenario where studying complex interplays between various, reciprocally entangled, quantum mechanical systems.

\appendix
\section{Appendix A }

\subsection{What is decoherence?}
In order to clarify better what decoherence is, let's consider a pedagogical introduction \footnote{See also the introduction of~\cite{nelson_quantum_2016} for a pedagogical presentation of decoherence.} to this concept. 
Let's assume that we have a set of system states, e.g. a basis  $|S_{i}(t_{0})\rangle$, to which we have full access. Instead, there are also some other degrees of freedom to which we can never access with measurements, but can affect the evolution of the system, e.g. by interactions or quantum correlations: we call an environment, initially in a state $|E(t_0)\rangle$. 
Initially, we have a system in a pure state, which is a coherent superposition of the basis states:
\begin{equation}
    |\psi(t_{0})\rangle=\sum_{i} c_{i} |S_{i}(t_{0}) \rangle.
\end{equation}
Initially, the system does not interact with the environment. Assume that at a certain instant of time we can switch on an interaction between the system and the environment. What an environment effective for decoherence 
is expected to do is to react dynamically to this coupling, in such a way that it evolves in a different way according to the different state of the system:
\begin{equation}
 \left|E\left(t_0\right)\right\rangle \sum_{i} c_{i}\left|S_i\left(t_0\right)\right\rangle \rightarrow\sum_{i}c_{i}\left|E_i(t)\right\rangle\left|S_i(t)\right\rangle.
 \label{eq:entanglementexample}
\end{equation}

Even if the state on the LHS of eq.(\ref{eq:entanglementexample}) is separable (i.e. can be written in a Hilbert state which is the product of the Hilbert state of the environment and the one of the system), the second state is a highly entangled state. The entanglement is caused by the interactions, but may keep being effective even after the interaction is switched off. Actually, entanglement keep memories of past interactions, if the auto-correlations in time of system or enviromnment do not decay in a short time.
Entanglement is fundamental for decoherence. On the other hand, both environmental and system states are written in specific basis. The basis are chosen by the interaction Hamiltonian between the two $H_{int, ij}$, specifically by requiring that the interaction Hamiltonian is diagonal in that basis. This basis is called ``pointer basis''; for example, for the inflation case, the pointer basis has been recognized to be the field basis, because the interactions are diagonal in that basis (see, e.g.,~\cite{kiefer_pointer_2007},~\cite{gong_quantum_2019}). 

A ``good'' environment has basis state which are orthogonal between themselves: 
\begin{equation}
    \langle E_{i}|E_{j}\rangle = \delta_{ij}.
\end{equation}
Our ignorance about the environment leads us to trace over these states. This operation changes the nature of the superposition, from coherent to incoherent. To see this, we introduce the reduced density matrix, which is a quantity we will deal with many times:
\begin{equation}
    \hat{\rho}_{\mathrm{R}} \equiv \operatorname{Tr}_{E} \hat{\rho} \equiv \sum_i\left\langle E_i|\hat{\rho}| E_i\right\rangle,
    \label{eq:reduceddensitymatrixexample}
\end{equation}
In the end we obtain that:
\begin{equation}
    \hat{\rho}_{\mathrm{R}} = \sum_{ij} c_{i} c^{*}_{j}| S_{i}\rangle \langle S_{j}| \langle E_{i}| E_{j} \rangle = \delta_{ij} \sum_{i} |S_{i}\rangle |c_{i}|^2\langle S_{i}| 
\end{equation}

So, the effect of decoherence is to make the reduced density matrix itself diagonal in the pointer basis. Indeed, the criterium we will use for decoherence is based on considering the suppression of the off diagonal elements in the pointer basis. In particular, we will consider the purity $\gamma$, defined as:
\begin{equation}
\gamma=\operatorname{Tr} (\rho)^2
\end{equation}
It is easy to show that $0<\gamma \leq 1$; in particular, if a state is completely coherent, the purity stays around 1; while, if it is decohered, the purity goes to zero rapidly; it is sensitive to the suppression of the interference terms.  
We now show a useful example taken from the literature, which will prove also useful next in the text. Mostly, in the paper we will have to deal with Gaussian states, since the free evolution of perturbations leads them to become squeezed two mode Gaussian states (~\cite{polarski_semiclassicality_1996}). For the interaction we will deal with, featuring just one system operator, the Gaussianity of the system density matrix is preserved also when considering the effect of the environment, i.e. after turning on the interaction with the environment. Consider, e.g., the expression for the off diagonal Gaussian states in~\cite{martin_observational_2018} for the Sasaki Mukhanov variable:
\begin{equation}
    \left|\left\langle v_{\boldsymbol{k}}+\frac{\Delta v_{\boldsymbol{k}}}{2}\left|\hat{\rho}_{R,\boldsymbol{k}}\right| v_{\boldsymbol{k}}-\frac{\Delta v_{\boldsymbol{k}}}{2}\right\rangle\right|=\left|\left\langle v_{\boldsymbol{k}}\left|\hat{\rho}_{R,\boldsymbol{k}}\right| v_{\boldsymbol{k}}\right\rangle\right| \exp \left[-\frac{\delta_{\boldsymbol{k}}+\frac{1}{4}}{2} \frac{\Delta v_{\boldsymbol{k}}^2}{P_{v v}(k)}\right].
\end{equation}
Here, the authors considered the off diagonal density matrix element between two different configuration for $v$, which have a distance $\Delta v/2$ from the diagonal. What we expect is that thanks to the decoherence process, these off diagonal elements get suppressed. Actually, these happens when the quantity $\delta_{k}$, determined dynamically by quantities related to the evolution of perturbations, becomes much bigger than one. In this way it is possible to show that purity is directly connected with the suppression of off diagonal elements:
\begin{equation}
\gamma_{k}=\operatorname{Tr}\left(\hat{\rho}_{R,\boldsymbol{k}}^2\right)=\int_{-\infty}^{\infty} \mathrm{d} v_{\boldsymbol{k}}^{(1)} \int_{-\infty}^{\infty} \mathrm{d} v_{\boldsymbol{k}}^{(2)}\langle v_{\boldsymbol{k}}^{(1)}\left|\hat{\rho}_{R,\boldsymbol{k}}\right| v_{\boldsymbol{k}}^{(2)}\rangle=\frac{1}{\sqrt{1+4 \delta_{\boldsymbol{k}}}}.
\end{equation}
So, for $\delta_{k} \gg 1$, the purity goes to zero and we can say that the Gaussian states have undergone decoherence. Actually, since Gaussian states are completely determined by their covariance matrix $\Sigma$\footnote{the matrix of all the possible two point correlation functions between field and momentum variables $\Sigma=\langle0| \chi_{i} \chi_{j} |0 \rangle$, $i=1,2$, $\chi_1=v,\, \chi_2=p$)}, it is possible to show that, for two mode squeezed states, purity can be expressed also as~\cite{Serafini:2003ke}:
\begin{equation}
     \gamma=\frac{1}{\sqrt{4 \operatorname{det}{\Sigma}}}
     \label{eq:purityforagaussianstate}
\end{equation}
This definition will prove useful in the text. 

What has changed in observables in the end?
Let's first assume to consider an observable $B$ defined on the system Hilbert space which is not diagonal in the pointer basis. In this case, the expectation value of the observable before decoherence is:
\begin{equation}
    \langle\psi| B |\psi\rangle= \sum_{ij} c^{*}_i c_j \langle S_i |B|S_j\rangle.
    \label{eq:interference}
\end{equation}
If instead we consider the expectation value after decoherence:
\begin{equation}
\operatorname{Tr}_{S} (B\rho_{r})= \sum_{i} |d_{i}|^2 \langle S_{i}|B|S_{i}\rangle,
\label{eq:noninterf}
\end{equation}
where notice that now we have written $d_{i}$ in place of $c_{i}$. In the general case, actually, the process of decoherence may change also the value of the diagonal terms in the density matrix, besides ``killing'' the off diagonal ones. This may happen if, for example, the unitary dynamics connects the off diagonal elements and the diagonal ones. In this way, the important change in the off-diagonal elements also drags perturbatively the diagonal ones, thus changing the values of observables~\cite{martin_observational_2018}. This effect is what we call non-unitary corrections to the power-spectrum in the main text. We have ignored for simplicity this effect in the previous eq.(\ref{eq:entanglementexample}). Let us keep ,as a first step, ignoring this effect since it is perturbatively small: we will consider $c_{i}=d_{i}$.

When computing the observables, we have in the first equation eq.(\ref{eq:interference}) some interference terms between different basis state which are non null, and can affect the final expectation value of the observable B. This is different with respect to eq.(\ref{eq:noninterf}) where, after decoherence, the diagonal terms are the only ones playing a role in the expectation value of B.

This difference is not there if the variable $A$ is diagonal in the $|S_{i}\rangle$ basis. In this case, actually, system states are eigenvectors and the off diagonal terms are null also in the quantum case in eq.(\ref{eq:interference}). The two cases are indistinguishable, since we have no sensitivity of the value of observables to off diagonal terms in the density matrix.

Unfortunately, this is exactly the case in cosmology, as the only observations we can do in CMB and LSS depend only (if are not directly proportional, see e.g.~\cite{Martin:2012pea}) from the field value, or from its correlation functions, while the momentum of the field is unobservable. This seems essentially to confirm the result obtained by the unitary evolution in the intrinsic decoherence formalism.  

The only chance we have to escape this paradigm seems to consider the change from $c_{i}$ to $d_{i}$ of the coefficients of the density matrix. As we will show explicitly in the text, also the unitary dynamics is changed, and this adds other distinguishable features in the primordial power-spectrum.

\section{Appendix B}
\subsection{Boundary terms and off diagonal coefficients}
\label{subsection:boundary terms}
In this appendix, we compute the results regarding the boundary terms of the coefficients of the $v$ $v$ terms in the quantum master equation, i.e. the $D_{11}$ coefficients in eq.(\ref{eq:Dmatrix}) for mixed derivative/derivativeless interactions. Then, we will compute the coefficients of the off diagonal $v$ $p$ terms in the quantum master equation, i.e. the off diagonal coefficients of the matrix in eq.(\ref{eq:mathcalDmatrix}) for the mixed derivative derivativeless interactions. In the end of the section we will also comment on the boundary terms associated to this $v$ $p$ terms in the quantum master equation, again for mixed interactions.
\subsubsection{Mixed terms: derivativeless memory}
We consider first the boundary terms for the $D_{11}$ coefficients in the quantum master equation, always for the mixed derivative-derivativeless interactions where the derivativeless interactions are integrated. Because of the coupling of derivativless interaction (see eqs.(\ref{eq:interactions}), this leads to a coefficient $1/x$ inside the memory integration which naturally cuts the memory effect, so we do not expect huge memory effects. The term from the integration by parts in eq.(\ref{eq:qme1-23beforepartint}) are:
\begin{equation}
    i\frac{H^2 \varepsilon \eta}{48 \pi^2 M_{\rm Pl} ^2} \left[\frac{d}{dx} \left(\frac{ e^{20 i (1-x) }(1-x)}{(1-x)^2+\epsilon^2}-c.c.\right)\frac{A(x,a)}{x}-\left(\frac{ e^{20 i (1-x) }(1-x)}{(1-x)^2+\epsilon^2}-c.c.\right)\frac{d}{dx}\left(\frac{A(x,a)}{x}\right)\right]_{a}^{1}.
    \label{eq:boundary1}
\end{equation}
We start from the terms in $x=1$. It is evident that:
\begin{equation}
   \left. \frac{d}{dx} \left(\frac{ e^{20 i (1-x) }(1-x)}{(1-x)^2+(\epsilon)^2}-c.c.\right) \right|_{x=1}=0,
\end{equation}
so the first term is null, while:
\begin{equation}
    \left. \frac{d}{dx} \left(\frac{A}{x}\right) \right|_{x=1}=1,
\end{equation}
but:
\begin{equation}
 \left.    \left(\frac{ e^{20 i (1-x) }(1-x)}{(1-x)^2+(\epsilon)^2}-c.c.\right) \right|_{x=1}=0,
\end{equation}
because we take the limit $\epsilon \rightarrow 0$ after substituting the value $x=1$. 
So, there is no coincidence limit divergence in the coefficient $D_{11}$. This is important because it is still under discussion in the literature how to properly consider a divergence in the non unitary part of the quantum master equation, since it cannot be reabsorbed by any  bare coefficient in a renormalization procedure in the Hamiltonian, being written in a non Hamiltonian term (see, e.g.~\cite{Agon:2014uxa}).  

Let's now consider the $x=a$ terms in eq.(\ref{eq:boundary1}). Various considerations are in order in this case. When solving the quantum master equation, the system mode is deep superhorizon. The $x=a$ term, in the quantum master equation, refers to the time in the integration when the mode of the system under consideration had just crossed the horizon. Thus, it is far in the memory of the environment and, probably, it is not legitimate to consider it. Our Born-Markov approximation
 (e.g. $\rho(\eta') \rightarrow \rho(\eta)$, see eq.(\ref{eq:markovianapprox}) probably would not be valid if solving the equation back in the far past, so we cannot rely on the results we obtain.
 Nevertheless, for completeness, we here investigate also these terms:
\begin{equation}
\left.    2i \frac{d}{dx}\left(\frac{\sin{20(x-1)}}{x-1}\right)\frac{A}{x}\right|_{x=a}\simeq \cos{20a}(\cos(-1)+\sin(-1))- \frac{\sin(20a)}{a}(\cos(-1)+\sin(-1)),
\end{equation}
while the second term goes to zero as $a \to \infty$, the first could in principle give a contribution. However, the oscillatory function has a really small amplitude, and it is centered around zero. We can thus repeat the argument applied for eq.(\ref{eq:averagingout}). Remembering that the principle behind the derivation of the quantum master equation is to coarse grain in time, if we average over a period the oscillatory function, we obtain 0 as a result. The period of the oscillations is, in fact, much shorter than the usual evolution time of the system $\simeq \eta$. Also:
\begin{equation}
    \left.    2i \left(\frac{\sin{20(x-1)}}{x-1}\right) \frac{d}{dx} \frac{A}{x}\right|_{x=a}\simeq  \frac{\sin(20a)}{a}(2\cos(-1)-\sin(1)),
\end{equation}
again negligible as $a \to \infty$. 

The coefficients for the off diagonal terms in the D matrix are, as defined in:
\begin{equation}
D^{int1-23}_{12} \propto \int_{1}^{a} B(x,a) K(x),
\label{eq:D123boundary}
\end{equation}
where we have omitted the coefficients upfront which are unimportant for our scopes. Consider first the imaginary part of eq.(\ref{eq:D123boundary}):
\begin{equation}
\int_{1}^{a} \frac{dx}{x} B \frac{d^2}{dx^2}\left(\frac{\cos(20(1-x))}{1-x}\right) .
\end{equation}

By integrating by parts twice, we ``move'' the two derivatives to $B/x$; however, the $1-x$ at the denominator gives also a divergence, so we integrate also it by parts, obtaining:
\begin{equation}
\begin{split}
   & \int_{1}^{a} \frac{d^2}{dx^2} \left(\frac{B}{x}\right) \frac{(x-1) \cos(20(x-1)) }{(x-1)^2+\epsilon^2}=- \int_{1}^{a} \frac{d}{dx}\left(\left(\frac{d^2}{dx^2}\frac{B}{x}\right)\cos(20(x-1))\right) \log((x-1) ^2+\epsilon^2)\\
   &+\left[\frac{d^2}{dx^2}\frac{B}{x} \cos[20(1-x)] \log[(x-1)^2+\epsilon^2]\right]_1^a.\\
    \end{split}
\end{equation}
Notice that in $x=1$ we have a divergence, so we instead apply the trick of integrating again by parts. The result of the integral has been obtained numerically for some trial values; for example, for $a=10^5$ it is going to be $\simeq 10^{-5}$, way beyond the values of the other coefficients in the D matrix and so well approximated by zero. This is due to the $1/a$ that suppresses the value of the integration. This suppressing factor also ensures that as $a\to\infty$ the value of the integration does not increase if we evaluate the series developing of:
\begin{equation}
    \frac{d^2}{dx^2}{\frac{B}{x}}\simeq \frac{8 (5 x \sin (20 (x-1))+\cos (20 (x-1)))}{a x^5},
\end{equation}
so it is proportional to $1/a$ and the integrand can only decrease as inflation ends. All the other terms are boundary terms:
\begin{equation}
    \left[\frac{B}{x}\frac{d}{dx}\frac{e^{20i(1-x)}}{1-x+i\epsilon}-\frac{d}{dx}\frac{B}{x} \frac{e^{20i(1-x)}}{1-x+i\epsilon}\right]_{1}^a,
    \label{eq:boundary2}
\end{equation}
in particular, by realizing that in $x=1$:
\begin{equation}
    \begin{split}
        &\frac{B}{x}=0,\\
        &\frac{d}{dx}\frac{B}{x}=\frac{1}{a},\\
        &\frac{d^2}{dx^2}\frac{B}{x}=\frac{-2}{a},\\
        \label{eq:bsuxbooundary}
    \end{split}
\end{equation}
 Every term is either zero or proportional to $1/a$, and that is the reason why we can actually neglect all these contributions. This is true, as can be expected, also, for the terms evaluated in eq.(\ref{eq:boundary2}) at $x=a$; in that case we have the denominator of the correlation function becoming of order $a$, so it is actually sufficient that the factors we now mention are at most constant as $a\to\infty$:
\begin{equation}
    \begin{split}
        &\frac{B}{x}\simeq \cos{1}-\sin{1}, \\
        &\frac{d}{dx}\frac{B}{x}\simeq\frac{2\cos{1}-\sin{1}}{a}+O(1/a),\\
        &\frac{d^2}{dx^2}\frac{B}{x}=\frac{3 \sin (1)}{a^2}-\frac{5 \cos (1)}{a^2}+O(1/a^2).\\
    \end{split}
\end{equation}
These values are obtained by taking the least suppressed order in $\frac{1}{a}$ of the considered functions as $a\to\infty$ applied to the functions evaluated in $x=a$.
In the end, all of the boundary terms are also subdominant and we can neglect them. 

The last term to check instead is the real part of the coefficient $D_{12}$ defined in eq.(\ref{eq:D123boundary}). The local part, proportional to the $\delta(x-1)$, can be checked if we recall the last equation in eqs.(\ref{eq:bsuxbooundary}), which actually shows that these local terms are also suppressed by one power of $a$, due to the value of $d^2/dx^2 B(x,a)$ in $x=1$, and thus negligible. Evaluating instead numerically the non local part:
\begin{equation}
    \Re D_{12}\propto \int_{1}^{a} dx \frac{d^2}{dx^2}\frac{B}{x} \frac{\sin{20(x-1)}}{x-1} .
\end{equation}
For a rather little value, $a=1000$, we obtain $\simeq 10^{-4}$; however, by using again the limit as $a\to\infty$,  this time for a generic x, we obtain that $\frac{d^2}{dx^2}\frac{B}{x} \propto \frac{1}{a}$ and so the integral should only decrease as $a\to\infty$.

Notice that no approximation was needed to get rid of these boundary terms. This is due to the presence of the derivativeless interaction in the integration, that by itself is an interaction which suppresses memory inside the intergration, being relevant as $\eta \to 0$. Thus, the system tends to be Markovian for this couple of interactions.

\subsubsection{Mixed terms: derivative memory}
We now perform analogous calculations for the boundary terms also for the case in which it is the derivative interaction which is integrated in $\eta'$, treated in detail in section\ref{sec:mixedterms231}. In this situation, the memory effects are far bigger, as the integrated interactions are proportional to $\eta'$, thus irrelevant as $\eta' \to 0$, and more important as we go back in time inside the integration. 

We start by considering the real part of the $D_{11}$ coefficient, as written in the eq.(\ref{eq:Dmatrix}). The boundary terms in eq.(\ref{eq:2,3-1mixedderivative}) are:
\begin{equation}
\begin{split}
D_{11}&=\frac{\varepsilon H^2}{24 \pi^2 M_{\rm Pl} ^2 \eta^2}  \left[x A(x,a) \frac{d}{dx} \left(\frac{\sin{20(1-x))} (1-x) } {(1-x)^2+ \epsilon^2}\right)-\right.\\
&\left.-\frac{d}{dx}(x A(x,a))  \left(\frac{\sin{20(1-x))} (1-x) } {(1-x)^2+ \epsilon^2}\right) \right]_1^a\\
\end{split}
\end{equation}
Computing the boundary terms in $x=1$ we have:
\begin{equation}
\begin{split}
&\frac{\sin{20(1-x))} (1-x) } {(1-x)^2+ \epsilon^2}=0,\\
&\frac{d}{dx} \left(\frac{\sin{20(1-x))} (1-x) } {(1-x)^2+ \epsilon^2}\right)=0.\\
\end{split}
\end{equation}
So, also in this case, we have no ultraviolet divergences in the non unitary term of the quantum master equation. In $x=a$ there are many terms; some of them are left with a positive power law dependence on $a$. This dependence on $a$ would end up in terms of the coefficients of the quantum master equation which dominate as $\eta \to 0$, and then overcome the finite terms found e.g. in  eq.(\ref{eq:derivativederivativelessD11negativememoty}), which source a decoherence proportional to $1/\eta^3$; in this sense, we call them ``dangerous terms''.  We remember that the terms associated to $x=a$ are related to the integration memory since when the mode crossed the horizon, and thus it seems unlikely that they are compatible with the Born-Markov approximation eq.(\ref{eq:markovianapprox}).  For the sake of simplicity, we just report these ``dangerous terms'':
\begin{equation}
\begin{split}
    &\frac{H^2 \varepsilon}{48 \pi^2 M_{\rm Pl} ^2 \eta^2} \left[20 a^2 
 \cos{20a} (\sin{1}-\cos{1})+ 20a \cos{20a}(-\cos{1}+\sin{1})+a \sin{20a}(\cos{1}-\sin{1})\right.\\
& \left.-a \sin{20a}\sin{1}\right].\\
\end{split}
\label{eq:int231oscillatory}
\end{equation}
Notice that we have excluded also the terms with no powers of a in front of the trigonometric function, since we have shown in eq.(\ref{eq:averagingout}) that we can get rid of them by averaging out the coefficients on the time scales on which the quantum master equation is coarse grained. On the other hand, the oscillatory character of the terms in eq.(\ref{eq:int231oscillatory}) and the observations on the independence on the details of the precise moment at which the system mode crosses the horizon would suggest to try to average out also the other terms as in eq.(\ref{eq:averagingout}), but the steeper power of $\eta$ around zero makes the procedure not valid.
A similar behaviour was noticed in~\cite{colas_benchmarking_2022}, in the terms dubbed ``spurious'', which also produced terms, depending on the initial time from the memory integration but dominating $D_{11}$ coefficients in an inflationary late time expansion ($\eta \to 0$). The proposed solution in~\cite{colas_benchmarking_2022} was to remove by hand the spurious contribution, since by doing that the results of the quantum master equations were impressively coincident to the exact solutions. One argument in favour of this prescription is that these terms are actually not present in the standard perturbation theory equation equivalent to the quantum master equation, but are a result of the partial resummation performed by the quantum master equation (\cite{colas_quantum_2023}). The fact that this kind of contributions appear also in our case, with more complicated interactions and a different environment, confirms that spurious contributions may be a generic feature of quantum master equations in time-dependent background, and not, as first suggested, a feature of the peculiar bilinear interaction adopted in \cite{colas_benchmarking_2022}. This confirms previous findings of \cite{Brahma:2024yor}. Since we recognize the nature of this terms, analogously to what done in the previous literature, also in our case, a possible way out of this problem could be just to remove also these terms by hand.  

Nevertheless, we remember that the Strong Markovian approximation appears to be needed in this case of derivative interactions. If not, the final result for the coefficient $D_{11}$ is negative, as developed in \ref{subsec:final results SMAR}. The boundary terms, after applying the Strong Markovian approximations, are:
\begin{equation}
    \frac{H^2 \varepsilon}{48 \pi^2 M_{\rm Pl} ^2\eta^2} \left[x \frac{d}{dx} \frac{(1-x)\sin{20(1-x)}}{(1-x)^2+\epsilon^2}-\frac{(1-x)\sin{20(1-x)}}{(1-x)^2+\epsilon^2}\right]_{1}^{a}.
    \label{eq:boundarytermsderivmemorysmar}
\end{equation}
 In $x=1$ all the terms are zero as before; in $x=a$, we have instead only one term which is not suppressed by some powers of $a$. This term is proportional at most to $\cos{20a}$ as $a \to \infty$. Such terms can be easily averaged out as shown in eq.(\ref{eq:averagingout}), so, if we perform the above procedure, we should not have that oscillating term there anymore. Thus, the Strong Markovian approximation erases the spurious contributions.

We now analyze the Lamb Shift Hamiltonian. The Strong Markovian approximation must be applied also in order to compute the Lamb Shift corrections in order to avoid divergences, as shown in the section \ref{sec:mixedterms231}. So we directly consider the boundary terms in the case of application of the Strong Markovian approximation. We could as well consider the same calculation without the Strong Markovian approximation, but we would have had a similar output as in eq.(\ref{eq:int231oscillatory}).
The boundary terms of eq.(\ref{eq:delta231afterSMAR}) are:
 \begin{equation}
     \frac{H^2\varepsilon}{48\pi^2 M_{\rm Pl} ^2} \left[x \frac{d}{dx} \frac{(1-x)\cos{20(1-x)}}{(1-x)^2+\epsilon^2}-\frac{(1-x)\cos{20(1-x)}}{(1-x)^2+\epsilon^2}\right]_{1}^{a}.
 \end{equation}
In $x=1$ we have now:
\begin{equation}
    \frac{H^2\varepsilon}{48\pi^2 M_{\rm Pl} ^2} \frac{1}{\epsilon^2}.
\end{equation}
This UV divergence is in the Hamiltonian corrections to the mass; thus, it can simply be reabsorbed by an infinite renormalization of the mass itself, analogously, e.g. to the logarithmic divergence in eq.(\ref{eq:renormalizationofthemass}) . In $x=a$, we now have just a term of the type $20 \sin{20a}$ which is not suppressed by any power of $a$, so we can simply average it out analogously to what was done with eq.(\ref{eq:averagingout}).

When considering the off diagonal term, as done for the previous case, we should integrate over $B(x,a)$ :
\begin{equation}
D^{int23-1}_{12} \propto \int_{1}^{a} x B(x,a) K(x).
\end{equation}
We start from the real part; of course, we have to ``move'' again the two derivatives to the $B$ coefficient, so obtaining:
\begin{equation}
 \int_{1}^{a} \frac{d^2}{dx^2} \left(B x\right) \frac{\sin(20(x-1)}{(x-1)}- \left[ B x 
 \frac{d}{dx}\left( \frac{sin(20(x-1))}{x-1}\right) -\left(\frac{d}{dx} B x\right)\frac{sin(20(x-1))}{x-1}\right]_{1}^{a}.
\end{equation}
The integral can be computed by considering the approximate form for B eq.(\ref{eq:Bapprox}) because the difference between B and the approximation is noticeable only around $x=a$, and we cannot believe well our results in that regime. By deriving it twice:
\begin{equation}
    D_{12}^{23-1}=\frac{\varepsilon H^2}{24 \pi^2 M_{\rm Pl} ^2 \eta^2} \int_{1}^{a} dx \frac{-2x}{a} \frac{sin{20(1-x)}}{1-x}=\frac{\varepsilon H^2}{24 \pi^2 M_{\rm Pl} ^2 \eta^2} \frac{1}{a} \left(\frac{1-\cos{20a}}{20}+\operatorname{Si}(20 (-1 + a))\right).
\end{equation}
Again, the suppression by the $1/a$ is much more important and drives the result to be subdominant around the end of inflation. This result has been also confirmed numerically, by integrating using the full, non approximated form for $B(x,a)$ as defined in \ref{eq:ABwithxanda}, for some specific values of a. 

The boundary terms in $x=1$ feature:
\begin{equation}
B x\Bigr|_{x=1}=0 \quad \quad \frac{d}{dx}Bx\Bigr|_{x=1}=\frac{-1}{a}, \frac{ d^2 Bx}{d x^2}\Bigr|_{x=1}=\frac{-1}{a}
\end{equation}
so they are subdominant for $a \to \infty$. The boundary terms in $x=a$ may instead be ``dangerous'', in the sense exposed in the discussion on eq.(\ref{eq:int231oscillatory}), if they are dominant with respect to the bulk contribution as $a \to \infty$. We report only those terms:
\begin{equation}
(\cos{20a}) \,\,20 a (\cos{1}-\sin{1}).
\label{eq:oscillations}
\end{equation}
These may be seen as spurious contributions, as in \cite{colas_benchmarking_2022} (see again the discussion after eq.(\ref{eq:int231oscillatory}). As before, these terms are irrelevant after applying the Strong Markovian approximation, since $B=0$ identically then and they disappear. 

For the imaginary part:
\begin{equation}
\begin{split}
 &\int_{1}^{a} \frac{d^2}{dx^2} \left(B x\right) \frac{\cos{20(x-1)}(x-1)}{((x-1)^2 +\epsilon^2)}- \left[ B x 
 \frac{d}{dx}\left( \frac{\cos{20(x-1)} (x-1)}{(x-1)^2+ \epsilon^2}\right) -\right.\\
 &\left.\left(\frac{d}{dx} B x\right)\frac{\cos{20(x-1)} (x-1)}{(x-1)^2 +\epsilon^2}\right]_{1}^{a}\\.
 \end{split}
\end{equation}
 Notice that in the boundary terms in $x=1$ we have no divergences, since $B=0$ in $x=1$. The boundary terms in $x=a$ of the first two integration by parts are absolutely similar to what seen in the other case, suffering for the presence of a term analogous to eq.(\ref{eq:int231oscillatory}) :
\begin{equation}
    (-20  a  (\cos{1} - \sin{1}))  \sin({20 - 20 a}),
\end{equation}
which can in principle dominate over the bulk term. However we can recognize it as a spurious term and discard it for this reason, analogously to what was done in \cite{colas_benchmarking_2022}. Instead, in the integration, we have:
\begin{equation}
   \int_{1}^{a} dx  \frac{-2x}{a}\frac{\cos{20(x-1)}(x-1)}{(x-1)^2+\epsilon^2}.
   \label{eq:BappImderivint}
\end{equation}
The divergence in $x=1$ can be cured, as already seen many times, by integrating by parts after writing $1/(x-1)$ as the derivative of a logarithm. The integral has a similar problem as the one found in eq.(\ref{eq:delta11derivativeless}): by integrating by parts, terms (artificially) divergent because of the logarithm as $a \to \infty$ appear; as well, this happens for the terms inside the residual integration, in spite of the fact that the integrand in eq.(\ref{eq:BappImderivint}) should be just oscillating at infinity. Both of these terms may be interpreted as spurious terms. The only finite term, in the end, will be suppressed by a factor $1/a$,  so we can neglect it with respect to the other terms in the matrix. Notice that if we apply the prescription of the Strong Markovian approximation $B$ is identically null. 

\subsubsection{Derivative interactions}
\label{par:boundarytermsfromderivative}
In the couples with only derivative interactions we have no finite terms, as written in the main text. It is interesting to consider also the boundary terms. The boundary terms from eq.(\ref{eq:derivderivreal}) are:
\begin{equation}
\begin{split}
    &\frac{\beta \varepsilon H^2}{8 M_{\rm Pl} ^2 \eta^2} \frac{1}{16} \left[(x A) \frac{d^3}{dx^3}\left(\frac{\sin{20(1-x)}}{1-x}\right)-\frac{d}{dx}(x A) \frac{d^2}{dx^2}\left(\frac{\sin{20(1-x)}}{1-x}\right)+\right.\\
    &\left.\frac{d^2}{dx^2}(x A) \frac{d}{dx}\left(\frac{\sin{20(1-x)}}{1-x}\right)-\frac{d^3}{dx^3}(x A)\left(\frac{\sin{20(1-x)}}{1-x}\right)\right]_{1}^a.\\
    \label{eq:boundarytermsderivativesreal}
    \end{split}
\end{equation}
The boundary terms from the imaginary part in eq.(\ref{eq:derivderivima}) are similar:
\begin{equation}
\begin{split}
    & -\frac{\beta \varepsilon H^2}{8 M_{\rm Pl} ^2 \eta^2} \frac{1}{32} \left[(x A) \frac{d^3}{dx^3}\left(\frac{\cos{20(1-x)}}{1-x}\right)-\frac{d}{dx}(x A) \frac{d^2}{dx^2}\left(\frac{\cos{20(1-x)}}{1-x}\right)+\right.\\
    &\left.+\frac{d^2}{dx^2}(x A) \frac{d}{dx}\left(\frac{\cos{20(1-x)}}{1-x}\right)-\frac{d^3}{dx^3}(x A)\left(\frac{\cos{20(1-x)}}{1-x}\right)\right]_{1}^a.\\
    \label{eq:boundarytermsderivatives}
    \end{split}
\end{equation}
In $x=a$, for simplicity, we only report the ``dangerous'' terms, which may diverge as $a \to \infty$:
\begin{equation}
    \frac{\beta \varepsilon H^2}{8 M_{\rm Pl} ^2 \eta^2} \frac{1}{16}((\cos{1}-\sin{1})(\cos{20a} (8000 a^2-8000a)-1200a \sin{20a} )+ 400a \sin{1} \sin{20a}).
    \label{eq:danferouscontr}
\end{equation}
The terms come just from the first two terms in eq.(\ref{eq:boundarytermsderivativesreal}). Similarly for the imaginary part we have:
\begin{equation}
    \frac{\beta \varepsilon H^2}{8 M_{\rm Pl} ^2 \eta^2} \frac{1}{16}((\cos{1}-\sin{1})(\sin{20a} (8000 a^2+8000a)-1200a \cos{20a} )+ 400a \sin{1} \cos{20 a})
\end{equation}
We can either consider these terms spurious and then remove them by hand; or we can apply the Strong Markovian approximation. In that case, with $A=1$, it is easy to see that no terms are dangerous; at most, we have a term proportional to $\cos{20a}$ or $\sin{20a}$, which can however be averaged out by eq.(\ref{eq:averagingout}).

In $x=1$ we have other interesting effects.  The imaginary part has:
\begin{equation}
    \begin{split}
        &\frac{d^3}{dx^3}\left(\frac{\cos{20(1-x)}(1-x)}{(1-x)^2+\epsilon^2}\right)=-\frac{6}{\epsilon^4}-\frac{1200}{\epsilon^2},\\
        &\frac{d^2}{dx^2}\left(\frac{\cos{20(1-x)}(1-x)}{(1-x)^2+\epsilon^2}\right)=0,\\
        &\frac{d}{dx}\left(\frac{\cos{20(1-x)}(1-x)}{(1-x)^2+\epsilon^2}\right)=\frac{1}{\epsilon^2},\\
        &\left(\frac{\cos{20(1-x)}(1-x)}{(1-x)^2+\epsilon^2}\right)=0.\\
        \end{split}
\end{equation}
The divergences can be reabsorbed as UV divergences in the mass terms, just as it was showed in eq.(\ref{eq:renormalizationofthemass}). This happens because these divergences occur in the Hamiltonian part of eq.(\ref{eq:canonicalqme}) and so it is easy to perform the usual renormalization procedure. When applying the ``Strong Markovian Approximation'', we have just the first line which is still there, while the others are erased. 

For the real part we have
\begin{equation}
     \begin{split}
        &\frac{d^3}{dx^3}\left(\frac{\sin{20(1-x)}(1-x)}{(1-x)^2+\epsilon^2}\right)=0,\\
        &\frac{d^2}{dx^2}\left(\frac{\sin{20(1-x)}(1-x)}{(1-x)^2+\epsilon^2}\right)=\frac{40}{\epsilon^2},\\
        &\frac{d}{dx}\left(\frac{\sin{20(1-x)}(1-x)}{(1-x)^2+\epsilon^2}\right)=0,\\
        &\left(\frac{\sin{20(1-x)}(1-x)}{(1-x)^2+\epsilon^2}\right)=0.\\
        \end{split}
        \label{eq:worryinguvdivergence}
\end{equation}
Notice that the first two lines are the only ones not erased also after applying the ``Strong Markovian approximation''. The ``worrying'' result is in the second line. Unlike all the other cases, here we have a divergence on the second line term which corresponds to a non Hamiltonian divergence, since it appears in front of the non unitary contribution in the quantum master equation eq.(\ref{eq:canonicalqme}). This is a known problem in the literature of the open quantum field theory; e.g., Agon et al. in~\cite{Agon:2014uxa} connects it to the renormalization of composite operators. More recently, \cite{Burgess:2024heo} argued that these divergences arising from the non-Hamiltonian part of the quantum master equation are expected to be there. Our calculations confirm this suggestion.  However, since there is still no definitive answer on how to renormalize them, we leave this point for future investigations.

%is true except for the possibility that $c_{i}$ is different from $d_{i}$. The process of decoherence changes, perturbatively but progressively, the value of the off diagonal elements of the density matrix, but the idea is that this changing can also affect in some way the values of the diagonal elements, e.g. because the free Hamiltonian couples the off diagonal elements with the diagonal ones (~\cite{martin_observational_2018}) . We can say that, in a certain sense, the entanglement with the environment dresses the system free Hamiltonian, renormalizing also the diagonal values, and so giving a perturbatively squared correction. This are the corrections we look for in the following of this paper.

\acknowledgments

The authors would like to thank Sabino Matarrese, Vincent Vennin, Takeshi Kobayashi, Jerome Martin, Valentina Danieli, Daniele Bertacca, Jesse de Kruijf, Flaminia Giacomini, Raul Jimenez for many stimulating discussions. Fp.L. is grateful to Vincent Vennin for his kind hospitality at LPENS, Paris, during the completion of this work. Fp.L. would like to thank Paolo Facchi, Angelo Bassi and Anirudh Gundhi for interesting discussions on the theory of open quantum systems.
NB acknowledges financial support from the INFN InDark initiative and from the COSMOS network (www.cosmosnet.it) through the ASI (Italian Space Agency) Grants 2016-24-H.0, 2016-24-H.1-2018 and 2020-9-HH.0. This work is supported in part by the MUR Departments of Excellence grant ``Quantum Frontiers'' of the Physics and Astronomy Department of Padova.

%\printbibliography

%\end{thebibliography}
\bibliographystyle{JHEP}
\bibliography{cautiouspaper}

\providecommand{\href}[2]{#2}\begingroup\raggedright\begin{thebibliography}{10}

\bibitem{mukhanov_quantum_1981}
V.F.~Mukhanov and G.V.~Chibisov, \emph{Quantum {Fluctuations} and a {Nonsingular} {Universe}}, {\emph{JETP Lett.} {\bfseries 33} (1981) 532}.

\bibitem{guth_inflationary_1981}
A.H.~Guth, \emph{The {Inflationary} {Universe}: {A} {Possible} {Solution} to the {Horizon} and {Flatness} {Problems}}, \href{https://doi.org/10.1103/PhysRevD.23.347}{\emph{Phys. Rev. D} {\bfseries 23} (1981) 347}.

\bibitem{linde_chaotic_1983}
A.D.~Linde, \emph{Chaotic {Inflation}}, \href{https://doi.org/10.1016/0370-2693(83)90837-7}{\emph{Phys. Lett. B} {\bfseries 129} (1983) 177}.

\bibitem{Starobinsky:1980te}
A.A.~Starobinsky, \emph{{A New Type of Isotropic Cosmological Models Without Singularity}}, \href{https://doi.org/10.1016/0370-2693(80)90670-X}{\emph{Phys. Lett. B} {\bfseries 91} (1980) 99}.

\bibitem{Albrecht:1982wi}
A.~Albrecht and P.J.~Steinhardt, \emph{{Cosmology for Grand Unified Theories with Radiatively Induced Symmetry Breaking}}, \href{https://doi.org/10.1103/PhysRevLett.48.1220}{\emph{Phys. Rev. Lett.} {\bfseries 48} (1982) 1220}.

\bibitem{Linde:1981mu}
A.D.~Linde, \emph{A new inflationary universe scenario: A possible solution of the horizon, flatness, homogeneity, isotropy and primordial monopole problems}, \href{https://doi.org/10.1016/0370-2693(82)91219-9}{\emph{Phys. Lett. B} {\bfseries 108} (1982) 389}.

\bibitem{akrami_planck_2020}
Y.~Akrami and {others}, \emph{Planck 2018 results. {IX}. {Constraints} on primordial non-{Gaussianity}}, \href{https://doi.org/10.1051/0004-6361/201935891}{\emph{Astron. Astrophys.} {\bfseries 641} (2020) A9}.

\bibitem{Planck:2018jri}
{\scshape Planck} collaboration, \emph{{Planck 2018 results. X. Constraints on inflation}}, \href{https://doi.org/10.1051/0004-6361/201833887}{\emph{Astron. Astrophys.} {\bfseries 641} (2020) A10} [\href{https://arxiv.org/abs/1807.06211}{{\ttfamily 1807.06211}}].

\bibitem{Acquaviva:2002ud}
V.~Acquaviva, N.~Bartolo, S.~Matarrese and A.~Riotto, \emph{{Second order cosmological perturbations from inflation}}, \href{https://doi.org/10.1016/S0550-3213(03)00550-9}{\emph{Nucl. Phys. B} {\bfseries 667} (2003) 119} [\href{https://arxiv.org/abs/astro-ph/0209156}{{\ttfamily astro-ph/0209156}}].

\bibitem{bartolo_non-gaussianity_2004}
N.~Bartolo, E.~Komatsu, S.~Matarrese and A.~Riotto, \emph{Non-{Gaussianity} from inflation: {Theory} and observations}, \href{https://doi.org/10.1016/j.physrep.2004.08.022}{\emph{Phys. Rept.} {\bfseries 402} (2004) 103}.

\bibitem{maldacena_non-gaussian_2003}
J.M.~Maldacena, \emph{Non-{Gaussian} features of primordial fluctuations in single field inflationary models}, \href{https://doi.org/10.1088/1126-6708/2003/05/013}{\emph{JHEP} {\bfseries 05} (2003) 013}.

\bibitem{guth_quantum_1985}
A.H.~Guth and S.-Y.~Pi, \emph{The {Quantum} {Mechanics} of the {Scalar} {Field} in the {New} {Inflationary} {Universe}}, \href{https://doi.org/10.1103/PhysRevD.32.1899}{\emph{Phys. Rev. D} {\bfseries 32} (1985) 1899}.

\bibitem{polarski_semiclassicality_1996}
D.~Polarski and A.A.~Starobinsky, \emph{Semiclassicality and decoherence of cosmological perturbations}, \href{https://doi.org/10.1088/0264-9381/13/3/006}{\emph{Class. Quant. Grav.} {\bfseries 13} (1996) 377}.

\bibitem{lesgourgues_quantum_1997}
J.~Lesgourgues, D.~Polarski and A.A.~Starobinsky, \emph{Quantum to classical transition of cosmological perturbations for nonvacuum initial states}, \href{https://doi.org/10.1016/S0550-3213(97)00224-1}{\emph{Nucl. Phys. B} {\bfseries 497} (1997) 479}.

\bibitem{albrecht_inflation_1994}
A.~Albrecht, P.~Ferreira, M.~Joyce and T.~Prokopec, \emph{Inflation and squeezed quantum states}, \href{https://doi.org/10.1103/PhysRevD.50.4807}{\emph{Phys. Rev. D} {\bfseries 50} (1994) 4807}.

\bibitem{kiefer_pointer_2007}
C.~Kiefer, I.~Lohmar, D.~Polarski and A.A.~Starobinsky, \emph{Pointer states for primordial fluctuations in inflationary cosmology}, \href{https://doi.org/10.1088/0264-9381/24/7/002}{\emph{Class. Quant. Grav.} {\bfseries 24} (2007) 1699}.

\bibitem{Danieli:2023dvu}
V.~Danieli, T.~Kobayashi, N.~Bartolo, S.~Matarrese and M.~Viel, \emph{{Anharmonic Effects on the Squeezing of Axion Perturbations}},  \href{https://arxiv.org/abs/2309.13112}{{\ttfamily 2309.13112}}.

\bibitem{kiefer_why_2009}
C.~Kiefer and D.~Polarski, \emph{Why do cosmological perturbations look classical to us?}, \href{https://doi.org/10.1166/asl.2009.1023}{\emph{Adv. Sci. Lett.} {\bfseries 2} (2009) 164}.

\bibitem{colas_decoherence_2024}
T.~Colas, C.~de~Rham and G.~Kaplanek, \emph{Decoherence out of fire: {Purity} loss in expanding and contracting universes},  Jan., 2024.

\bibitem{Brahma:2024ycc}
S.~Brahma, J.~Calder\'on-Figueroa, X.~Luo and D.~Seery, \emph{{The special case of slow-roll attractors in de Sitter: Non-Markovian noise and evolution of entanglement entropy}},  \href{https://arxiv.org/abs/2411.08632}{{\ttfamily 2411.08632}}.

\bibitem{burgess_minimal_2023}
C.P.~Burgess, R.~Holman, G.~Kaplanek, J.~Martin and V.~Vennin, \emph{Minimal decoherence from inflation}, \href{https://doi.org/10.1088/1475-7516/2023/07/022}{\emph{JCAP} {\bfseries 07} (2023) 022}.

\bibitem{nelson_quantum_2016}
E.~Nelson, \emph{Quantum {Decoherence} {During} {Inflation} from {Gravitational} {Nonlinearities}}, \href{https://doi.org/10.1088/1475-7516/2016/03/022}{\emph{JCAP} {\bfseries 03} (2016) 022}.

\bibitem{gong_quantum_2019}
J.-O.~Gong and M.-S.~Seo, \emph{Quantum non-linear evolution of inflationary tensor perturbations}, \href{https://doi.org/10.1007/JHEP05(2019)021}{\emph{JHEP} {\bfseries 05} (2019) 021}.

\bibitem{martineau_decoherence_2007}
P.~Martineau, \emph{On the decoherence of primordial fluctuations during inflation}, \href{https://doi.org/10.1088/0264-9381/24/23/006}{\emph{Class. Quant. Grav.} {\bfseries 24} (2007) 5817}.

\bibitem{Colas:2022kfu}
T.~Colas, J.~Grain and V.~Vennin, \emph{{Quantum recoherence in the early universe}}, \href{https://doi.org/10.1209/0295-5075/acdd94}{\emph{EPL} {\bfseries 142} (2023) 69002} [\href{https://arxiv.org/abs/2212.09486}{{\ttfamily 2212.09486}}].

\bibitem{Colas:2024ysu}
T.~Colas, J.~Grain, G.~Kaplanek and V.~Vennin, \emph{{In-in formalism for the entropy of quantum fields in curved spacetimes}},  \href{https://arxiv.org/abs/2406.17856}{{\ttfamily 2406.17856}}.

\bibitem{Breuer:2007juk}
H.-P.~Breuer and F.~Petruccione, \emph{{The Theory of Open Quantum Systems}}, Oxford University Press (1, 2007), \href{https://doi.org/10.1093/acprof:oso/9780199213900.001.0001}{10.1093/acprof:oso/9780199213900.001.0001}.

\bibitem{colas_benchmarking_2022}
T.~Colas, J.~Grain and V.~Vennin, \emph{Benchmarking the cosmological master equations}, \href{https://doi.org/10.1140/epjc/s10052-022-11047-9}{\emph{Eur. Phys. J. C} {\bfseries 82} (2022) 1085}.

\bibitem{Burgess:2024eng}
C.P.~Burgess, T.~Colas, R.~Holman, G.~Kaplanek and V.~Vennin, \emph{{Cosmic purity lost: perturbative and resummed late-time inflationary decoherence}}, \href{https://doi.org/10.1088/1475-7516/2024/08/042}{\emph{JCAP} {\bfseries 08} (2024) 042} [\href{https://arxiv.org/abs/2403.12240}{{\ttfamily 2403.12240}}].

\bibitem{Shandera:2017qkg}
S.~Shandera, N.~Agarwal and A.~Kamal, \emph{{Open quantum cosmological system}}, \href{https://doi.org/10.1103/PhysRevD.98.083535}{\emph{Phys. Rev. D} {\bfseries 98} (2018) 083535} [\href{https://arxiv.org/abs/1708.00493}{{\ttfamily 1708.00493}}].

\bibitem{Zarei:2021dpb}
M.~Zarei, N.~Bartolo, D.~Bertacca, A.~Ricciardone and S.~Matarrese, \emph{Non-markovian open quantum system approach to the early universe: Damping of gravitational waves by matter}, \href{https://doi.org/10.1103/PhysRevD.104.083508}{\emph{Phys. Rev. D} {\bfseries 104} (2021) 083508} [\href{https://arxiv.org/abs/2104.04836}{{\ttfamily 2104.04836}}].

\bibitem{Pueyo:2024twm}
C.D.~Pueyo, H.~Goodhew, C.~McCulloch and E.~Pajer, \emph{{Perturbative unitarity bounds from momentum-space entanglement}},  \href{https://arxiv.org/abs/2410.23709}{{\ttfamily 2410.23709}}.

\bibitem{Green:2024cmx}
D.~Green and G.~Sun, \emph{{Effective Field Theory and In-In Correlators}},  \href{https://arxiv.org/abs/2412.02739}{{\ttfamily 2412.02739}}.

\bibitem{Alicki:2023tfz}
R.~Alicki, G.~Barenboim and A.~Jenkins, \emph{{The irreversible relaxation of inflation}},  \href{https://arxiv.org/abs/2307.04803}{{\ttfamily 2307.04803}}.

\bibitem{Alicki:2023rfv}
R.~Alicki, G.~Barenboim and A.~Jenkins, \emph{{Quantum thermodynamics of de Sitter space}}, \href{https://doi.org/10.1103/PhysRevD.108.123530}{\emph{Phys. Rev. D} {\bfseries 108} (2023) 123530} [\href{https://arxiv.org/abs/2307.04800}{{\ttfamily 2307.04800}}].

\bibitem{deKruijf:2024ufs}
J.~de~Kruijf and N.~Bartolo, \emph{{The effect of quantum decoherence on inflationary gravitational waves}}, \href{https://doi.org/10.1088/1475-7516/2024/11/041}{\emph{JCAP} {\bfseries 11} (2024) 041} [\href{https://arxiv.org/abs/2408.02563}{{\ttfamily 2408.02563}}].

\bibitem{martin_observational_2018}
J.~Martin and V.~Vennin, \emph{Observational constraints on quantum decoherence during inflation}, \href{https://doi.org/10.1088/1475-7516/2018/05/063}{\emph{JCAP} {\bfseries 05} (2018) 063}.

\bibitem{martin_non_2018}
J.~Martin and V.~Vennin, \emph{Non {Gaussianities} from {Quantum} {Decoherence} during {Inflation}}, \href{https://doi.org/10.1088/1475-7516/2018/06/037}{\emph{JCAP} {\bfseries 06} (2018) 037}.

\bibitem{daddi_hammou_cosmic_2023}
A.~Daddi~Hammou and N.~Bartolo, \emph{Cosmic decoherence: primordial power spectra and non-{Gaussianities}}, \href{https://doi.org/10.1088/1475-7516/2023/04/055}{\emph{JCAP} {\bfseries 04} (2023) 055}.

\bibitem{Arkani-Hamed:2015bza}
N.~Arkani-Hamed and J.~Maldacena, \emph{{Cosmological Collider Physics}},  \href{https://arxiv.org/abs/1503.08043}{{\ttfamily 1503.08043}}.

\bibitem{Micheli:2022tld}
A.~Micheli and P.~Peter, \emph{{Quantum Cosmological Gravitational Waves?}},  \href{https://arxiv.org/abs/2211.00182}{{\ttfamily 2211.00182}}.

\bibitem{Maldacena:2015bha}
J.~Maldacena, \emph{{A model with cosmological Bell inequalities}}, \href{https://doi.org/10.1002/prop.201500097}{\emph{Fortsch. Phys.} {\bfseries 64} (2016) 10} [\href{https://arxiv.org/abs/1508.01082}{{\ttfamily 1508.01082}}].

\bibitem{Bhattacharyya:2024duw}
A.~Bhattacharyya, S.~Brahma, S.S.~Haque, J.S.~Lund and A.~Paul, \emph{{The early universe as an open quantum system: complexity and decoherence}}, \href{https://doi.org/10.1007/JHEP05(2024)058}{\emph{JHEP} {\bfseries 05} (2024) 058} [\href{https://arxiv.org/abs/2401.12134}{{\ttfamily 2401.12134}}].

\bibitem{Campo:2005qn}
D.~Campo and R.~Parentani, \emph{{Quantum correlations in inflationary spectra and violation of bell inequalities}}, \href{https://doi.org/10.1590/S0103-97332005000700016}{\emph{Braz. J. Phys.} {\bfseries 35} (2005) 1074} [\href{https://arxiv.org/abs/astro-ph/0510445}{{\ttfamily astro-ph/0510445}}].

\bibitem{martin_real-space_2021}
J.~Martin and V.~Vennin, \emph{Real-space entanglement in the {Cosmic} {Microwave} {Background}}, \href{https://doi.org/10.1088/1475-7516/2021/10/036}{\emph{Journal of Cosmology and Astroparticle Physics} {\bfseries 2021} (2021) 036}.

\bibitem{Tejerina-Perez:2024opu}
P.~Tejerina-P\'erez, D.~Bertacca and R.~Jimenez, \emph{{An Entangled Universe}},  \href{https://arxiv.org/abs/2403.15742}{{\ttfamily 2403.15742}}.

\bibitem{dePutter:2019xxv}
R.~de~Putter and O.~Dor\'e, \emph{In search of an observational quantum signature of the primordial perturbations in slow-roll and ultraslow-roll inflation}, \href{https://doi.org/10.1103/PhysRevD.101.043511}{\emph{Phys. Rev. D} {\bfseries 101} (2020) 043511} [\href{https://arxiv.org/abs/1905.01394}{{\ttfamily 1905.01394}}].

\bibitem{Martin:2021znx}
J.~Martin, A.~Micheli and V.~Vennin, \emph{{Discord and decoherence}}, \href{https://doi.org/10.1088/1475-7516/2022/04/051}{\emph{JCAP} {\bfseries 04} (2022) 051} [\href{https://arxiv.org/abs/2112.05037}{{\ttfamily 2112.05037}}].

\bibitem{Martin:2022kph}
J.~Martin, A.~Micheli and V.~Vennin, \emph{{Comparing quantumness criteria}}, \href{https://doi.org/10.1209/0295-5075/acc3be}{\emph{EPL} {\bfseries 142} (2023) 18001} [\href{https://arxiv.org/abs/2211.10114}{{\ttfamily 2211.10114}}].

\bibitem{Espinosa-Portales:2022yok}
L.~Espinosa-Portal\'es and V.~Vennin, \emph{{Real-space Bell inequalities in de~Sitter}}, \href{https://doi.org/10.1088/1475-7516/2022/07/037}{\emph{JCAP} {\bfseries 07} (2022) 037} [\href{https://arxiv.org/abs/2203.03505}{{\ttfamily 2203.03505}}].

\bibitem{gomez_quantum_2023}
C.~Gomez and R.~Jimenez, \emph{The quantum origin of quasi de {Sitter}: a model independent quantum cosmological tilt}, \href{https://doi.org/10.1088/1475-7516/2023/01/036}{\emph{JCAP} {\bfseries 01} (2023) 036}.

\bibitem{brahma_universal_2022}
S.~Brahma, A.~Berera and J.~Calder\`on-Figueroa, \emph{Universal signature of quantum entanglement across cosmological distances}, \href{https://doi.org/10.1088/1361-6382/aca066}{\emph{Classical and Quantum Gravity} {\bfseries 39} (2022) 245002}.

\bibitem{brahma_quantum_2022}
S.~Brahma, A.~Berera and J.~Calder\`on-Figueroa, \emph{Quantum corrections to the primordial tensor spectrum: open {EFTs} \& {Markovian} decoupling of {UV} modes}, \href{https://doi.org/10.1007/JHEP08(2022)225}{\emph{Journal of High Energy Physics} {\bfseries 2022} (2022) 225}.

\bibitem{boyanovsky_effective_2015}
D.~Boyanovsky, \emph{Effective field theory during inflation: {Reduced} density matrix and its quantum master equation}, \href{https://doi.org/10.1103/PhysRevD.92.023527}{\emph{Phys. Rev. D} {\bfseries 92} (2015) 023527}.

\bibitem{Weinberg:2005vy}
S.~Weinberg, \emph{{Quantum contributions to cosmological correlations}}, \href{https://doi.org/10.1103/PhysRevD.72.043514}{\emph{Phys. Rev. D} {\bfseries 72} (2005) 043514} [\href{https://arxiv.org/abs/hep-th/0506236}{{\ttfamily hep-th/0506236}}].

\bibitem{Seery:2010kh}
D.~Seery, \emph{{Infrared effects in inflationary correlation functions}}, \href{https://doi.org/10.1088/0264-9381/27/12/124005}{\emph{Class. Quant. Grav.} {\bfseries 27} (2010) 124005} [\href{https://arxiv.org/abs/1005.1649}{{\ttfamily 1005.1649}}].

\bibitem{Green:2020txs}
D.~Green and A.~Premkumar, \emph{{Dynamical RG and Critical Phenomena in de Sitter Space}}, \href{https://doi.org/10.1007/JHEP04(2020)064}{\emph{JHEP} {\bfseries 04} (2020) 064} [\href{https://arxiv.org/abs/2001.05974}{{\ttfamily 2001.05974}}].

\bibitem{Boyanovsky:1998aa}
D.~Boyanovsky, H.J.~de~Vega, R.~Holman and M.~Simionato, \emph{{Dynamical renormalization group resummation of finite temperature infrared divergences}}, \href{https://doi.org/10.1103/PhysRevD.60.065003}{\emph{Phys. Rev. D} {\bfseries 60} (1999) 065003} [\href{https://arxiv.org/abs/hep-ph/9809346}{{\ttfamily hep-ph/9809346}}].

\bibitem{Burgess:2009bs}
C.P.~Burgess, L.~Leblond, R.~Holman and S.~Shandera, \emph{{Super-Hubble de Sitter Fluctuations and the Dynamical RG}}, \href{https://doi.org/10.1088/1475-7516/2010/03/033}{\emph{JCAP} {\bfseries 03} (2010) 033} [\href{https://arxiv.org/abs/0912.1608}{{\ttfamily 0912.1608}}].

\bibitem{Boyanovsky:2015xoa}
D.~Boyanovsky, \emph{{Effective Field Theory out of Equilibrium: Brownian quantum fields}}, \href{https://doi.org/10.1088/1367-2630/17/6/063017}{\emph{New J. Phys.} {\bfseries 17} (2015) 063017} [\href{https://arxiv.org/abs/1503.00156}{{\ttfamily 1503.00156}}].

\bibitem{Brahma:2024yor}
S.~Brahma, J.~Calder\`on-Figueroa and X.~Luo, \emph{Time-convolutionless cosmological master equations: Late-time resummations and decoherence for non-local kernels},  \href{https://arxiv.org/abs/2407.12091}{{\ttfamily 2407.12091}}.

\bibitem{Breuer_2016}
H.-P.~Breuer, E.-M.~Laine, J.~Piilo and B.~Vacchini, \emph{Colloquium: Non-markovian dynamics in open quantum systems}, \href{https://doi.org/10.1103/revmodphys.88.021002}{\emph{Reviews of Modern Physics} {\bfseries 88} (2016) }.

\bibitem{Hall_2014}
M.J.W.~Hall, J.D.~Cresser, L.~Li and E.~Andersson, \emph{Canonical form of master equations and characterization of non-markovianity}, \href{https://doi.org/10.1103/physreva.89.042120}{\emph{Physical Review A} {\bfseries 89} (2014) }.

\bibitem{Kaplanek:2022xrr}
G.~Kaplanek and E.~Tjoa, \emph{{Effective master equations~for two accelerated qubits}}, \href{https://doi.org/10.1103/PhysRevA.107.012208}{\emph{Phys. Rev. A} {\bfseries 107} (2023) 012208} [\href{https://arxiv.org/abs/2207.13750}{{\ttfamily 2207.13750}}].

\bibitem{burgess_gravity_2023}
C.P.~Burgess and G.~Kaplanek, \emph{Gravity, {Horizons} and {Open} {EFTs}},  Jan., 2023.

\bibitem{Salcedo:2024smn}
S.A.~Salcedo, T.~Colas and E.~Pajer, \emph{{The open effective field theory of inflation}}, \href{https://doi.org/10.1007/JHEP10(2024)248}{\emph{JHEP} {\bfseries 10} (2024) 248} [\href{https://arxiv.org/abs/2404.15416}{{\ttfamily 2404.15416}}].

\bibitem{Kodama:1984ziu}
H.~Kodama and M.~Sasaki, \emph{{Cosmological Perturbation Theory}}, \href{https://doi.org/10.1143/PTPS.78.1}{\emph{Prog. Theor. Phys. Suppl.} {\bfseries 78} (1984) 1}.

\bibitem{Ford:1977dj}
L.H.~Ford and L.~Parker, \emph{{Quantized Gravitational Wave Perturbations in Robertson-Walker Universes}}, \href{https://doi.org/10.1103/PhysRevD.16.1601}{\emph{Phys. Rev. D} {\bfseries 16} (1977) 1601}.

\bibitem{ning_decoherence_2023}
S.~Ning, C.M.~Sou and Y.~Wang, \emph{On the decoherence of primordial gravitons}, \href{https://doi.org/10.1007/JHEP06(2023)101}{\emph{Journal of High Energy Physics} {\bfseries 2023} (2023) 101}.

\bibitem{Hollowood:2017bil}
T.J.~Hollowood and J.I.~McDonald, \emph{{Decoherence, discord and the quantum master equation for cosmological perturbations}}, \href{https://doi.org/10.1103/PhysRevD.95.103521}{\emph{Phys. Rev. D} {\bfseries 95} (2017) 103521} [\href{https://arxiv.org/abs/1701.02235}{{\ttfamily 1701.02235}}].

\bibitem{Colas:2023wxa}
T.~Colas, \emph{Open Effective Field Theories for primordial cosmology : dissipation, decoherence and late-time resummation of cosmological inhomogeneities}, Ph.D. thesis, Institut d'astrophysique spatiale, France, AstroParticule et Cosmologie, France, APC, Paris, 2023.

\bibitem{albrecht_cosmological_2014}
A.~Albrecht, N.~Bolis and R.~Holman, \emph{Cosmological {Consequences} of {Initial} {State} {Entanglement}}, \href{https://doi.org/10.1007/JHEP11(2014)093}{\emph{JHEP} {\bfseries 11} (2014) 093}.

\bibitem{Tasaki_2007}
S.~Tasaki, K.~Yuasa, P.~Facchi, G.~Kimura, H.~Nakazato, I.~Ohba et~al., \emph{On the assumption of initial factorization in the master equation for weakly coupled systems i: General framework}, \href{https://doi.org/10.1016/j.aop.2006.06.004}{\emph{Annals of Physics} {\bfseries 322} (2007) 631–656}.

\bibitem{Lindblad:1975ef}
G.~Lindblad, \emph{{On the Generators of Quantum Dynamical Semigroups}}, \href{https://doi.org/10.1007/BF01608499}{\emph{Commun. Math. Phys.} {\bfseries 48} (1976) 119}.

\bibitem{Kossakowski:1972sbn}
A.~Kossakowski, \emph{{On quantum statistical mechanics of non-Hamiltonian systems}}, \href{https://doi.org/10.1016/0034-4877(72)90010-9}{\emph{Rept. Math. Phys.} {\bfseries 3} (1972) 247}.

\bibitem{Gorini:1975nb}
V.~Gorini, A.~Kossakowski and E.C.G.~Sudarshan, \emph{{Completely Positive Dynamical Semigroups of N Level Systems}}, \href{https://doi.org/10.1063/1.522979}{\emph{J. Math. Phys.} {\bfseries 17} (1976) 821}.

\bibitem{Rivas_2014}
A.~Rivas, S.F.~Huelga and M.B.~Plenio, \emph{Quantum non-markovianity: characterization, quantification and detection}, \href{https://doi.org/10.1088/0034-4885/77/9/094001}{\emph{Reports on Progress in Physics} {\bfseries 77} (2014) 094001}.

\bibitem{Lidar:2019qog}
D.A.~Lidar, \emph{{Lecture Notes on the Theory of Open Quantum Systems}},  \href{https://arxiv.org/abs/1902.00967}{{\ttfamily 1902.00967}}.

\bibitem{colas_quantum_2023}
T.~Colas, J.~Grain and V.~Vennin, \emph{Quantum recoherence in the early universe}, \href{https://doi.org/10.1209/0295-5075/acdd94}{\emph{Europhysics Letters} {\bfseries 142} (2023) 69002}.

\bibitem{Danielson:2022sga}
D.L.~Danielson, G.~Satishchandran and R.M.~Wald, \emph{{Killing horizons decohere quantum superpositions}}, \href{https://doi.org/10.1103/PhysRevD.108.025007}{\emph{Phys. Rev. D} {\bfseries 108} (2023) 025007} [\href{https://arxiv.org/abs/2301.00026}{{\ttfamily 2301.00026}}].

\bibitem{Frob:2012ui}
M.B.~Fr\"ob, A.~Roura and E.~Verdaguer, \emph{{One-loop gravitational wave spectrum in de Sitter spacetime}}, \href{https://doi.org/10.1088/1475-7516/2012/08/009}{\emph{JCAP} {\bfseries 08} (2012) 009} [\href{https://arxiv.org/abs/1205.3097}{{\ttfamily 1205.3097}}].

\bibitem{Whitney_2008}
R.S.~Whitney, \emph{Staying positive: going beyond lindblad with perturbative master equations}, \href{https://doi.org/10.1088/1751-8113/41/17/175304}{\emph{Journal of Physics A: Mathematical and Theoretical} {\bfseries 41} (2008) 175304}.

\bibitem{burgess_eft_2015}
C.P.~Burgess, R.~Holman, G.~Tasinato and M.~Williams, \emph{{EFT} {Beyond} the {Horizon}: {Stochastic} {Inflation} and {How} {Primordial} {Quantum} {Fluctuations} {Go} {Classical}}, \href{https://doi.org/10.1007/JHEP03(2015)090}{\emph{JHEP} {\bfseries 03} (2015) 090}.

\bibitem{Galloni:2022mok}
G.~Galloni, N.~Bartolo, S.~Matarrese, M.~Migliaccio, A.~Ricciardone and N.~Vittorio, \emph{{Updated constraints on amplitude and tilt of the tensor primordial spectrum}}, \href{https://doi.org/10.1088/1475-7516/2023/04/062}{\emph{JCAP} {\bfseries 04} (2023) 062} [\href{https://arxiv.org/abs/2208.00188}{{\ttfamily 2208.00188}}].

\bibitem{Lidar_2001}
D.A.~Lidar, Z.~Bihary and K.~Whaley, \emph{From completely positive maps to the quantum markovian semigroup master equation}, \href{https://doi.org/10.1016/s0301-0104(01)00330-5}{\emph{Chemical Physics} {\bfseries 268} (2001) 35–53}.

\bibitem{Agon:2014uxa}
C.~Agon, V.~Balasubramanian, S.~Kasko and A.~Lawrence, \emph{{Coarse Grained Quantum Dynamics}}, \href{https://doi.org/10.1103/PhysRevD.98.025019}{\emph{Phys. Rev. D} {\bfseries 98} (2018) 025019} [\href{https://arxiv.org/abs/1412.3148}{{\ttfamily 1412.3148}}].

\bibitem{Burgess:2024heo}
C.P.~Burgess, T.~Colas, R.~Holman and G.~Kaplanek, \emph{{Does decoherence violate decoupling?}},  \href{https://arxiv.org/abs/2411.09000}{{\ttfamily 2411.09000}}.

\bibitem{Serafini:2003ke}
A.~Serafini, F.~Illuminati and S.~De~Siena, \emph{{Von Neumann entropy, mutual information and total correlations of Gaussian states}}, \href{https://doi.org/10.1088/0953-4075/37/2/L02}{\emph{J. Phys. B} {\bfseries 37} (2004) L21} [\href{https://arxiv.org/abs/quant-ph/0307073}{{\ttfamily quant-ph/0307073}}].

\bibitem{Martin:2012pea}
J.~Martin, V.~Vennin and P.~Peter, \emph{{Cosmological Inflation and the Quantum Measurement Problem}}, \href{https://doi.org/10.1103/PhysRevD.86.103524}{\emph{Phys. Rev. D} {\bfseries 86} (2012) 103524} [\href{https://arxiv.org/abs/1207.2086}{{\ttfamily 1207.2086}}].

\end{thebibliography}\endgroup

\end{document}